\documentclass[aps,preprint,floats,nofootinbib]{revtex4}
\usepackage{graphicx}

\begin{document}

\preprint{
\hfill
\begin{minipage}[t]{3in}
\begin{flushright}
\vspace{-0.2in}
FERMILAB--PUB--06--212--A
\end{flushright}
\end{minipage}
}
\bigskip

\title{The intergalactic propagation of ultra-high energy cosmic ray nuclei\\}
\author{Dan Hooper$^{1}$, Subir Sarkar$^2$ and Andrew M.~Taylor$^3$}
\address{
$^1$Theoretical Astrophysics Group, Fermilab, Batavia, IL  60510, USA\\
$^2$ Rudolf Peierls Centre for Theoretical Physics, University of Oxford, 
     1 Keble Road, Oxford OX1 3NP, UK\\
$^3$ Astrophysics, University of Oxford, Denys Wilkinson Building,
     Keble Road, Oxford OX1 3RH, UK}
\date{\today}

\bigskip

\begin{abstract}
We investigate the propagation of ultra-high energy cosmic ray nuclei
($A=1-56$) from cosmologically distant sources through the cosmic
radiation backgrounds. Various models for the injected composition and
spectrum and of the cosmic infrared background are studied using
updated photodisintegration cross-sections. The observational data on
the spectrum and the composition of ultra-high energy cosmic rays are
jointly consistent with a model where all of the injected primary
cosmic rays are iron nuclei (or a mixture of heavy and light nuclei).
\end{abstract}

\maketitle

\section{Introduction}
The origin of the highest energy cosmic rays is among the most
pressing questions in astroparticle physics
\cite{Nagano:ve,Anchordoqui:2002hs}. The ultra-high energy cosmic ray
(UHECR) spectrum has been measured, using both air shower and
atmospheric fluorescence techniques, to energies beyond $10^{20}$~eV
\cite{Lawrence:1991cc,Bird:1994uy,Takeda:1998ps,Abu-Zayyad:2002sf}. Protons
of such high energies are not expected to be deflected significantly
by galactic or extragalactic magnetic fields \cite{Dolag:2004kp}, but
should scatter inelastically on the cosmic microwave background (CMB)
with an attenuation length of $\lesssim 30$ Mpc, i.e. about the size
of the local supercluster of galaxies. The observed UHECRs do not
point back to any plausible sources within this range and have an
isotropic sky distribution, so their sources, if astrophysical, must
be cosmologically distant. If so the spectrum should be suppressed at
energies above $E_{\rm GZK} \sim 6\times 10^{19}$~eV (the
``Greisen-Zatsepin-Kuzmin cutoff''), if the primaries are indeed
protons \cite{Greisen:1966jv,Zatsepin:1966jv}. Presently, there are
conflicting claims concerning the existence of this spectral
feature. Data from AGASA (Akeno Giant Air Shower Array) show no
indication of any suppression \cite{Takeda:1998ps}, while data from
the HiRes (High Resolution Fly's Eye) air fluorescence detectors are
consistent with the expected cutoff \cite{Abu-Zayyad:2002sf}. The
Pierre Auger Observatory, which employs both techniques, has
accumulated an exposure comparable to AGASA and about half that of
HiRes. Its results are presently consistent with either possibility
\cite{Sommers:2005vs} and forthcoming data should be able to settle
the issue definitively. The AGASA observation that the spectrum
continues smoothly beyond the GZK cutoff has motivated many
suggestions for the origin of the trans-GZK events such as decaying
superheavy dark matter particles in the Galactic halo
\cite{Berezinsky:1997hy,Birkel:1998nx} and ultra-high energy neutrinos
creating `Z-bursts' in local interactions with the cosmic neutrino
background \cite{Weiler:1997sh,Fargion:1997ft}. It has also been
suggested that the primaries may be neutrinos with large interaction
cross-sections at ultra-high energies
\cite{Domokos:1998ry,Jain:2000pu}, or perhaps new stable strongly
interacting particles heavier than the proton
\cite{Albuquerque:1998va}.

Rather less exotic ways of evading the GZK cutoff are also
possible. In particular, if UHECRs are not protons but consist of
heavy nuclei \cite{heavycase}, then the spectrum will be altered from
the usual expectation. Although evidence for the
presence of heavy nuclei in UHECRs has been around for quite some time
\cite{Bird:1993yi,yakutsk,earlyevidence}, the information available on UHECR
composition is still rather imprecise~\cite{Watson:2004ew}, the
most reliable result being that the primaries are not mostly photons
but hadrons
\cite{Halzen:1994gy,Shinozaki:2002ve,Ave:2000nd,Risse:2005hi}. Despite
the widespread impression to the contrary, however, there is little
reason to believe that these particles are protons rather than heavier
nuclei.

Determination of the mass composition of UHECRs has been attempted
using a variety of variables such as the rate of change with primary
energy of the depth of shower maximum $X_\mathrm{max}$, the degree of
fluctuation in the depth of the shower maximum, the number of muons
which reach the Earth's surface, the lateral density profile with
respect to the shower axis, the width of cosmic ray shower disks,
etc. Such measurements must be compared to Monte Carlo simulations,
typically conducted with programs such as QGSJET, SIBYLL and DPMJET
\cite{Knapp:2002vs}.  The mass composition inferred is quite sensitive
to the simulation program used. Furthermore, data from different
experiments can yield rather different results regarding cosmic ray
composition even when the same shower simulation is used. For example,
at energies near $10^{18}$~eV, the mass composition determined from
the lateral density profile of Volcano Ranch and Haverah Park data are
in disagreement at $>2\sigma$ significance \cite{Dova:2004nq}. The
situation does not improve much at higher energies. Above
$10^{19}$~eV, only the Fly's Eye/HiRes and AGASA experiments have
reported any information pertaining to mass composition, and these
results extend to only about $2.5 \times 10^{19}$~eV --- in this
energy range the data suggest a predominantly light composition
although the uncertainties are large \cite{Abbasi:2004nz}.

So far no individual sources of UHECRs have been identified. This can
be understood if there are a very large number of faint sources or,
alternatively, if there are large-scale magnetic fields which deflect
UHECRs sufficiently so as to to conceal their origin. As mentioned
above, protons at trans-GZK energies are not expected to be
significantly deflected \cite{Dolag:2004kp} (but see
ref.\cite{Sigl:2003ay}). Heavy nuclei, by contrast, have greater
electric charge and may thus have their arrival directions
isotropised, even at such high energies.

There is also a theoretical prejudice to expect the UHECR spectrum to
be dominated by heavy nuclei, since the maximum energy to which
particles can be accelerated scales with the electric charge. Thus
astrophysical candidates for UHECR accelerators such as active
galactic nuclei and gamma-ray bursts, which barely meet the ``Hillas
criterion''~\cite{hillas} for containment and acceleration of
$10^{20}$~eV protons, can in principle easily accelerate heavy nuclei
to such energies.

Given these arguments, and the inconclusive nature of the experimental
data regarding UHECR composition, it is rather surprising that most
studies in this field have usually {\em assumed} that UHECRs are
protons. By contrast, relatively little attention has been paid to the
propagation of ultra-high energy nuclei
\cite{Anchordoqui:1997rn,Epele:1998ia,Stecker:1998ib} although there
has been a resurgence of interest in recent years
\cite{Bertone:2002ks,Yamamoto:2003tn,Khan:2004nd,Armengaud:2004yt,Allard:2005ha,Sigl:2005md,Allard:2005cx}. We
revisit this problem, using up-to-date nuclear physics data, and
considering a wide range of nuclei as possible UHECR primaries. We
also explore different models for the cosmic infrared background
(CIB), as well as the effects of possible intergalactic magnetic
fields. We pay particular attention to the relationship between the
composition of UHECRs at Earth and that injected at source.

We begin by considering the intergalactic propagation of ultra-high
energy (UHE) protons (Section~\ref{protons}). In
Sections~\ref{photodis}, \ref{irb} and \ref{prop}, we study the
photodisintegration of cosmic ray nuclei, including the effect of
uncertainties in the cross-sections and the CIB model used. We then
proceed to discuss the composition of UHECRs observed at Earth
(Section~\ref{comp}), and the role of magnetic fields in the
propagation of cosmic ray nuclei (Section~\ref{bfields}). We propose
some representative models of UHECR primary composition and discuss
$X_\mathrm{max}$ measurements in Sections~\ref{mixedsec} and
\ref{X_{max}}. Our conclusions are presented in Section~\ref{conc}.

\section{Propagation of Ultra-High Energy Protons}
\label{protons}

Before discussing the propagation of intermediate mass and heavy
nuclei, we first recapitulate the physics governing the propagation of
UHE protons.

\begin{figure}[tbh]
\centering\leavevmode
\mbox{
\includegraphics[width=2.2in,angle=-90]{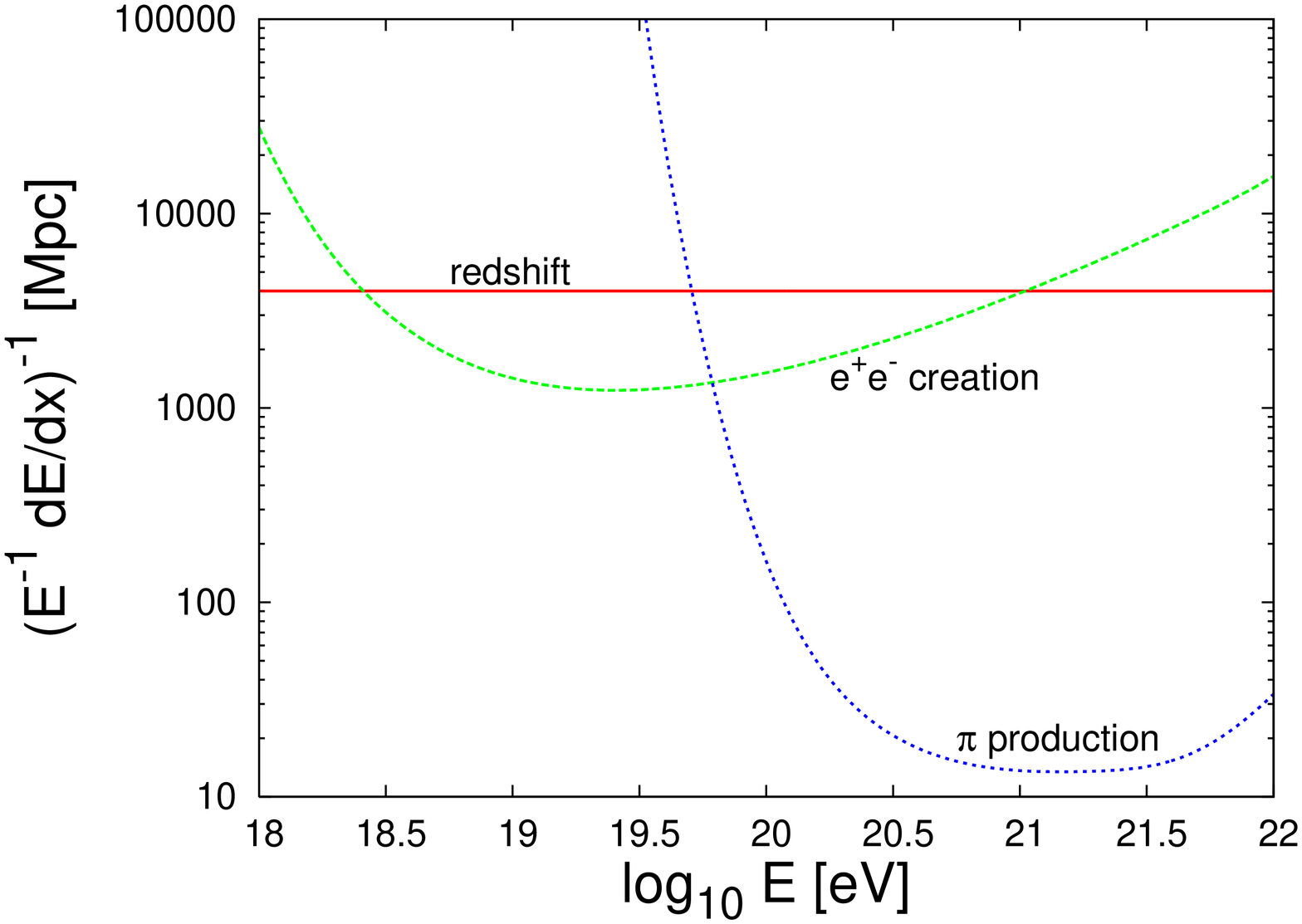}
\includegraphics[width=2.2in,angle=-90]{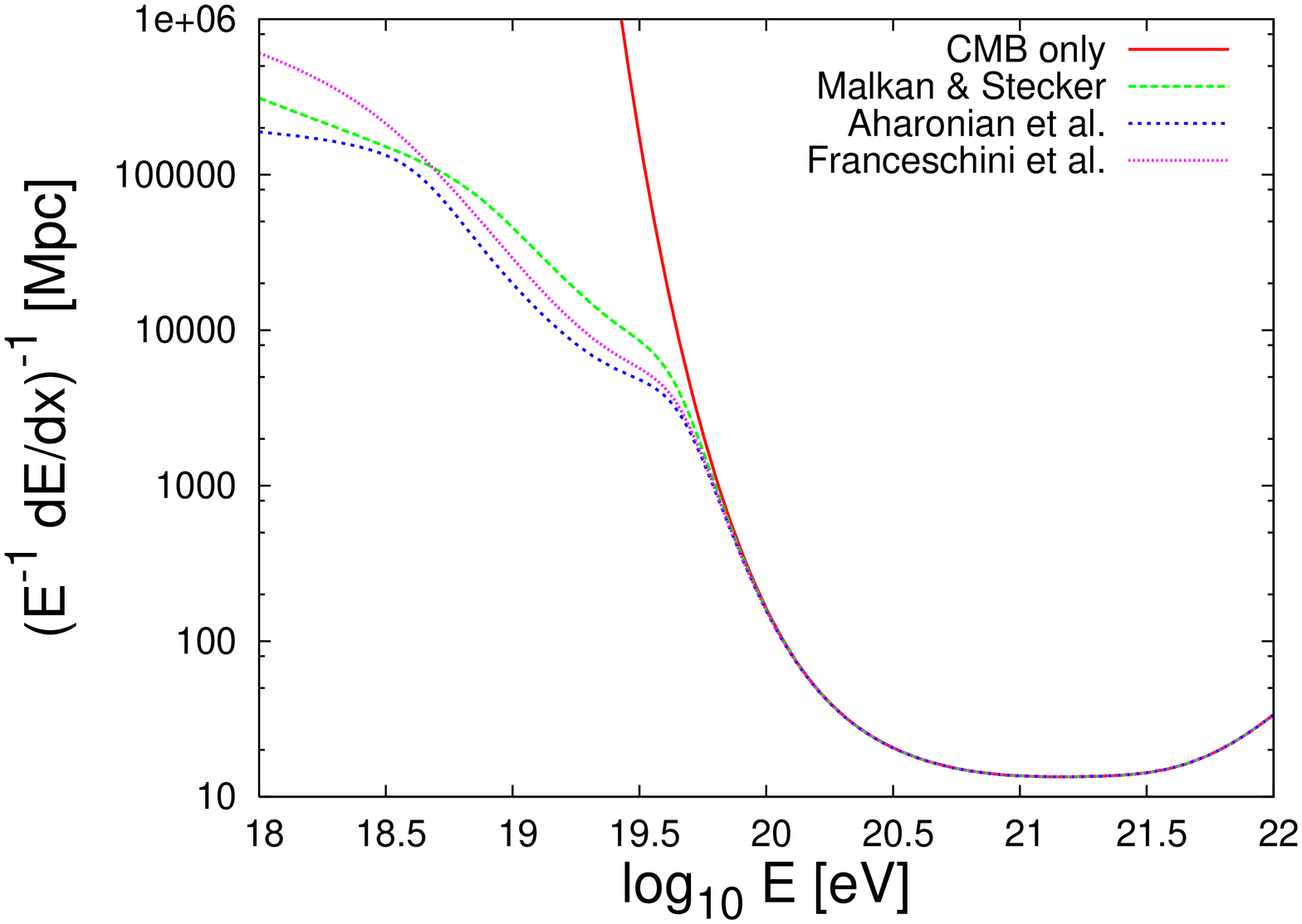}
}
\caption{Energy loss lengths for UHE protons propagating through the
universe. The left frame shows the results for the processes $p +
\gamma_{\rm CMB} \rightarrow p + e^+ + e^-$ (``$e^+ e^-$ creation''),
$p + \gamma_{\rm CMB} \rightarrow p + \pi^0$, $p + \gamma_{\rm CMB}
\rightarrow n + \pi^+$ and multi-pion production (``$\pi$
production''), as well as due to the Hubble expansion
(``redshift''). In the right frame, only pion production processes are
considered and the effects of the cosmic infrared background (CIB) are
shown in addition to the CMB (the three CIB models are discussed in
Section~\ref{irb}).}
\label{protonrates}
\end{figure}

Over cosmological distances, the dominant processes effecting UHE
proton propagation are interactions with the CMB producing pions or
electron-positron pairs. Pair production ($p +\gamma_{\rm CMB}
\rightarrow p + e^+ + e^-$) occurs sufficiently frequently that it can
be treated as a continuous energy loss process \cite{Blumenthal}; as
shown in Figure~\ref{protonrates} this effect dominates over the
energy loss due to the Hubble expansion for proton energies $\sim
10^{18}-10^{21}$~eV and peaks at $\sim 3 \times10^{19}$~eV.

%%%%%%%%%%%%%%%%%%%%%%%%%%%%%%%%%%%%%
%%%%%%%%%%%%%%%%%%%%%%%%%%%%%%%%%%%%%%%%%%%%%%%%%%

Pion production is not as simple to implement. Individual occurrences
of the processes $p + \gamma_{\rm CMB} \rightarrow p +\pi^0$ and $p +
\gamma_{\rm CMB} \rightarrow n + \pi^+$ cause the primary proton to
lose a considerable fraction of its energy. Therefore these cannot be
treated as a continuous processes and it becomes necessary to use
Monte Carlo techniques \cite{Mucke:1999yb}. Furthermore, if enough
energy is available, multi-pion production can be important; these
non-perturbative processes take place through the near-resonance
exchange of the 1.232~GeV $\Delta^+$-hadron \cite{pionsigma}. The
associated energy loss lengths are shown in Figure~\ref{protonrates}.

It is also possible for UHE protons to produce pions through
interactions with the cosmic infrared background (CIB) as shown in the
right frame of Figure~\ref{protonrates}. We show results for three
models of the CIB which are discussed further in
Section~\ref{irb}. Although the rates for these interactions are
sub-dominant in comparison to energy losses from pair production, they
can be important in determining the spectrum of ``cosmogenic"
neutrinos produced in UHE proton propagation \cite{Stanev:2004kz}.

%%%%%%%%%%%%%%%%%%%%%%%%%%%%%%%%%%%%%
\begin{figure}[tbh]
\centering\leavevmode
\mbox{
\includegraphics[width=3.2in,angle=-90]{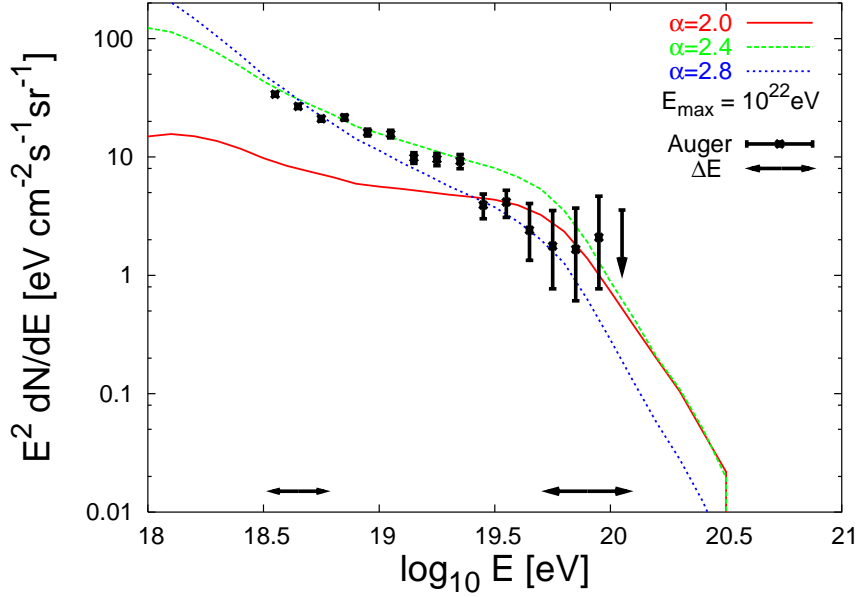}
}
\caption{The UHECR spectrum at Earth for purely proton primaries
injected by homogeneously distributed extragalactic sources with power
law spectral indices of $\alpha =$ 2.0, 2.4 and 2.8 up to a maximum
energy of $(E_{\rm max}/26) = (10^{22}$ eV/26) $\approx 10^{20.5}$
eV. The overall flux has in each case been normalized to the Auger
data \cite{Sommers:2005vs} (the $\chi^2$/d.o.f. for the fits are 9.97,
4.36, and 1.49 respectively). The effects of magnetic fields are {\em
not} included.}
\label{protonspec}
\end{figure}
%%%%%%%%%%%%%%%%%%%%%%%%%%%%%%%%%%%%%%%%%%%%%%%%%%

For the injection spectrum, we follow previous work in adopting a
simple power law with a cutoff:
\begin{equation}
\frac{d N_p}{d E_p} \propto E_p^{-\alpha}, \qquad E_p < (E_{\rm max}/26).
\end{equation}
Later in this paper, the quantity $E_\mathrm{max}$ will be used to
describe the maximum energy to which iron nuclei can be accelerated by
astrophysical sources. Since this maximum energy scales with the
electric charge of the accelerated particle, protons have a cutoff
energy 26 times smaller.

We also assume a constant comoving density of sources:
\begin{equation}
\frac{d N}{d V} \propto (1 + z)^3.
\end{equation}
The possible effects of source number evolution has also been
considered, typically by adopting a distribution proportional to $(1 +
z)^4$. We find however that this has a negligible impact as UHECRs
typically propagate over distances of only 10-100 Mpc (i.e. $z \ll
1$). In Figure~\ref{protonspec}, we show the UHECR spectrum expected
at Earth for proton primaries --- the ``GZK cutoff'' is clearly
seen. It is also evident that none of these models fit the data well.

\section{photodisintegration of Ultra-High Energy Nuclei}
\label{photodis}

The propagation of UHECRs is quite different in the case of heavy
nuclei. These undergo photodisintegration in scattering off the CMB
and/or CIB at a rate:
\begin{equation}
R_{A, Z, i_p, i_n} = \frac{A^2 m^2_p c^2}{2 E^2} \int^{\infty}_{0} 
\frac{d \epsilon\, n (\epsilon)}{\epsilon^2} \int^{2 E \epsilon/A m_p c}_{0} 
d \epsilon^\prime \epsilon^\prime \sigma_{A, Z, i_p, i_n} (\epsilon^\prime),
\end{equation}
where $A$ and $Z$ are the atomic number and charge of the nucleus,
$i_p$ and $i_n$ are the numbers of protons and neutrons broken off
from a nucleus in the interaction, $n (\epsilon)$ is the density of
background photons of energy $\epsilon$, and $\sigma_{A, Z, i_p,
i_n}(\epsilon^\prime)$ is the appropriate cross-section. (Note that
$\epsilon$ is measured here in the laboratory rest frame, but in
subsequent equations it refers to the energy measured in the rest
frame of the nucleus.)

The cross-sections for photodisintegration have often been modelled
using the parameterization of Stecker and collaborators
\cite{Stecker:1969fw,Puget:1976nz,Stecker:1998ib}:
\begin{equation} 
\sigma_{A,i}(\epsilon)=\left\{\begin{array}{l@{\quad}l}
\xi_{i}\Sigma_{\rm d}
W_{i}^{-1}e^{-2(\epsilon-\epsilon_{p,i})^{2}/\Delta_{i}^{2}}
\Theta_{+}(\epsilon_{\rm thr})\Theta_{-}
(\epsilon_{1}), & \epsilon_{\rm thr}\le \epsilon 
\le \epsilon_{1},\quad i=1,2 \\
\zeta\Sigma_{\rm d}
\Theta_{+}(\epsilon_{\rm max})\Theta_{-}(\epsilon_{1})/
(\epsilon_{\rm max}-\epsilon_{1}), &  \epsilon_{1}< \epsilon \le
\epsilon_{\rm max} \\
0, & \epsilon> \epsilon_{\rm max} 
\end{array} \right.
\end{equation}
where $\xi_{i}$, $\zeta$, $\epsilon_{p,i}$ and $\Delta_i$ are
parameters whose values are obtained by fitting to nuclear data.
Here, $i$ is the total number of nucleons broken off from the nucleus
in the interaction and the integrated cross-section is
\begin{equation}
\Sigma_{\rm d} \equiv \int_{0}^{\infty} \sigma(\epsilon)\,d\epsilon 
= \frac{2\pi^2 e^2 \hbar}{m_p c}\frac{(A-Z)Z}{A} = 60 \frac{(A-Z)Z}{A}
\mbox{mb-MeV},
\end{equation}
while the function $W_i$ is given by
\begin{equation}
W_i = \Delta_i \sqrt{\frac{\pi}{8}}
\left[\mbox{erf}\left(\frac{\epsilon_{\rm max} - \epsilon_{p,i}}
{\Delta_i/\sqrt{2}}\right) + \mbox{erf}\left(\frac{\epsilon_{p,i}-\epsilon_{1}}
{\Delta_{i}/\sqrt{2}}\right)\right].
\end{equation}
Here $\Theta_{+}(x)$ and $\Theta_{-}(x)$ are the Heaviside step
functions, $\epsilon_1$ = 30 MeV, $\epsilon_{\rm max}$ = 150 MeV, and
the threshold energy for a given process is in most cases
$\epsilon_{\rm thr} \approx i \times 10$ MeV (values are tabulated in
Ref.~\cite{Stecker:1998ib}). These cross-sections are dominated by the
giant dipole resonance which peaks in the energy range $\sim 10-30$
MeV; at higher energies, quasi-deuteron emission is the main process.

This model incorporates two major approximations. First, a simple
Gaussian form is assumed for the photodisintegration cross-section,
cut off abruptly below the theoretical reaction threshold. Second, a
fixed choice of $Z$ is made for each given atomic number,$A$, thus
neglecting the many possible isotopes which may be generated in such
interactions.

Although this parametrization does a fairly good job of reproducing
the measured cross-sections, possible improvements have been
proposed. In particular, a generalized Lorentzian model, which can be
fitted to the available nuclear data, has been used to parametrize a
wide range of photodisintegration cross-sections \cite{Khan:2004nd}:
\begin{eqnarray}
\sigma_{A,Z,i_p,i_n} (\epsilon_\gamma) = 
\frac{\Gamma_{A,Z,i_p,i_n} (\epsilon_\gamma) \, 
\epsilon_\gamma^4}{(\epsilon_\gamma^2 - 
E_{A,Z,i_p,i_n}^2)^2 + \Gamma_{A,Z,i_p,i_n}^2 
(\epsilon_{\gamma})\,\epsilon_\gamma^2}.
\end{eqnarray}
Here $E_{A,Z,i_p,i_n}$ is the position of the giant dipole resonance
and $\Gamma_{A,Z,i_p,i_n}$ is the width of that resonance, given by
\begin{eqnarray}
\Gamma_{A,Z,i_p,i_n} (\epsilon_{\gamma})=\Gamma_{A,Z,i_p,i_n}
(E_{A,Z,i_p,i_n})\, \frac{\epsilon_\gamma^2}{E_{A,Z,i_p,i_n}^2}.
\end{eqnarray}

%%%%%%%%%%%%%%%%%%%%%%%%%%%%%%%%%%%%%
\begin{figure}[t]
\centering\leavevmode
\mbox{
\includegraphics[width=2.2in,angle=-90]{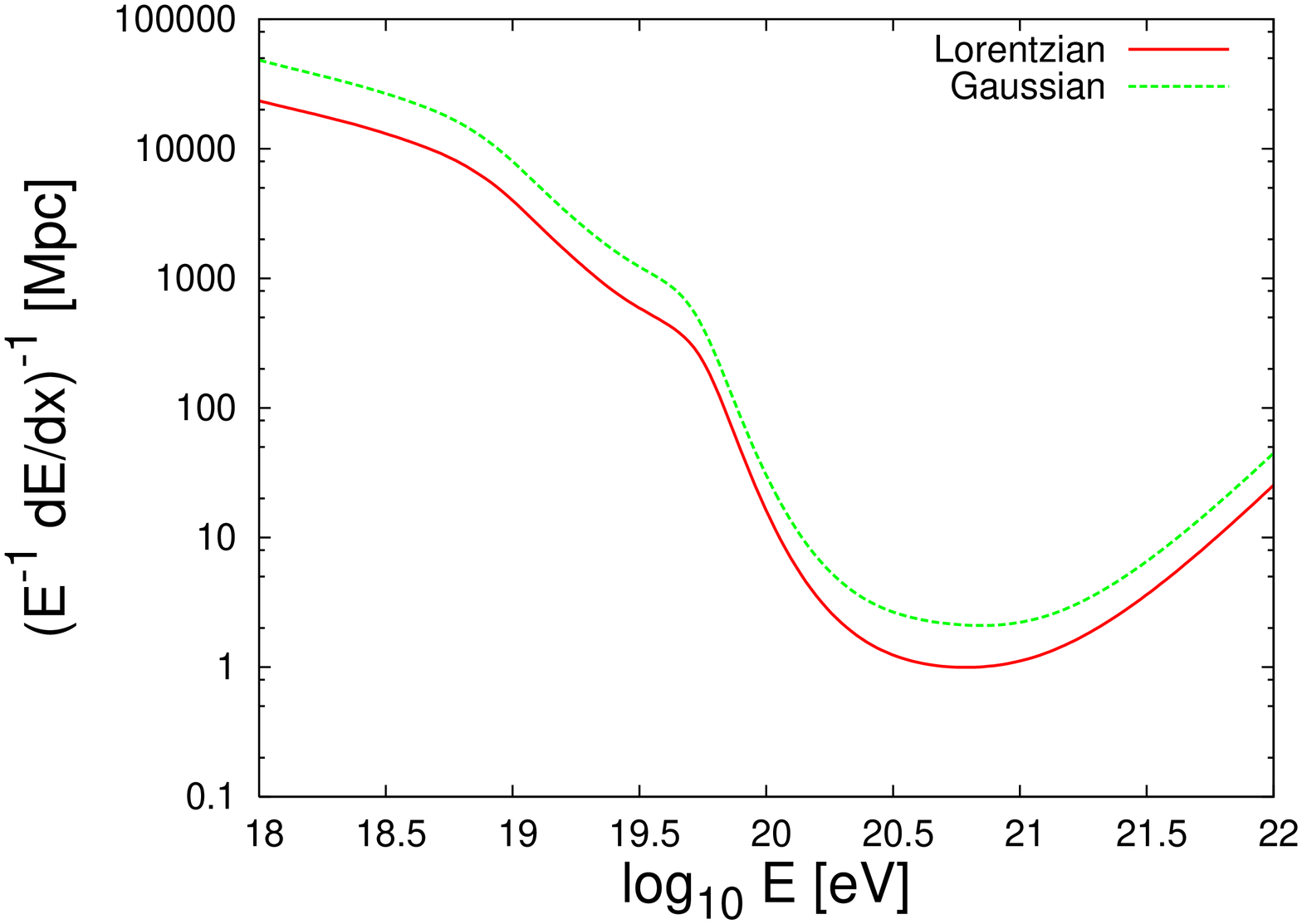}
\includegraphics[width=2.2in,angle=-90]{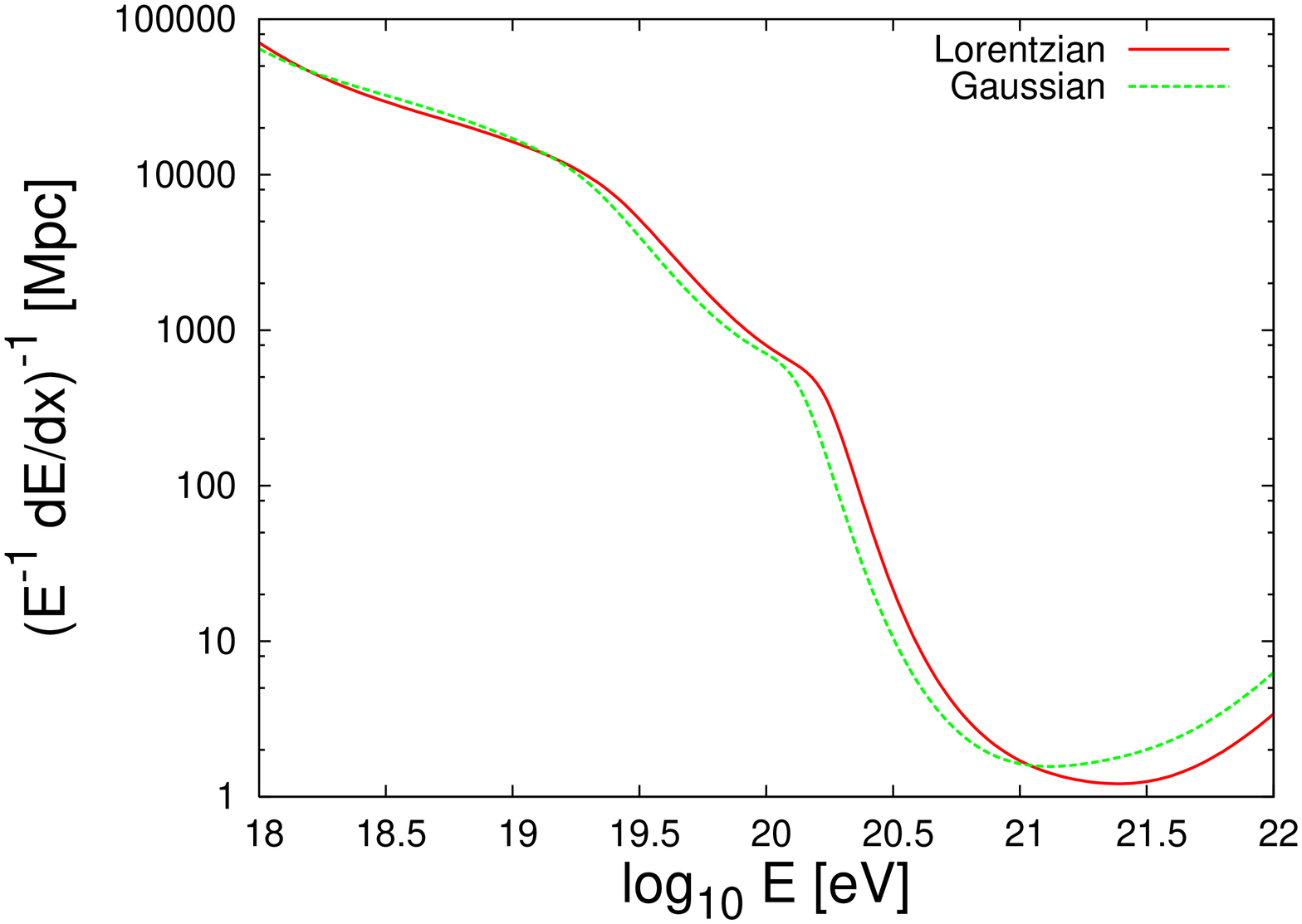}
}
\caption{Energy loss lengths for photodisintegration of oxygen (left)
and iron (right) nuclei on the CMB and CIB for both Gaussian and
Lorentzian parameterizations of the cross-section. The CIB has been
modelled according to Malkan \& Stecker \cite{Malkan:2000gu} (see
Section~\ref{irb}).}
\label{sigmavary}
\end{figure}
%%%%%%%%%%%%%%%%%%%%%%%%%%%%%%%%%%%%%%%%%%%%%%%%%%

In some cases, the Lorentzian form can be quite different from the
Gaussian parameterization. In particular, the latter often
overestimates the width of the giant dipole resonance. Despite these
differences we find, as shown in Figure~\ref{sigmavary}, that the
energy loss rates due to photodisintegration change little whether we
use the Gaussian model \cite{Stecker:1998ib} or the Lorentzian model
\cite{Khan:2004nd} for nuclei between $A = 11$ and 56 --- henceforth
we adopt the Lorentzian model. (Below this mass range, we use the
Gaussian parameterization.)

\section{The Cosmic Infrared Background}
\label{irb}
Since photodisintegration processes occur most efficiently when the
Lorentz-boosted target photon can excite the giant dipole resonance at
$\sim 10$ MeV, an UHE nucleus with an energy of $10^{19}$ eV (i.e. a
Lorentz factor of $\sim 10^8-10^9$) needs to collide with a $\sim
0.01-0.1$ eV background photon in order to most efficiently undergo
photodisintegration. Photons of this energy, corresponding to
wavelengths of $\lambda \sim 10-100 \, \mu$m, are present in the CIB
rather than in the CMB.

%%%%%%%%%%%%%%%%%%%%%%%%%%%%%%%%%%%%%
\begin{figure}[t]
\centering\leavevmode
\mbox{
\includegraphics[width=3.2in,angle=-90]{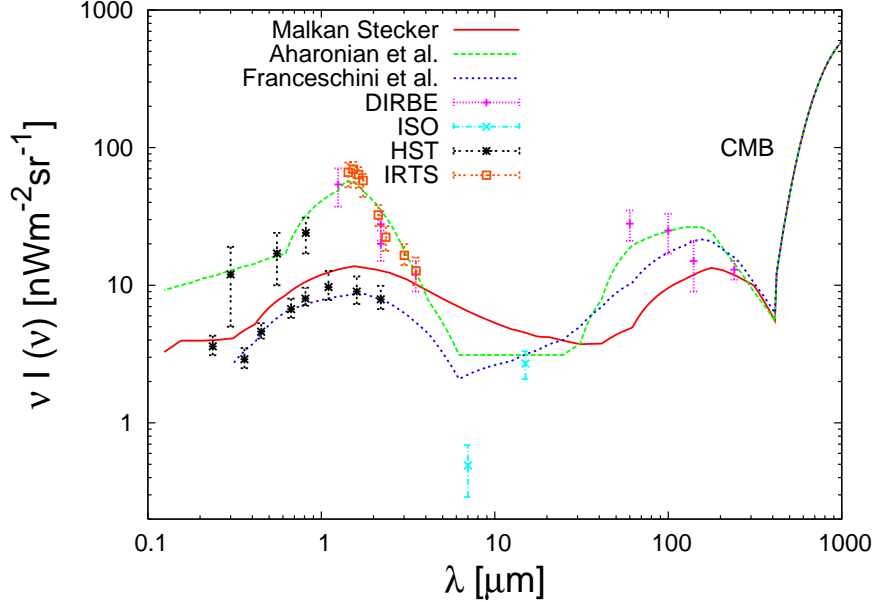}
}
\caption{Representative models of the CIB spectrum from Malkan \&
Stecker \cite{Malkan:2000gu}, Aharonian {\em et al.}
\cite{Aharonian:2003wu} and Franceschini {\em et al.}
\cite{Franceschini:2001yg}. Data shown are from
DIRBE~\cite{Hauser:2001xs}, ISO~\cite{Metcalfe:2003zi},
HST~\cite{Madau:1999yh,Bernstein:2001rz,Gard} and IRTS~\cite{irts}.}
\label{irbplot}
\end{figure}

The CIB is an expected relic of the cosmological structure formation
processes \cite{Hauser:2001xs}. The assembly of baryonic matter into
stars and galaxies and the subsequent evolution of such systems is
accompanied by the release of radiant energy; cosmic expansion and the
absorption of short wavelength radiation by dust and re-emission at
longer wavelengths shifts a significant part of this radiant energy
into the infrared: $\lambda \sim$ 1-1000 $\mu$m. Thus the CIB spectrum
depends on the luminosity and evolution of the sources and
distribution of dust from which it is scattered.

%%%%%%%%%%%%%%%%%%%%%%%%%%%%%%%%%%%%%
\begin{figure}[t]
%\centering\leavevmode
%\mbox{
\includegraphics[width=2.2in,angle=-90]{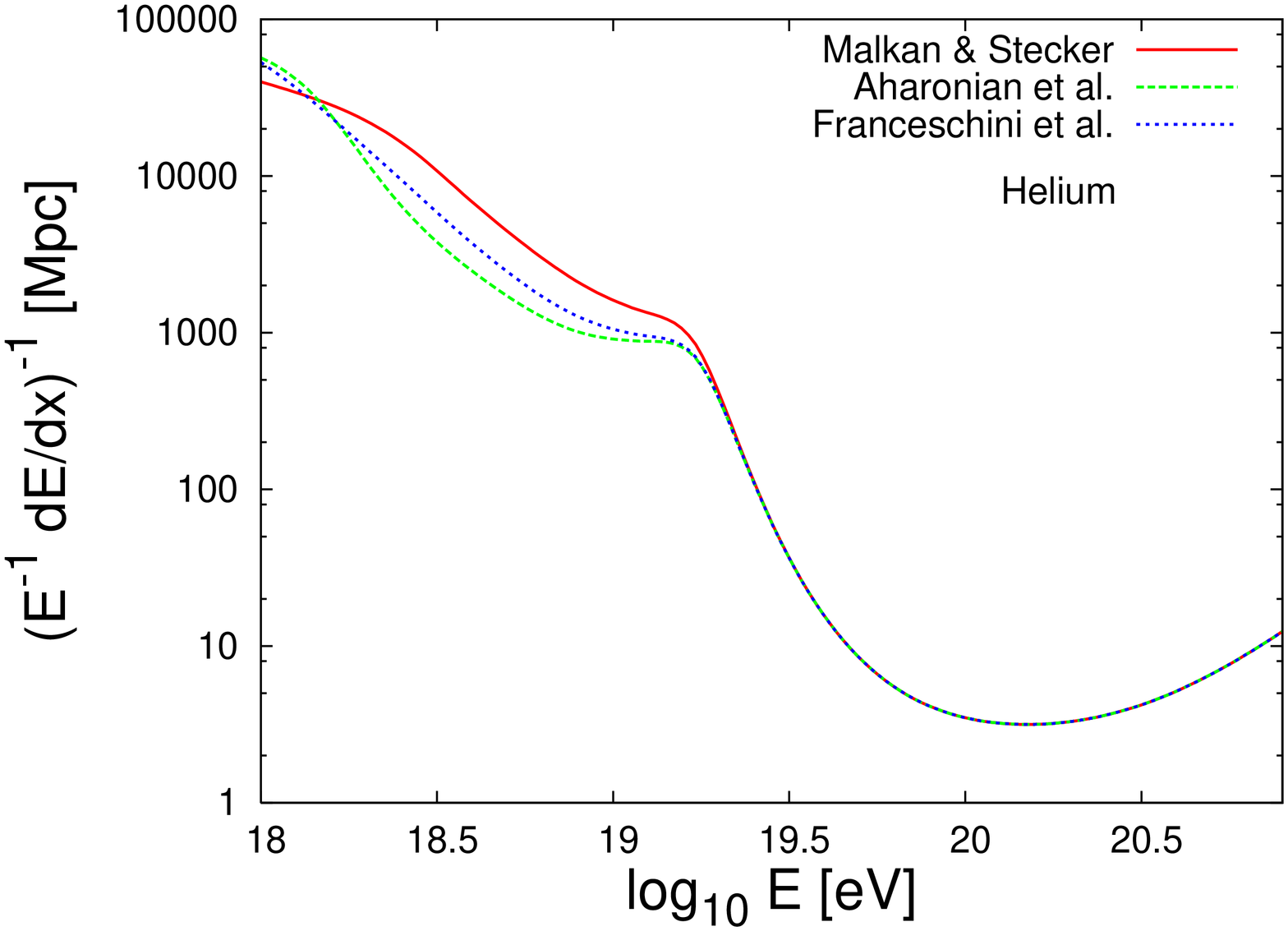}
\includegraphics[width=2.2in,angle=-90]{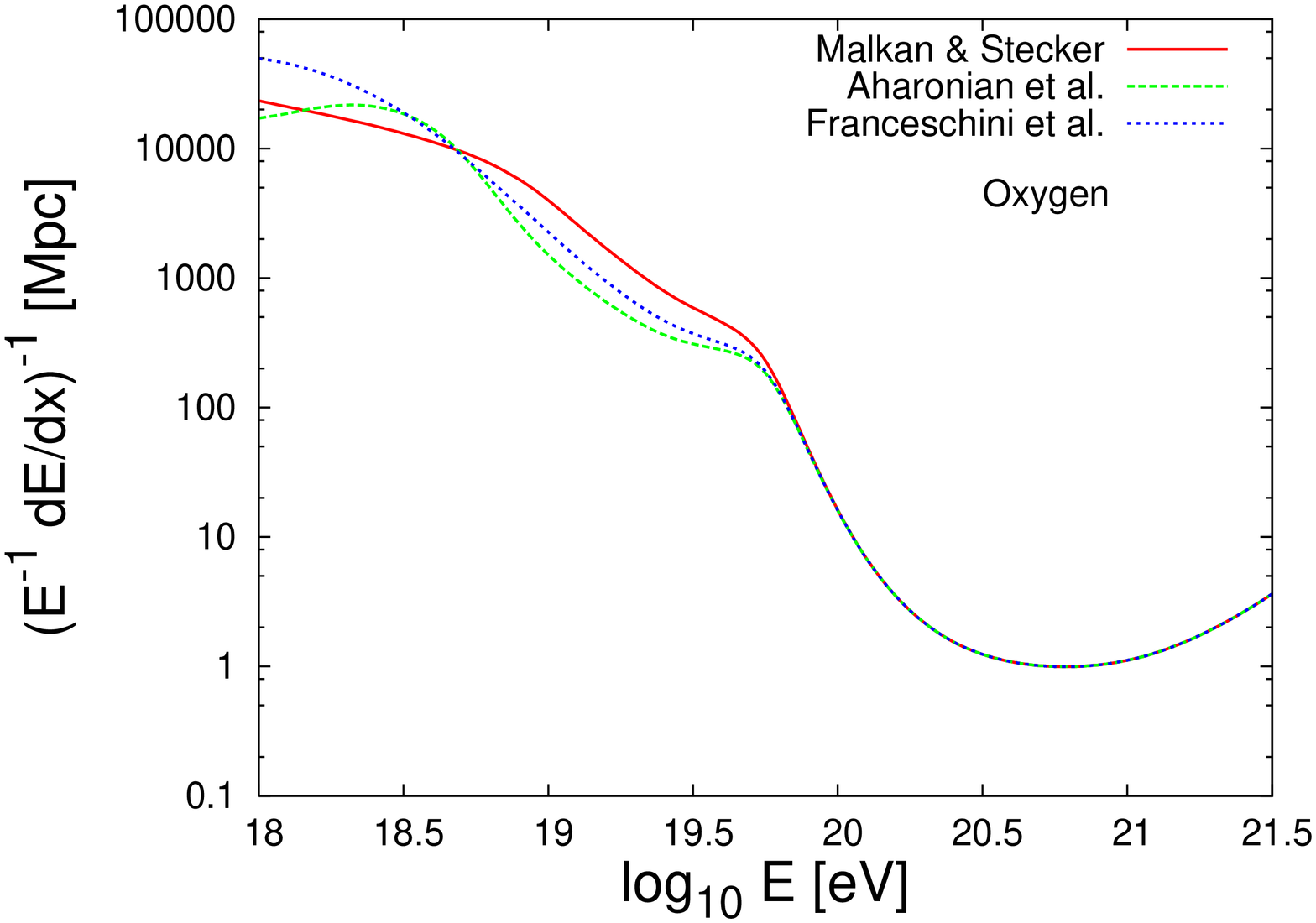} \\
\includegraphics[width=2.2in,angle=-90]{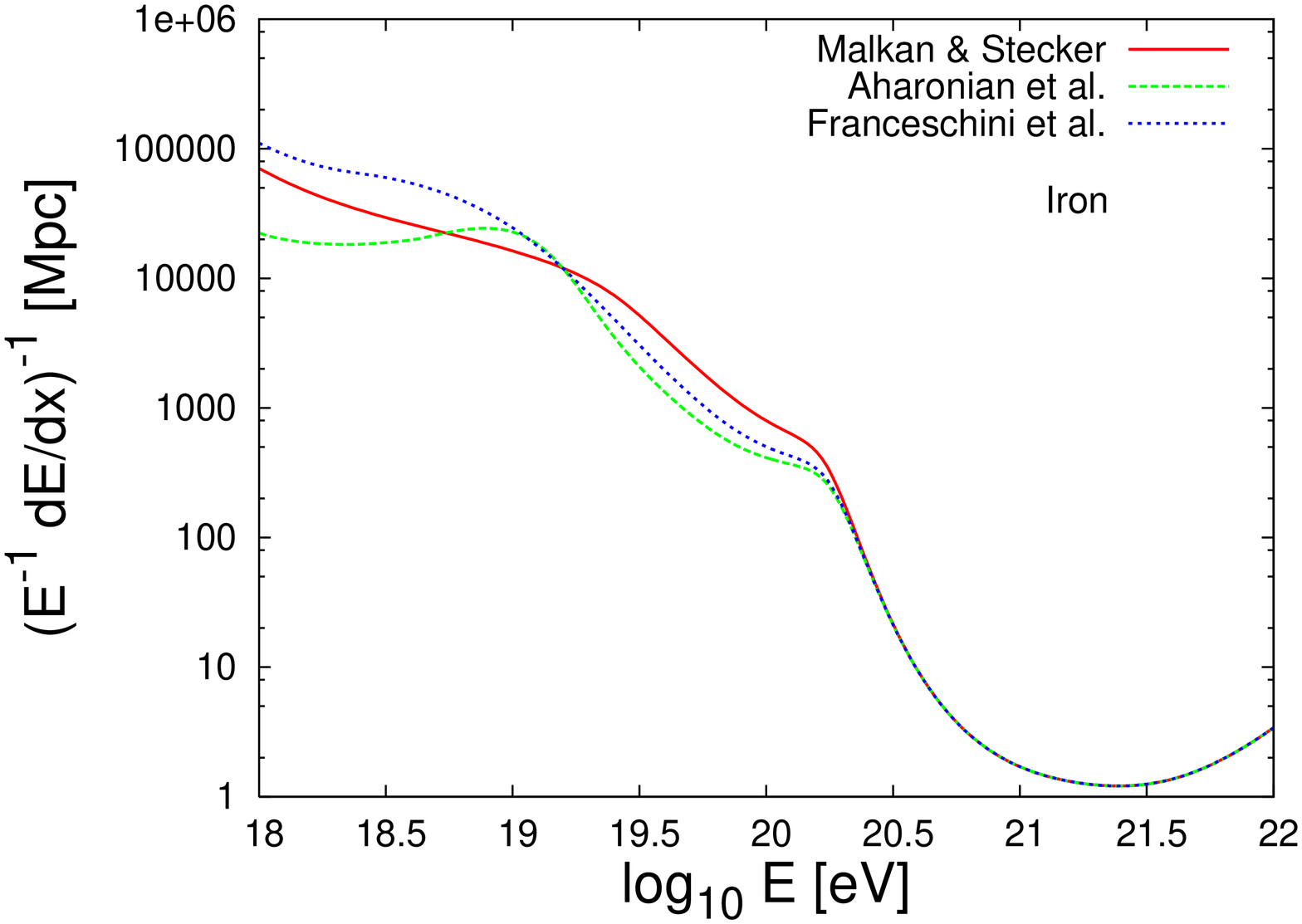}
%}
\caption{Energy loss lengths due to photodisintegration for helium,
oxygen and iron nuclei for three models of the CIB spectrum. The
Lorentzian model \cite{Khan:2004nd} for photodisintegration
cross-sections, been used.}
\label{irbrates}
\end{figure}
%%%%%%%%%%%%%%%%%%%%%%%%%%%%%%%%%%%%%%%%%%%%%%%%%%

%Direct measurement of the CIB requires removal of the foreground
%such as emission from the telescope, Earth's atmosphere and local bright
%sources such as the Sun and zodiacal (scattered) light from
%interplanetary dust.
%
%Over much of the wavelength range we are interested in, the dominant
%contribution to the CIB comes from interplanetary dust. Between 1.25 
%and 3.5 $\mu$m, starlight is also a substantial
%contributor, and above 60$\mu$m interstellar dust emission becomes 
%significant. At wavelengths $\gtrsim$400$\mu$m, the CMB
%becomes dominant.

Direct measurement of the CIB has been performed by several
satellites, in particular the COsmic Background Explorer (COBE) and
the InfraRed Telescope inSpace (IRTS). DIRBE, an instrument aboard the
COBE satellite \cite{Hauser:1998ri}, has provided measurements in the
wavelength range 1.25--240 $\mu$m. FIRAS, another instrument aboard
COBE, covered the range 25 $\mu$m -- 1 mm. The ISO satellite carried
two instruments which were employed in the indirect measurement of the
CIB: ISOCAM at 7 and 15 $\mu$m and ISOPHOT at 170 $\mu$m
\cite{Metcalfe:2003zi}.

In addition to these measurements, telescopes such as the Hubble Space
Telescope's (HST) wide field planetary camera, combined with
spectrophotometry from the duPont telescope and the HST Faint Object
Spectrograph, were used to measure the CIB at 0.3, 0.55 and
0.8~$\mu$m. Galaxy counts made using the HST Northern and Southern
Deep Fields, between 0.36 and 2.2 $\mu$m, supplemented with shallower
ground based observations, have also been used to set lower limits on
the CIB \cite{Madau:1999yh,Bernstein:2001rz,Gard}.

We have considered three forms for the CIB spectrum which bracket the
range of possibilities, The first is based on a compilation of the
direct observations by Aharonian {\it et al} \cite{Aharonian:2003wu}
which is the uppermost curve at 1 $\mu$m in Figure~\ref{irbplot}. The
second, from Franceschini {\it et al.} \cite{Franceschini:2001yg},
corresponds roughly to the lower limit set by galaxy counts. The third
is the empirical model by Malkan \& Stecker \cite{Malkan:2000gu} which
lies between the other two curves (although at 100 $\mu$m it goes
below the Franceschini {\it et al.} model). The observational data are
also shown for comparison.

In Figure~\ref{irbrates}, we show the effect of our choice of the CIB
spectrum on the energy loss rates of cosmic ray nuclei through
photodisintegration. It is clear that the choice makes a substantial
difference only for particularly heavy nuclei. Henceforth, unless
otherwise stated, we will adopt the Malkan \& Stecker model
\cite{Malkan:2000gu} for the CIB spectrum. We note that this spectrum
is consistent with recent observations by HESS of TeV $\gamma$-rays
from distant blazars which imply restrictive upper bounds on the CIB
and suggest that the direct observations in the $\sim 1-10~\mu$m range
may well have been contaminated by zodiacal light
\cite{Aharonian:2006fe}.

\section{Propagation of Ultra-High Energy Nuclei}
\label{prop}

%%%%%%%%%%%%%%%%%%%%%%%%%%%%%%%%%%%%%
\begin{figure}[t]
\centering\leavevmode
\mbox{
\includegraphics[width=2.2in,angle=-90]{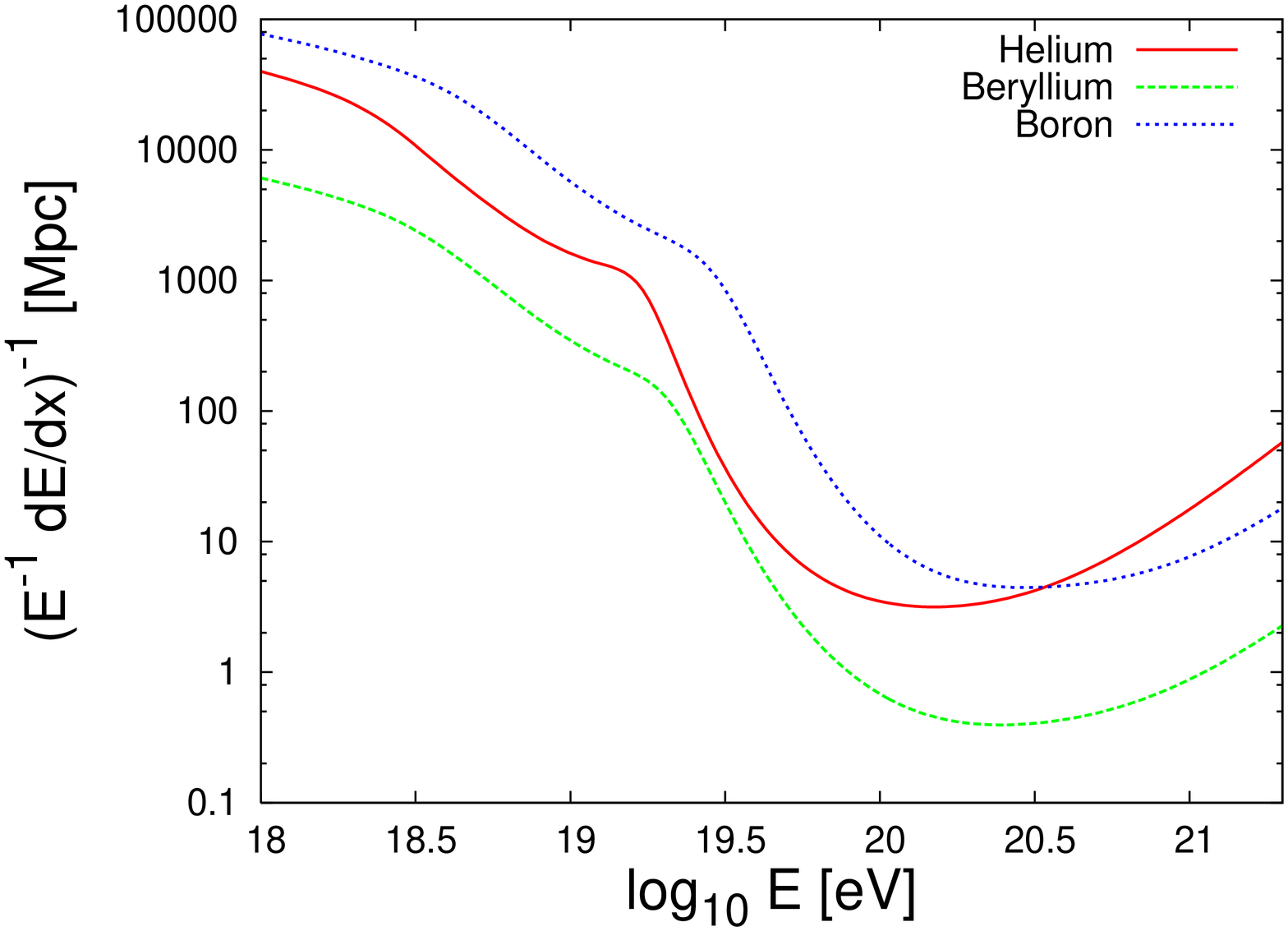}
\includegraphics[width=2.2in,angle=-90]{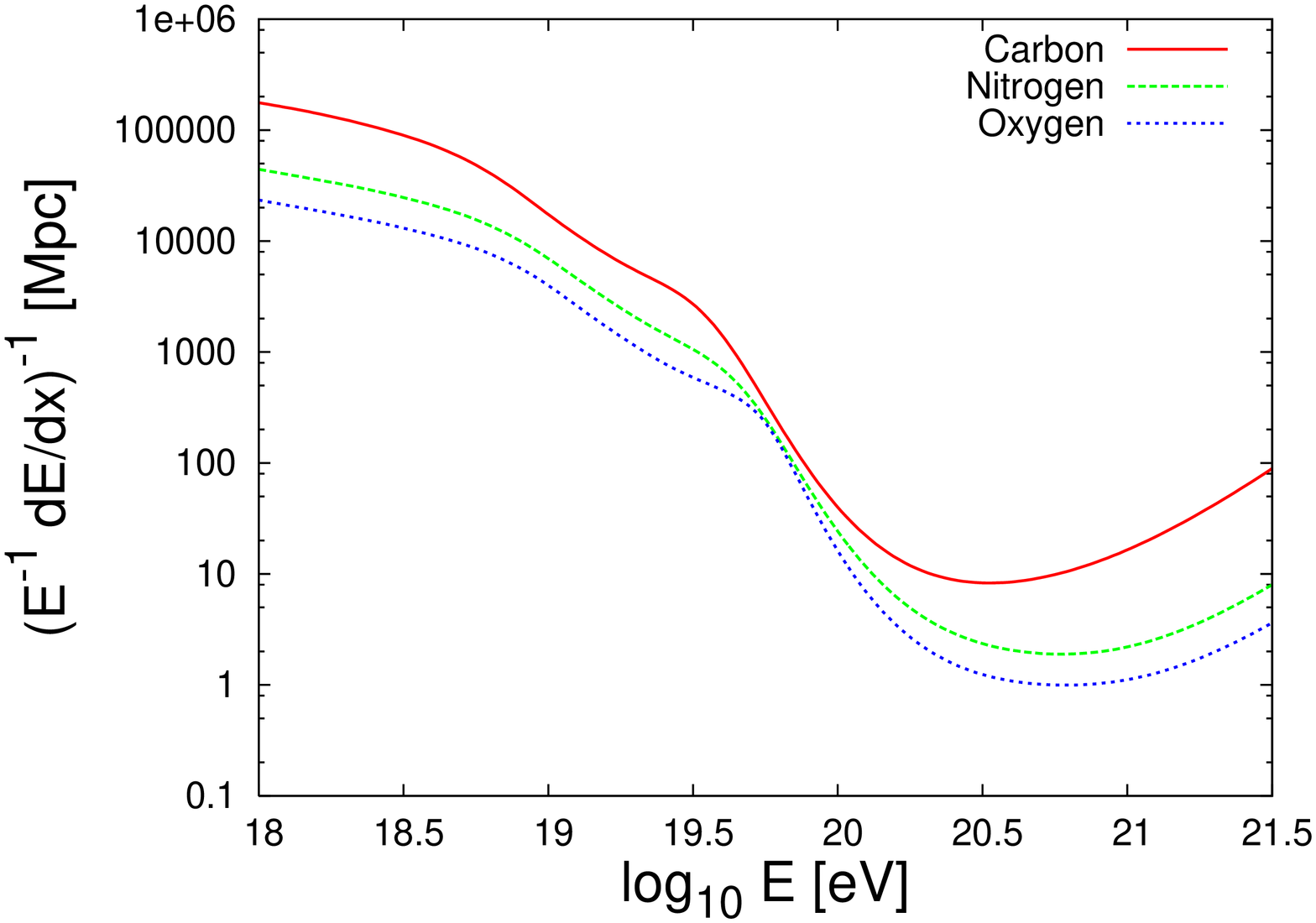}
}
\mbox{
\includegraphics[width=2.2in,angle=-90]{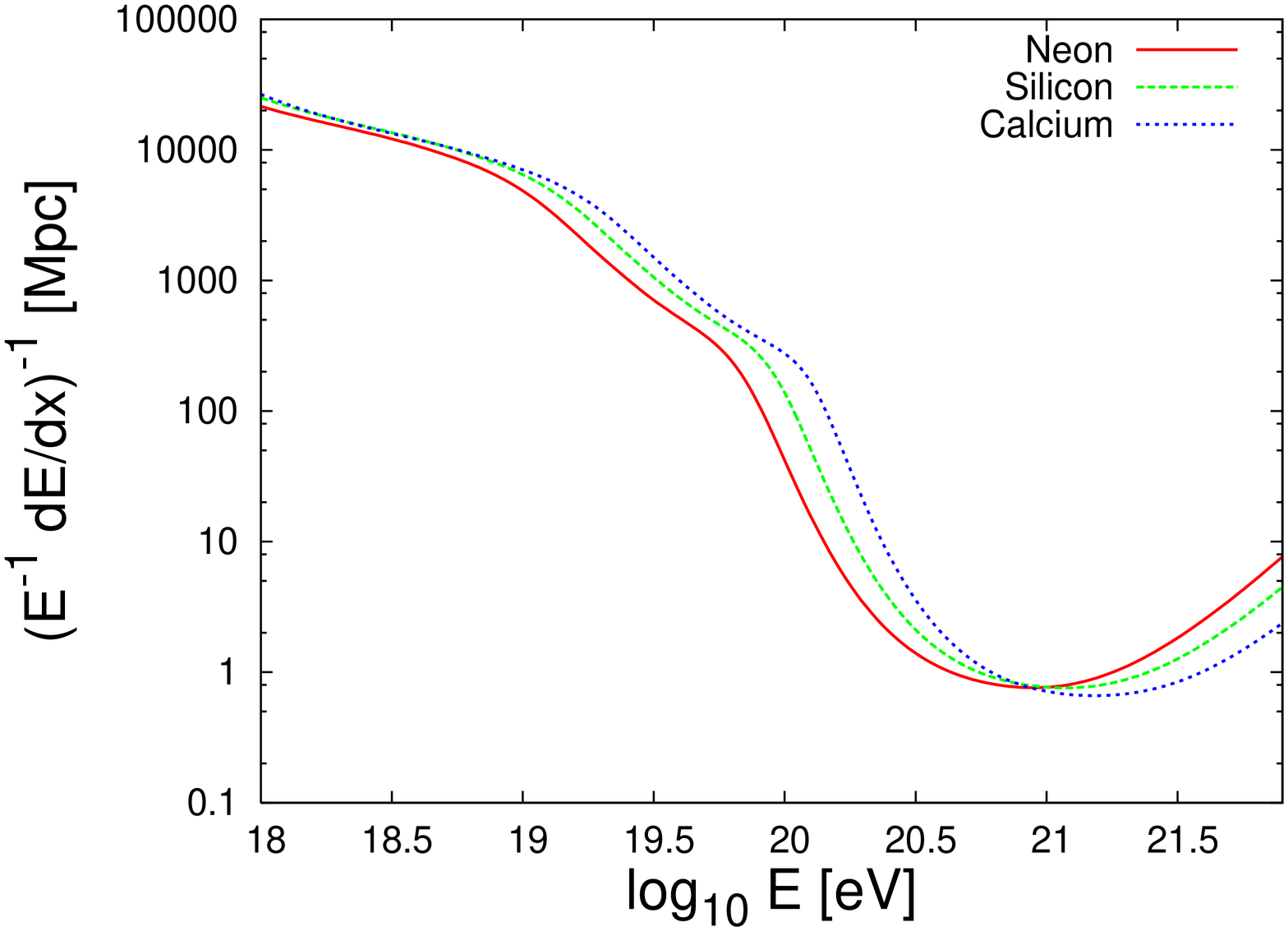}
\includegraphics[width=2.2in,angle=-90]{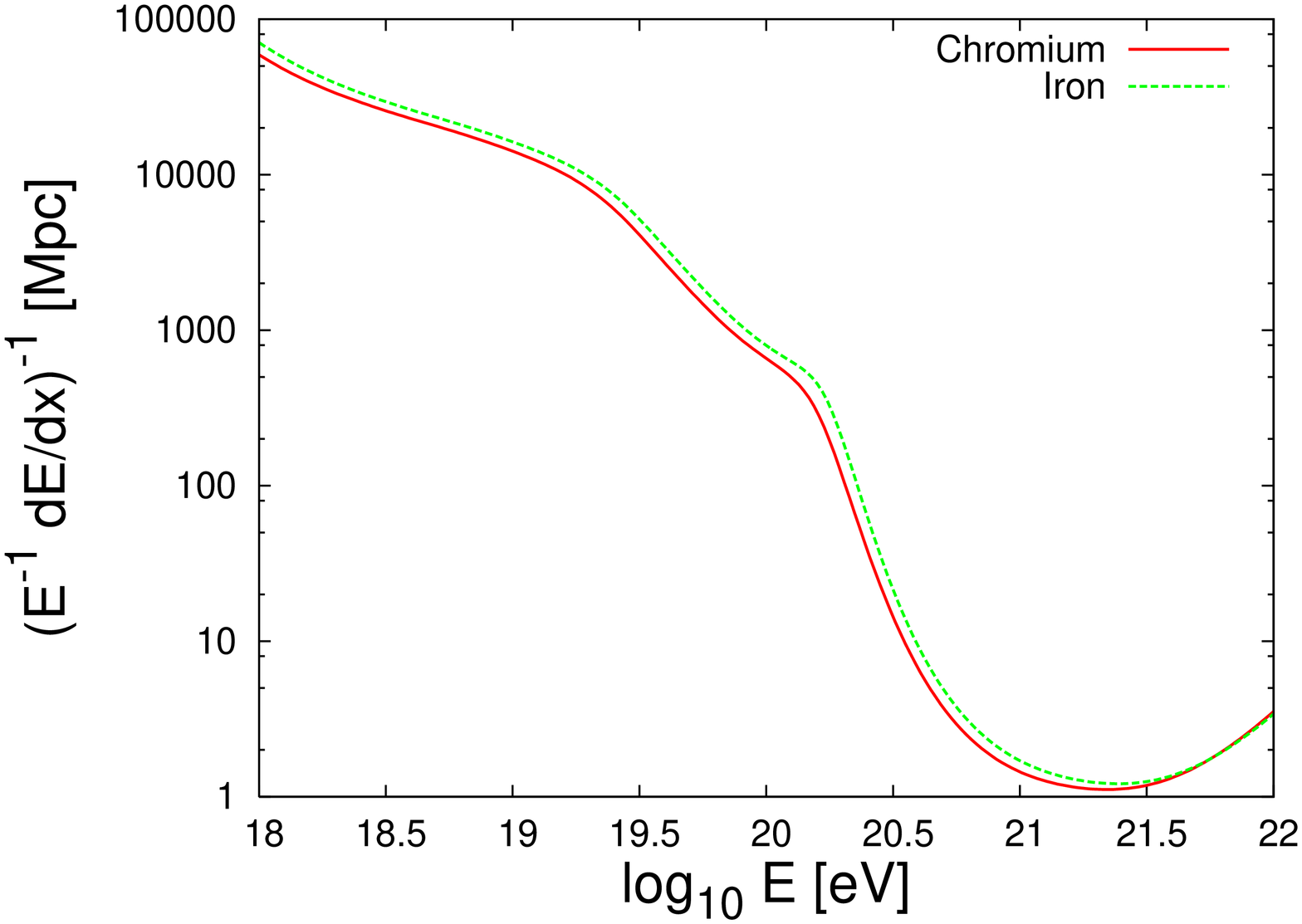}
}
\caption{Energy loss lengths due to photodisintegration for a range
of intermediate mass and heavy nuclei. The Malkan \& Stecker CIB model
\cite{Malkan:2000gu} and the Lorentzian model \cite{Khan:2004nd} for
photodisintegration cross-sections have been used.}
\label{heavyrates}
\end{figure}
%%%%%%%%%%%%%%%%%%%%%%%%%%%%%%%%%%%%%%%%%%%%%%%%%%

Now we can study the intergalactic propagation of UHE nuclei and
determine the UHECR spectrum at Earth for various types of nuclei
injected at source. In Figure~\ref{heavyrates} we plot the energy loss
length due to photodisintegration for several nuclei species. It is
seen that there are significant variations, for example, carbon nuclei
are relatively robust to photodisintegration with an energy loss
length of $\sim 50$ Mpc at $10^{20}$ eV and several thousand Mpc at $3
\times 10^{19}$ eV, while beryllium nuclei are highly fragile, with an
energy loss length smaller by a factor of $\sim 10-100$. Very heavy
nuclei are generally quite stable up to energies $\sim 10^{20}$ eV.

\begin{figure}[!]
\centering\leavevmode
\mbox{
\includegraphics[width=2.1in,angle=-90]{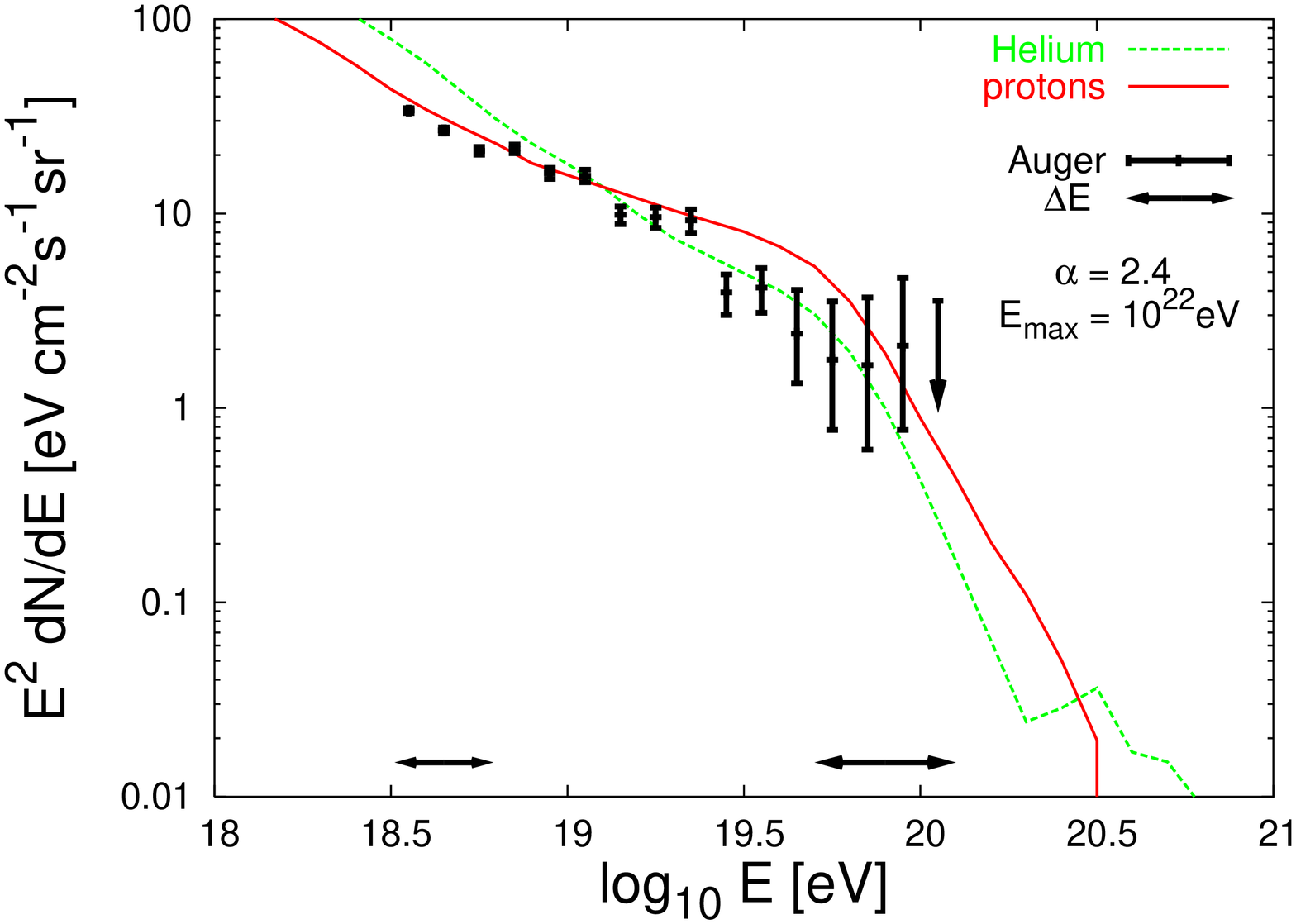}
\includegraphics[width=2.1in,angle=-90]{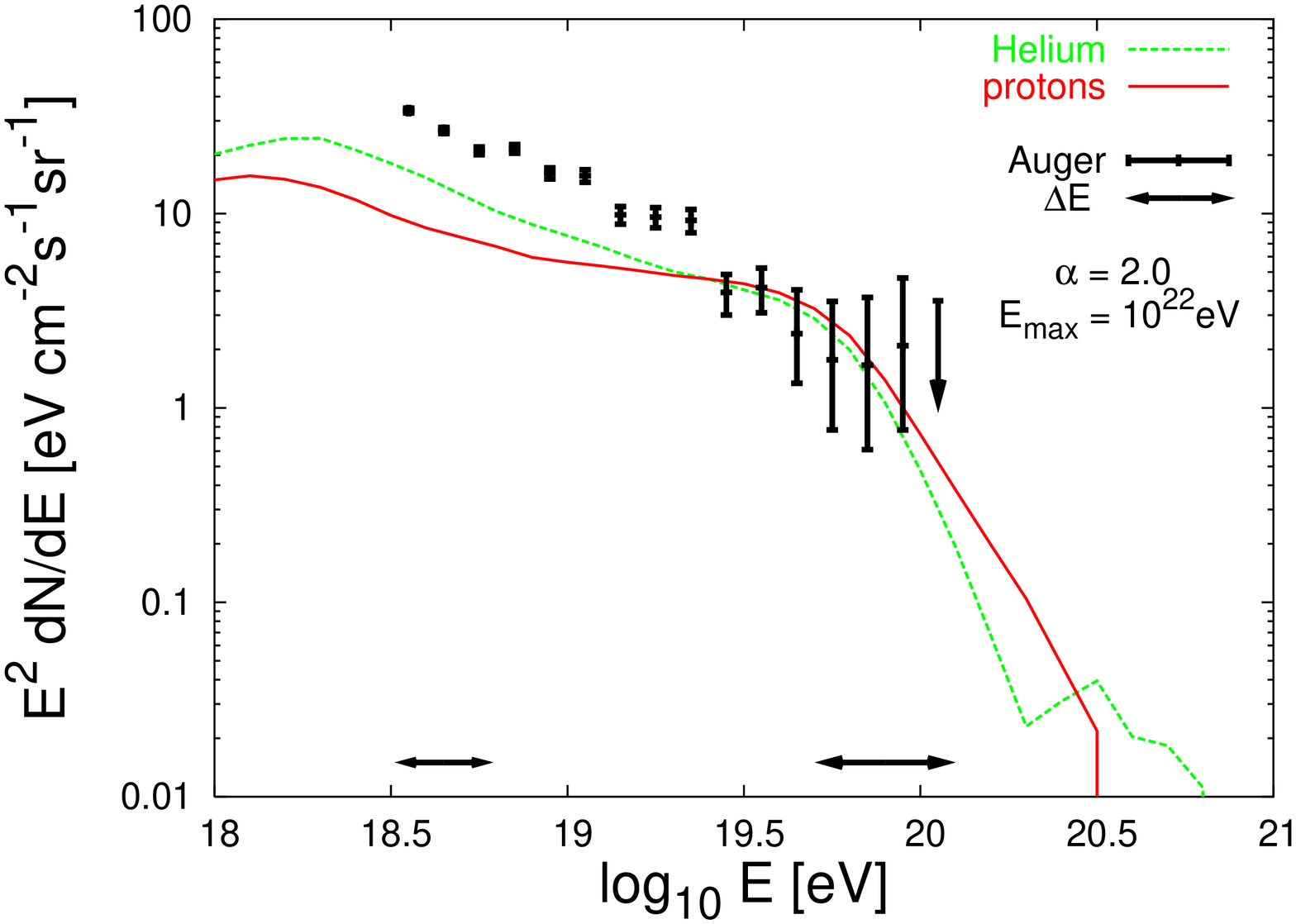}
}
\mbox{
\includegraphics[width=2.1in,angle=-90]{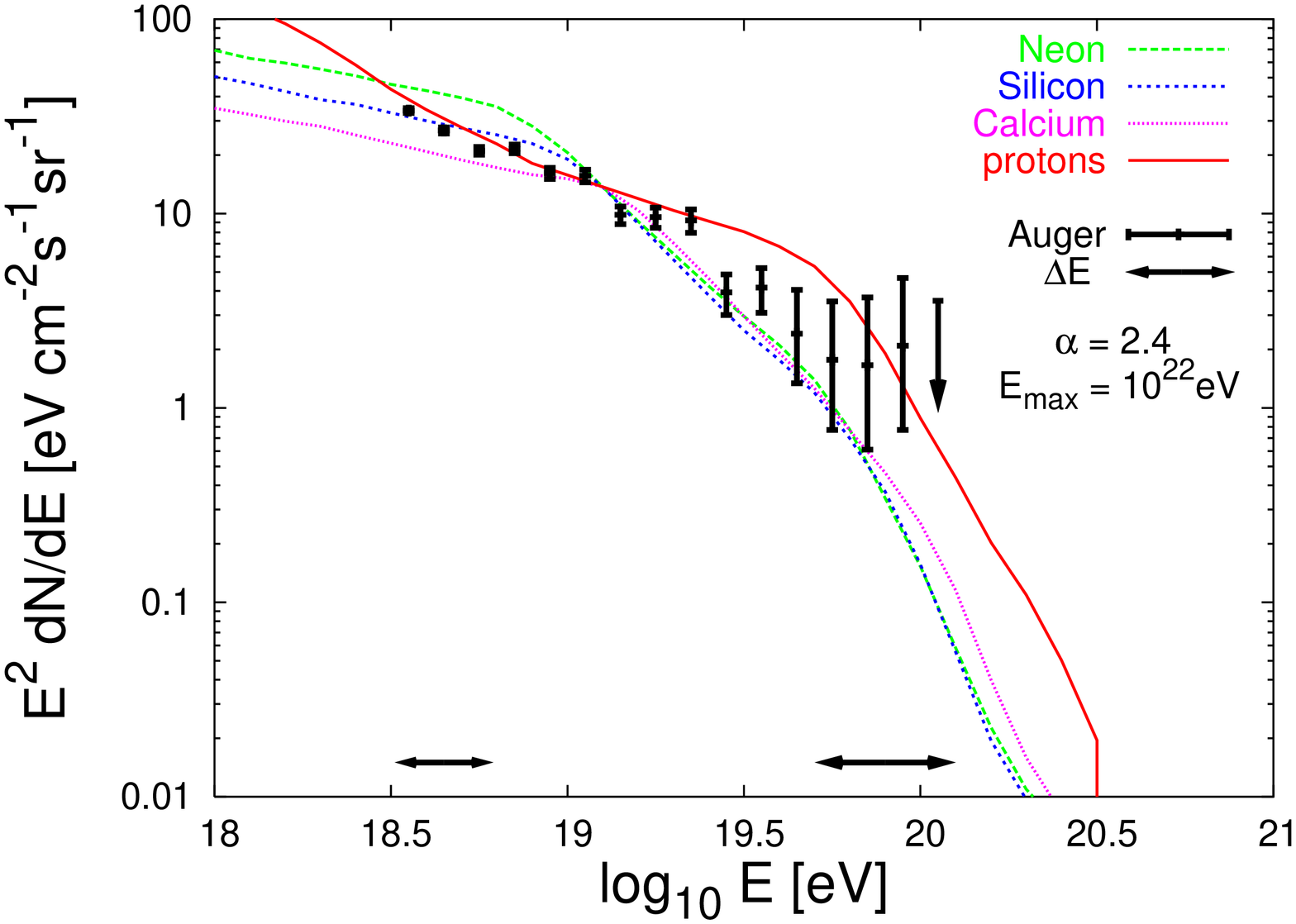}
\includegraphics[width=2.1in,angle=-90]{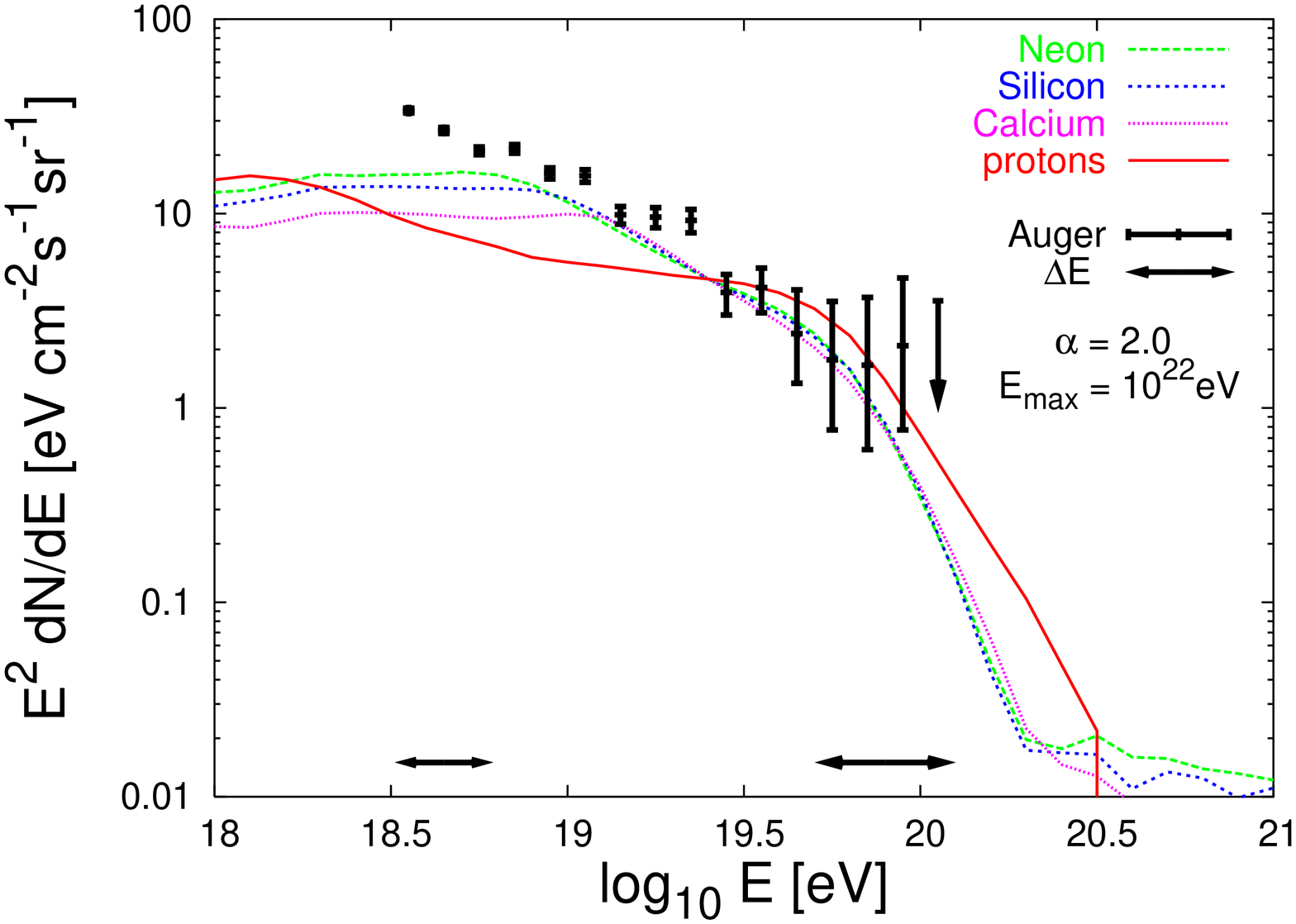}
}
\mbox{
\includegraphics[width=2.1in,angle=-90]{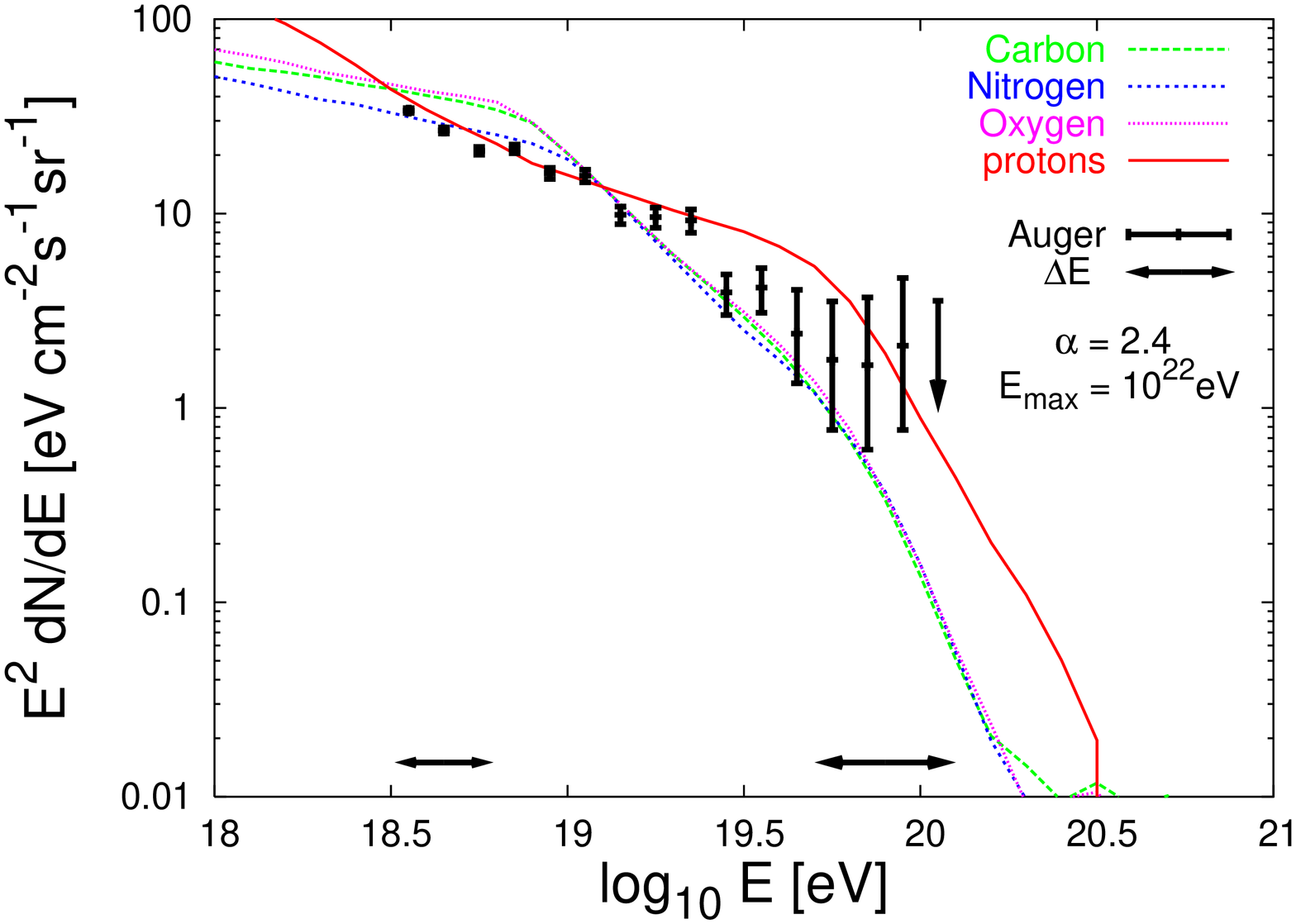}
\includegraphics[width=2.1in,angle=-90]{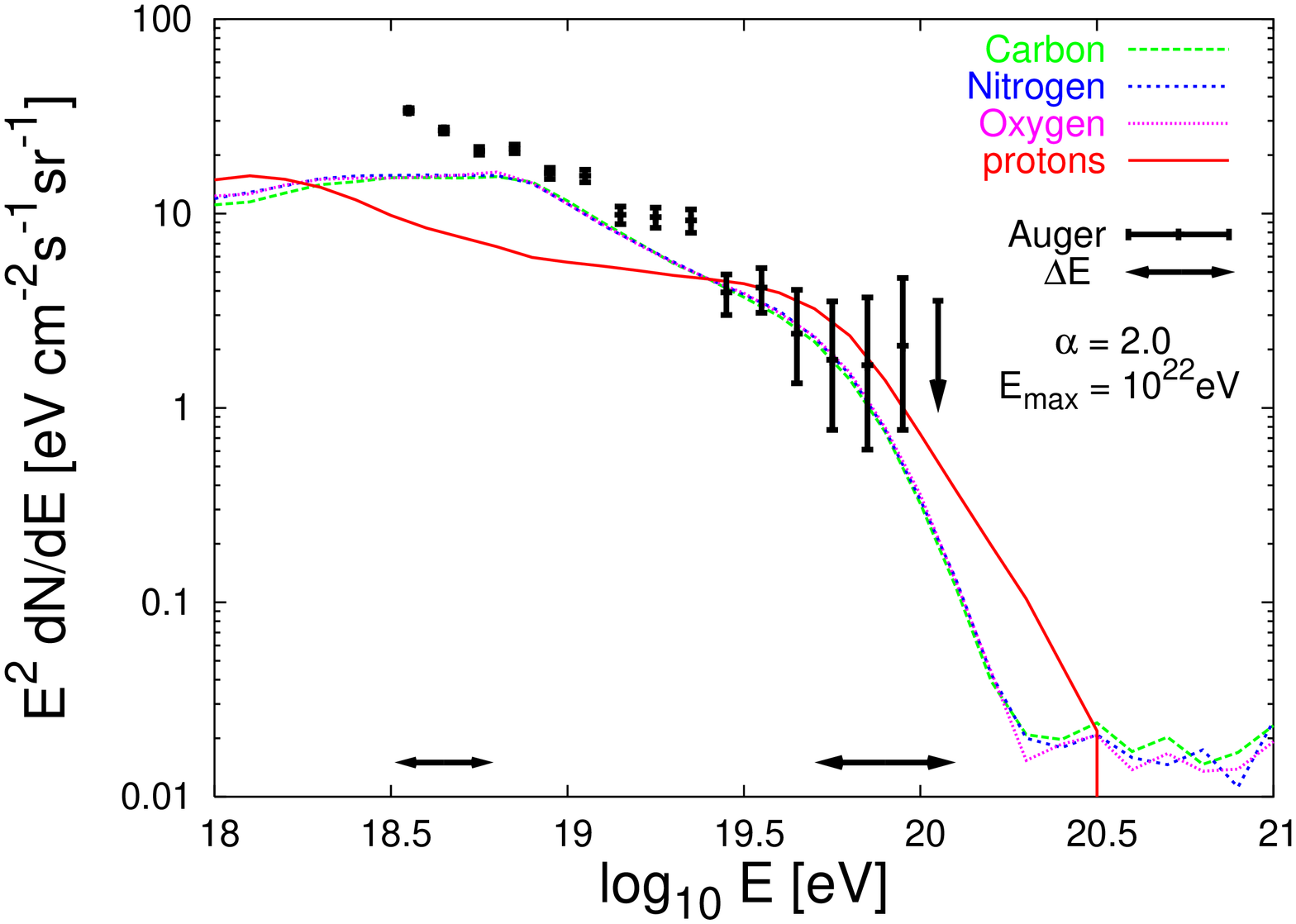}
}
\mbox{
\includegraphics[width=2.1in,angle=-90]{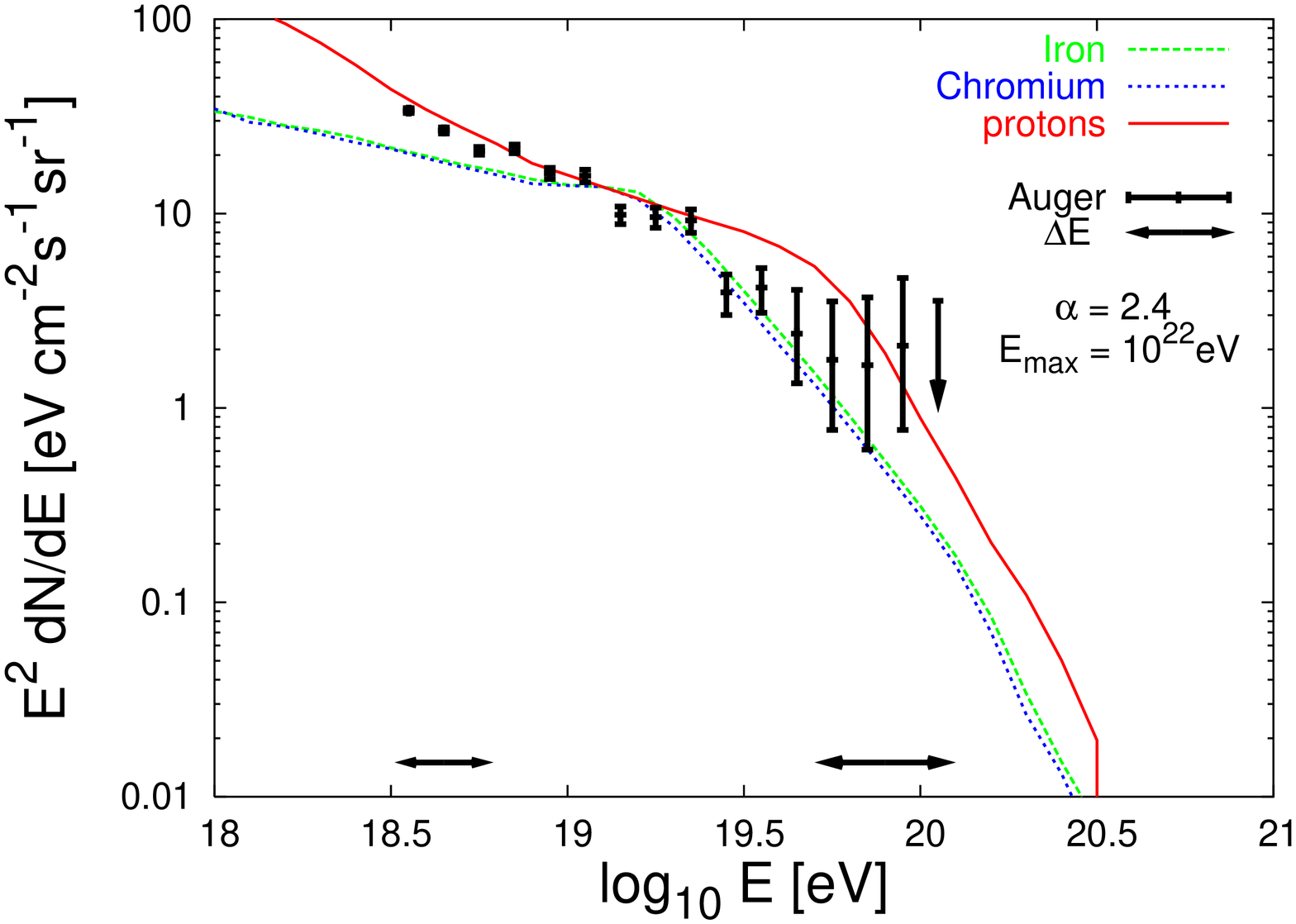}
\includegraphics[width=2.1in,angle=-90]{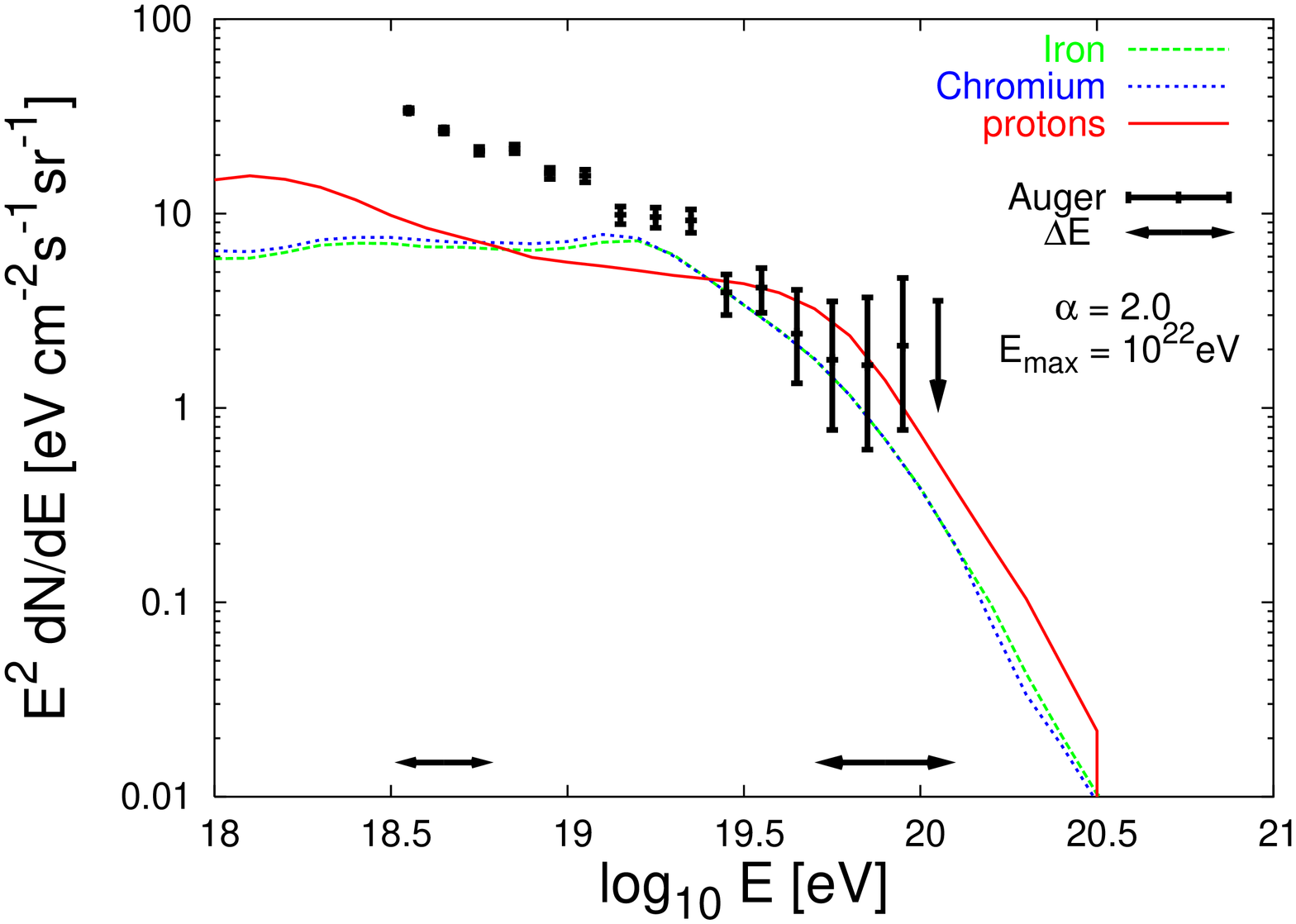}
}
\caption{The spectrum of UHECRs observed at Earth for a range of
injected heavy nuclei with power-law spectral index $\alpha = 2.4$ or
2.0 and $E_{\rm max}=10^{22}$ eV. The overall flux has in each case
been normalized to the Auger data \cite{Sommers:2005vs}. The Malkan \&
Stecker CIB model \cite{Malkan:2000gu} and the Lorentzian model
\cite{Khan:2004nd} for photodisintegration cross-sections have been
used. The effects of magnetic fields have {\em not} been included.}
\label{heavyspec}
\end{figure}

In practice, a heavy nucleus would undergo many photodisintegration
reactions, cascading down in atomic number and charge. Thus over a
given trajectory, an injected particle will take on the identity of
many species of nuclei, generating UHE protons, neutrons and alpha
particles along the way, each of which will continue to propagate and
contribute to the UHECR spectrum. In Figure~\ref{heavyspec} we show
the spectrum observed at Earth for various species of injected nuclei,
assuming an injection spectrum:
\begin{equation}
\frac{d N}{d E} \propto E^{-\alpha}, \qquad E < (E_{\rm max}\times Z\,/\,26),
\end{equation}
where $Z$ is the charge of the nuclei species under consideration.

%%%%%%%%%%%%%%%%%%%%%%%%%%%%%%%%%%%%%
%%%%%%%%%%%%%%%%%%%%%%%%%%%%%%%%%%%%%%%%%%%%%%%%%%
%%%%%%%%%%%%%%%%%%%%%%%

\section{The Composition at Earth}
\label{comp}

If intermediate mass or heavy nuclei are injected in a distant cosmic
ray accelerator, these particles will gradually disintegrate into
lighter nuclei and nucleons as they propagate through intergalactic
space. Depending on the distance to the sources, the cosmic ray
composition observed at Earth may be quite different from that at
injection.

\begin{figure}[!]
\centering\leavevmode
\mbox{
\includegraphics[width=2.1in,angle=-90]{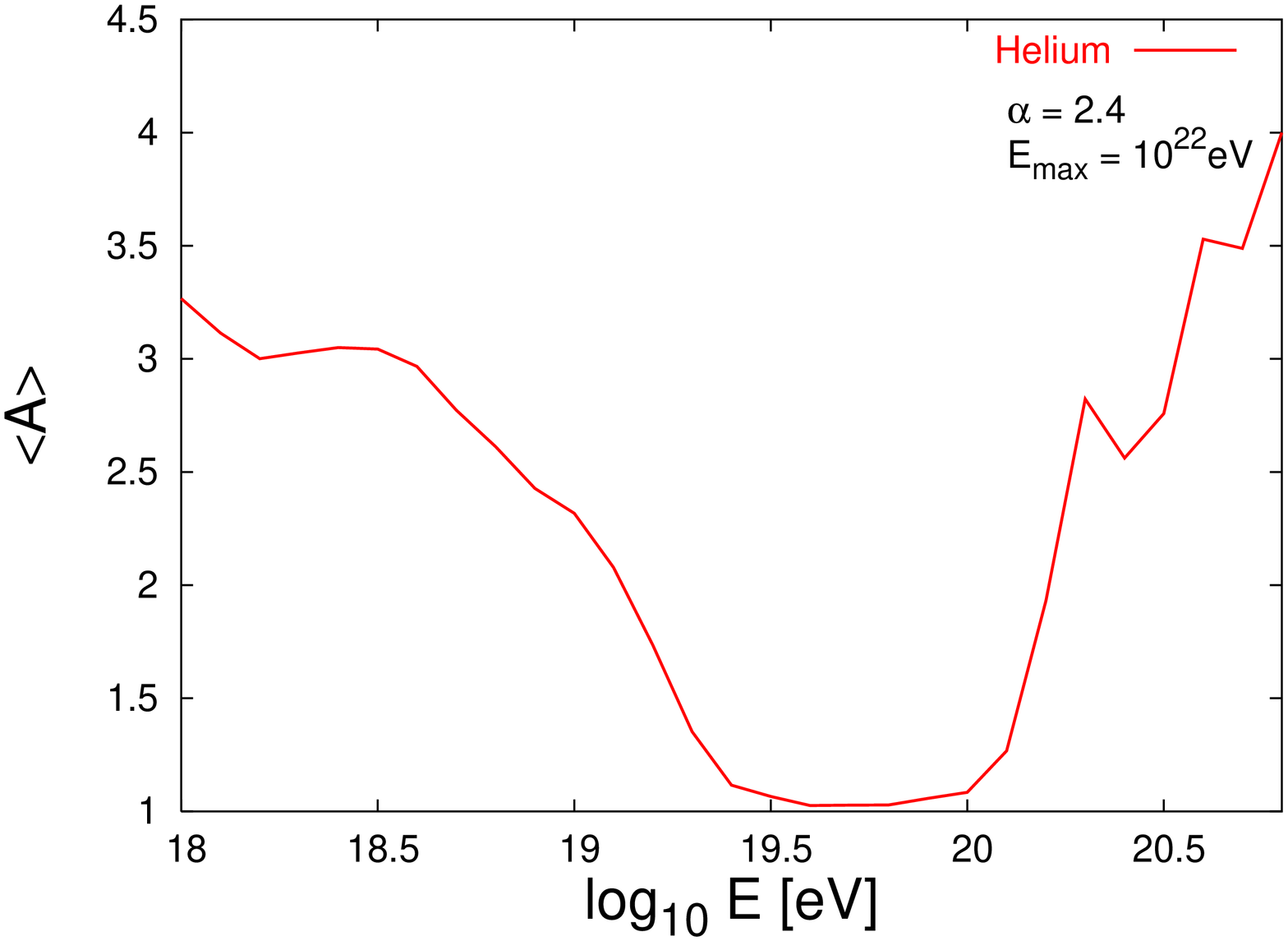}
\includegraphics[width=2.1in,angle=-90]{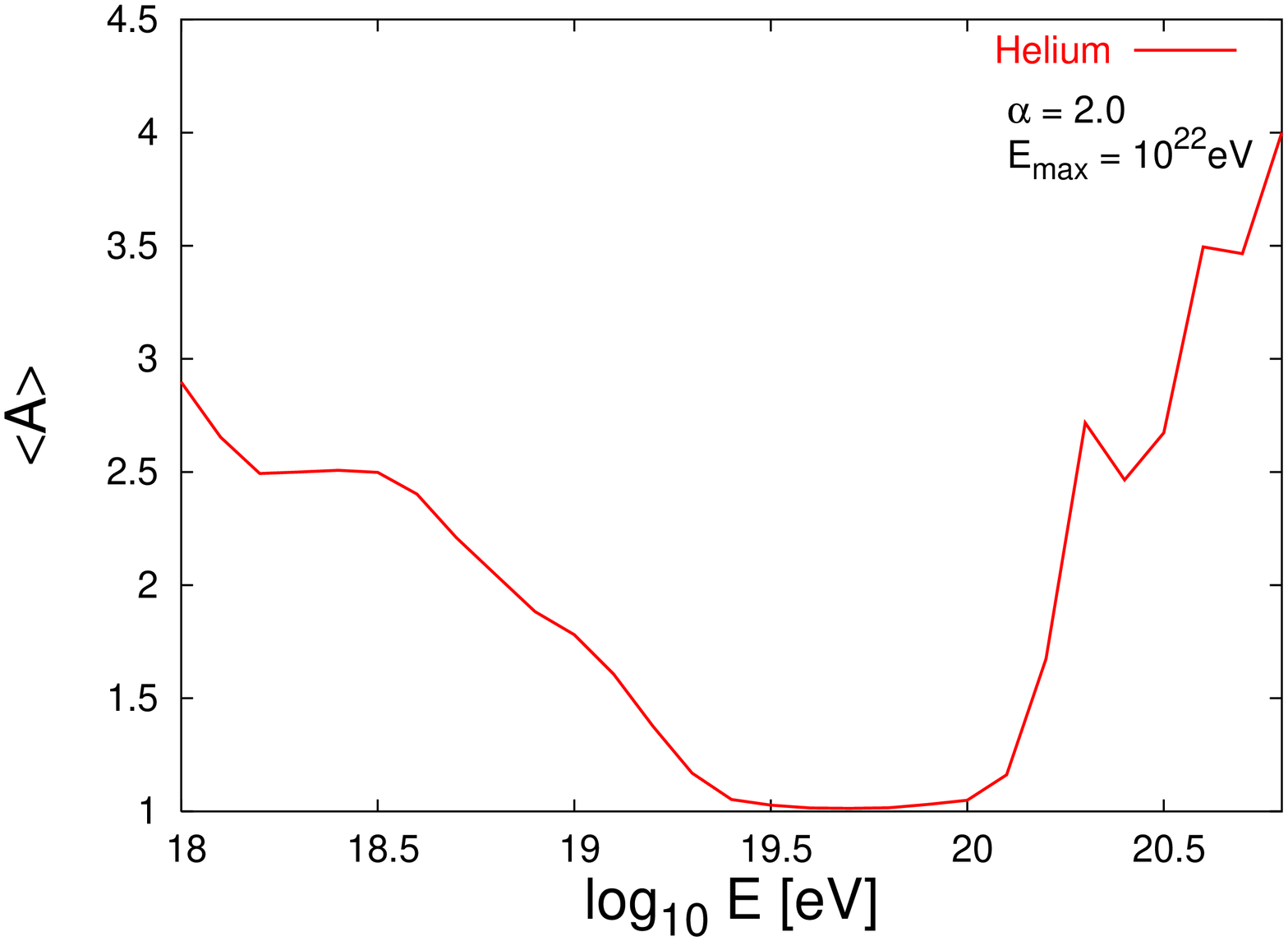}
}
\mbox{
\includegraphics[width=2.1in,angle=-90]{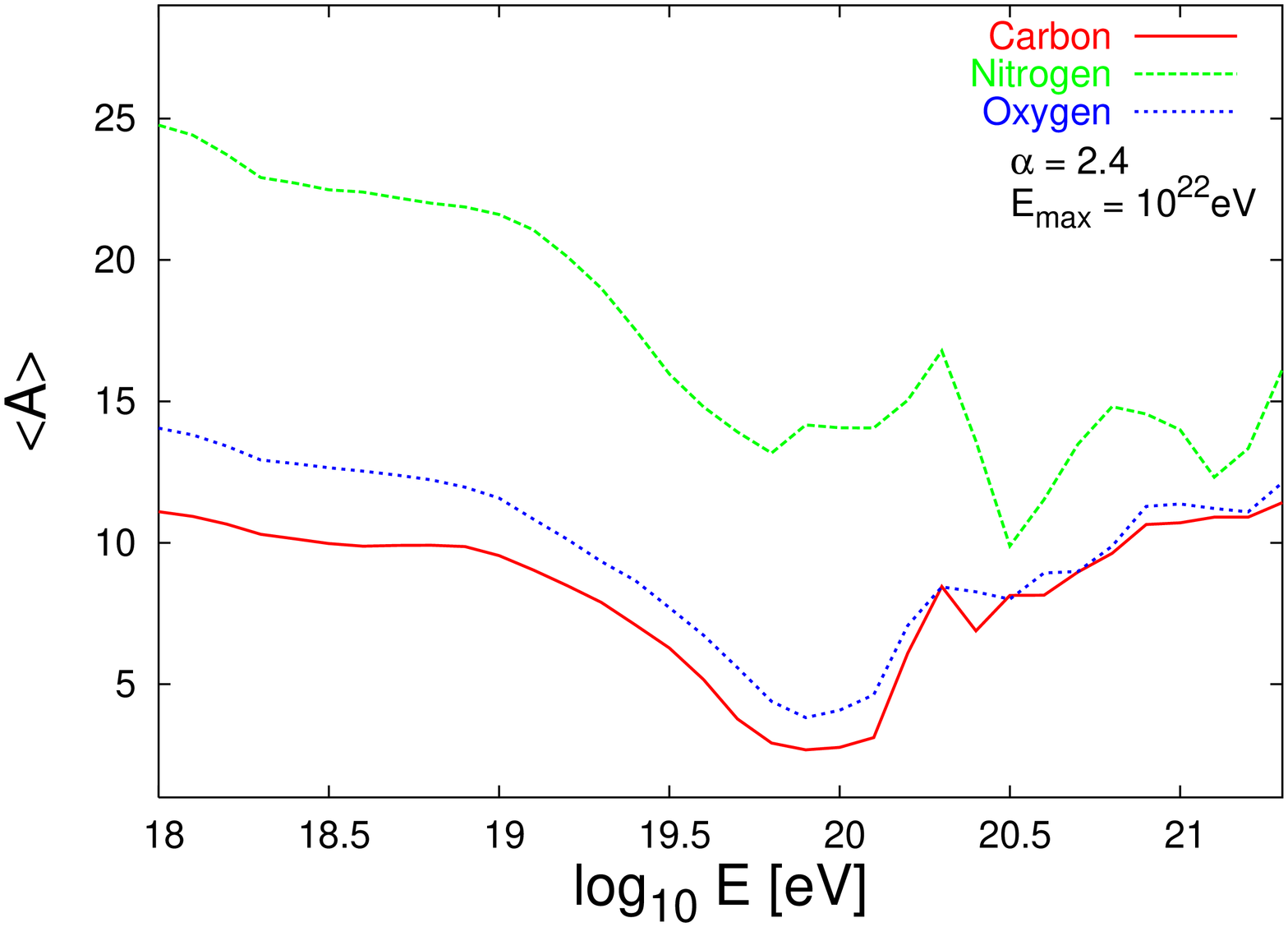}
\includegraphics[width=2.1in,angle=-90]{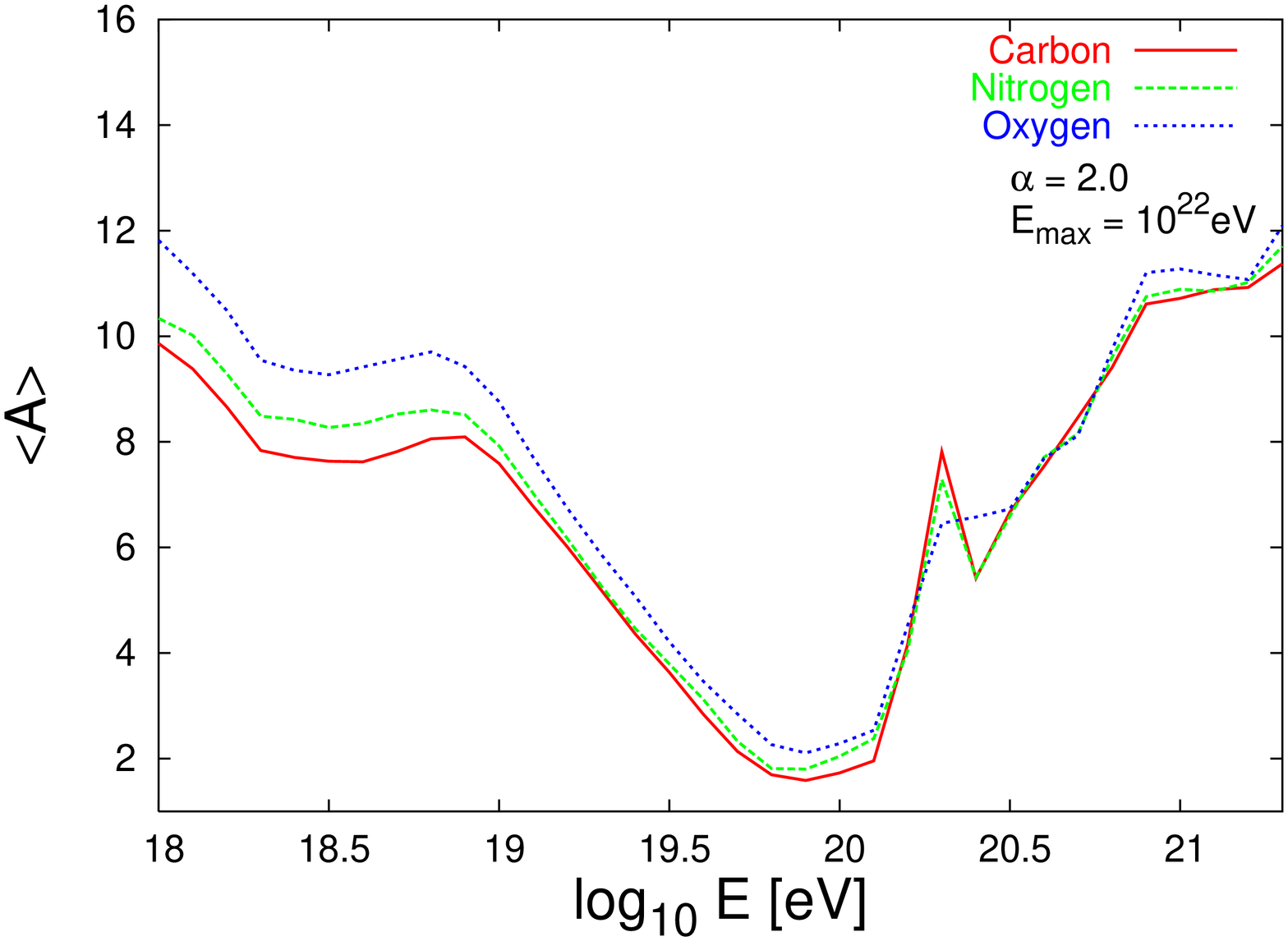}
}
\mbox{
\includegraphics[width=2.1in,angle=-90]{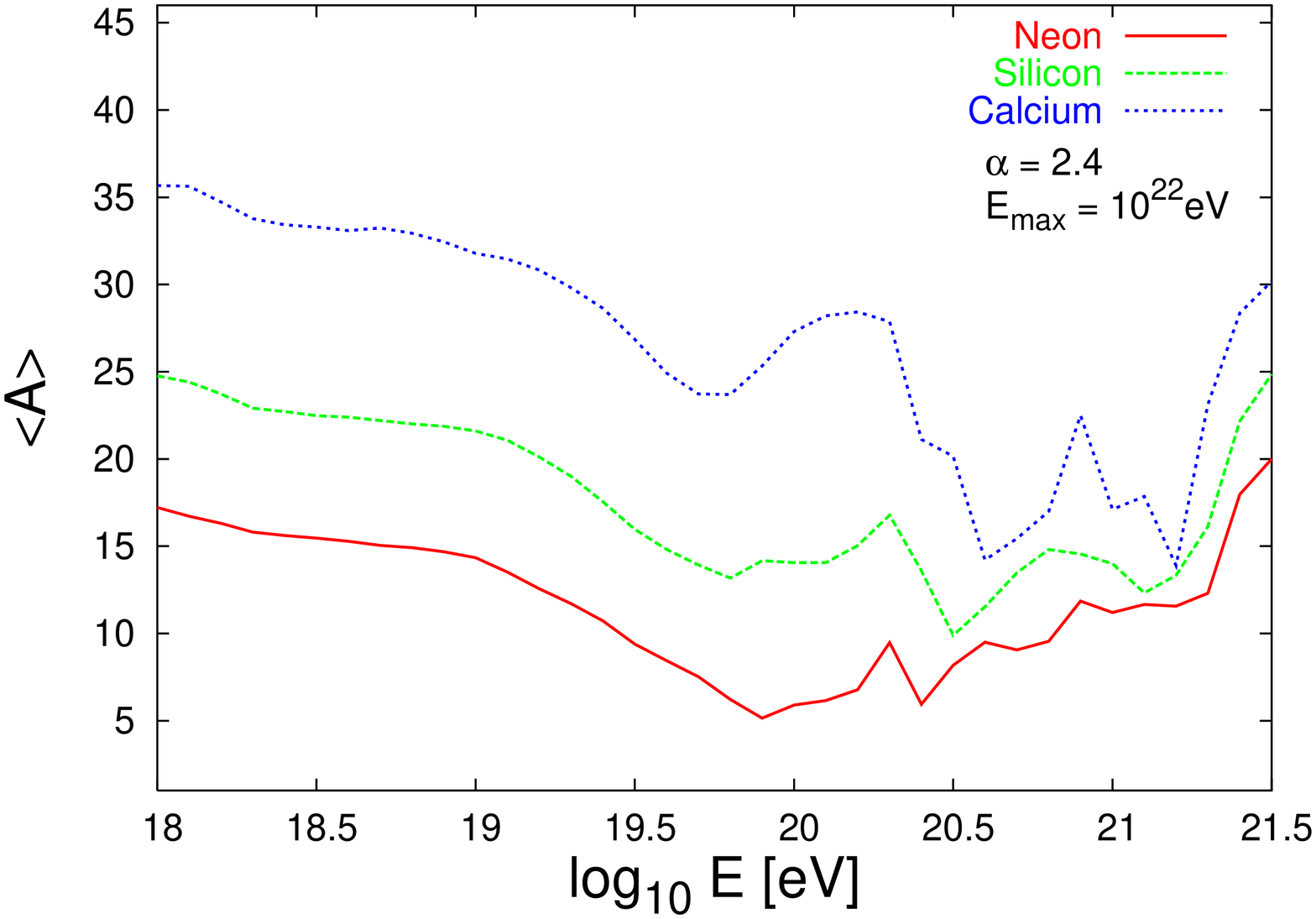}
\includegraphics[width=2.1in,angle=-90]{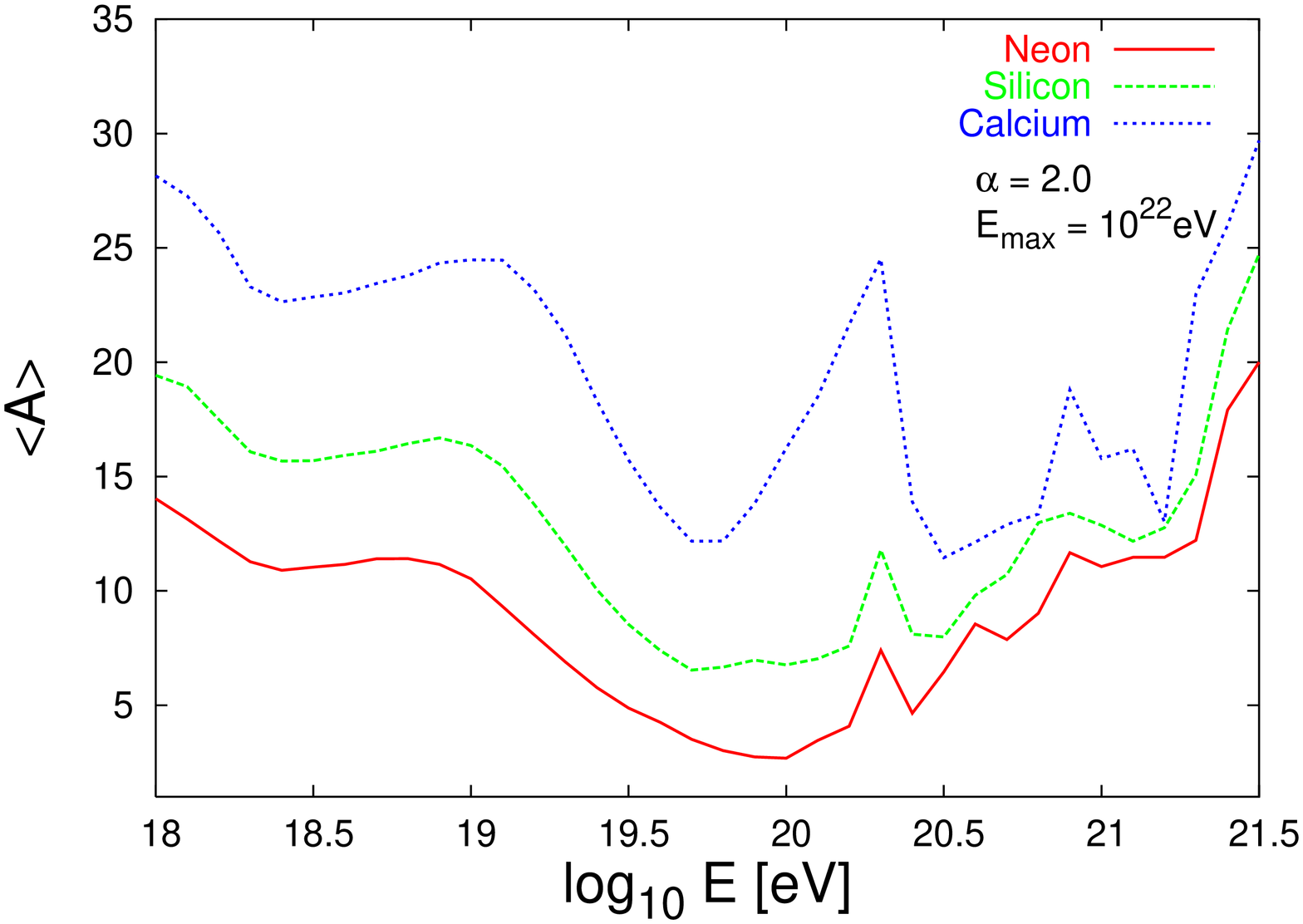}
}
\mbox{
\includegraphics[width=2.1in,angle=-90]{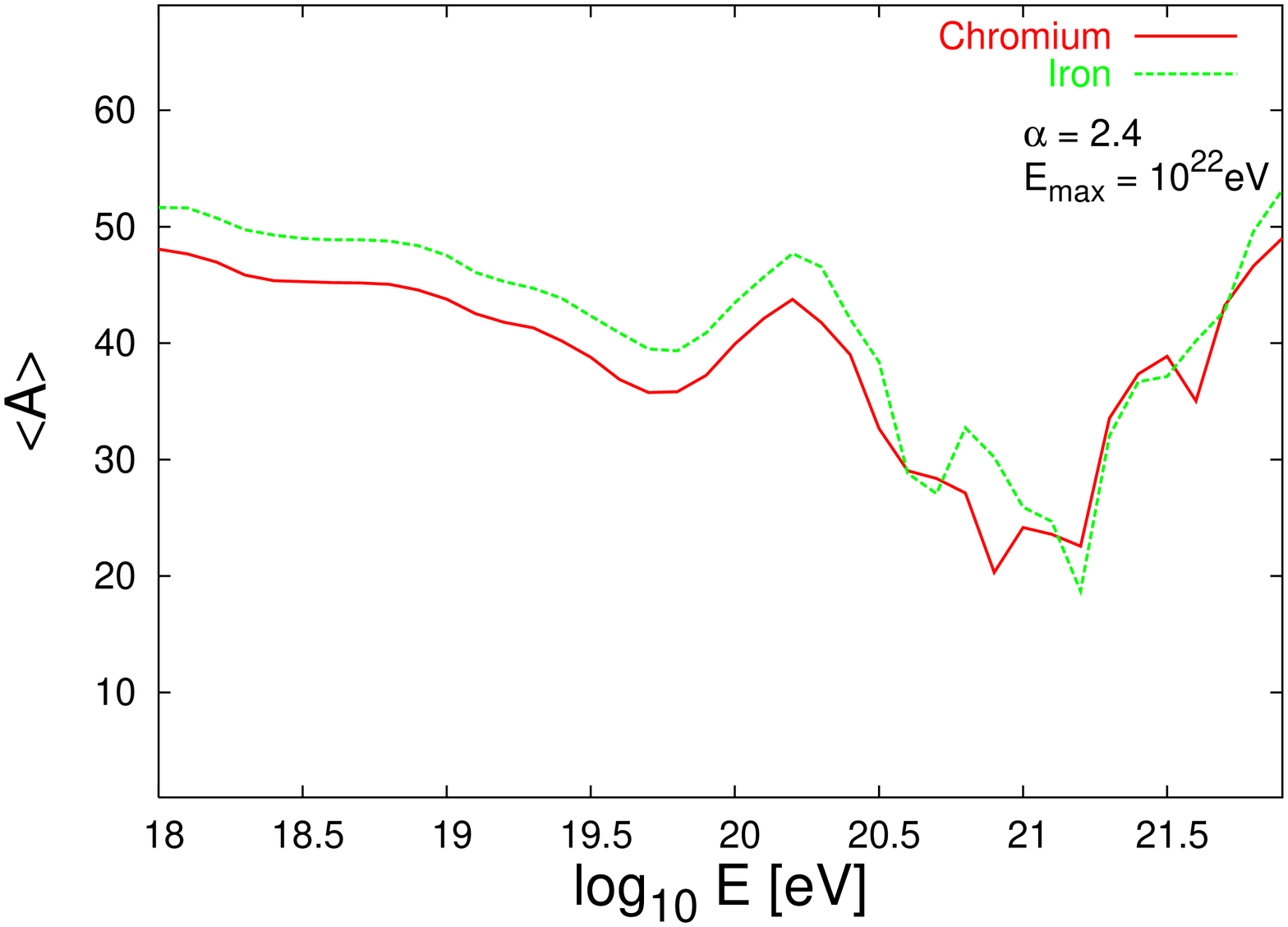}
\includegraphics[width=2.1in,angle=-90]{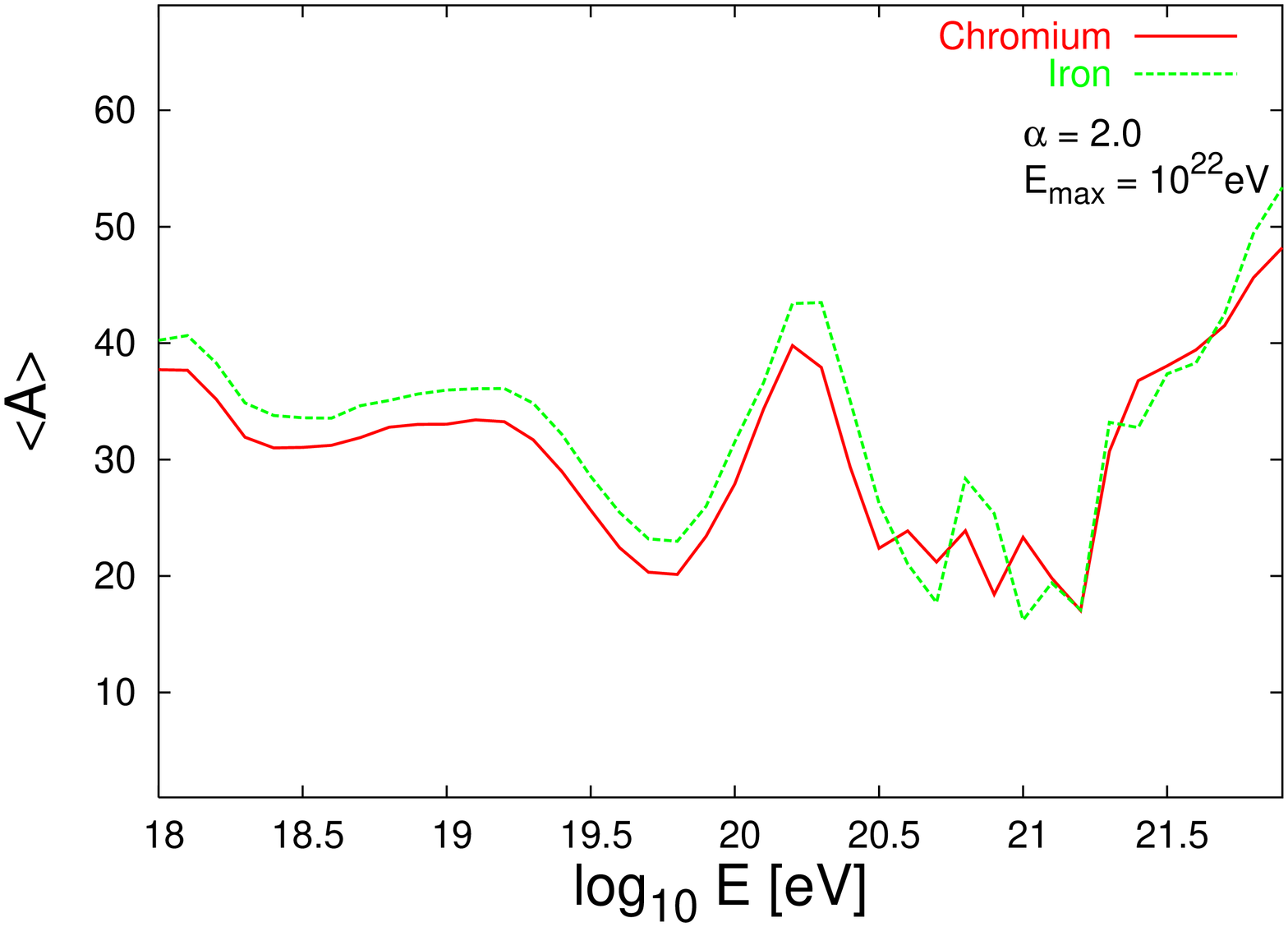}
}
\caption{The mean atomic mass of cosmic rays arriving at Earth for a
range of injected heavy nuclei with power-law spectral index $\alpha =
2.4$ or 2.0 and $E_\mathrm{max}=10^{22}$ eV. The Malkan \& Stecker CIB
model \cite{Malkan:2000gu} and the Lorentzian model \cite{Khan:2004nd}
for photodisintegration cross-sections have been used. The effects of
magnetic fields have {\em not} been included.}
\label{intave}
\end{figure}

The observed composition at Earth has a distinctive dependence on the
energy as can be seen in Figure~\ref{intave}. In the interesting
energy range $\sim3 \times 10^{19} - 10^{20}$~eV, i.e. where the GZK
suppression is expected for proton primaries, photodisintegration is
most effective. For sources injecting intermediate mass nuclei, the
effective mass number at Earth reaches a well-defined minimum,
probably indistinguishable from proton primaries. This minimum is less
pronounced for injection of very heavy nuclei and the effective
composition at Earth is distinctly heavier than protons.

These issues are of critical importance, observationally speaking. The
injected composition of the UHECR spectrum is not directly accessible
experimentally, and can only be reconstructed from the composition
observed at Earth. As stated earlier, the present observational status
is rather uncertain. Future data will hopefully reach a level of
quality which makes it possible to reliably infer the approximate
composition at Earth. With such data, general trends such as those
seen in Figure~\ref{intave} would aid in estimating the composition of
cosmic rays at the sources.

%%%%%%%%%%%%%%%%%%%%%%%%%%%%%%%%%%%%%
%%%%%%%%%%%%%%%%%%%%%%%%%%%%%%%%%%%%%%%%%%%%%%%%%%

\section{Effects of Intergalactic Magnetic Fields}
\label{bfields}

So far in this study, we have neglected the effects of magnetic fields
on the propagation of UHECRs. For protons or nuclear primaries,
however, such effects can play an important role in determining the
cosmic ray spectrum. The importance of these effects depend, of
course, on the strength of the extragalactic magnetic fields which is
currently a subject of some debate with contrary conclusions drawn in
Ref.~\cite{Dolag:2004kp} and in
Refs.~\cite{Sigl:2003ay,Armengaud:2004yt}.

A charged particle moving through a uniform magnetic field undergoes
an angular deflection upon traversing a distance, $L_{\rm{coh}}$, of
$\alpha=L_{\rm{coh}}/R_{L}$, where $R_{L}$ is the Larmor radius of the
particle. Therefore a particle traversing a distance, $L$, through a
series of $L/L_{\rm{coh}}$ randomly orientated uniform magnetic field
regions of length $L_{\rm{coh}}$, suffers an overall angular
deflection given by
\begin{equation}
\theta(E,Z) \approx \bigg(\frac{L}{L_{\rm{coh}}}\bigg)^{0.5} \,\alpha
\approx 0.8^{\circ} \, \bigg(\frac{10^{20} \, \rm{eV}}{E}\bigg) \,
\bigg(\frac{L}{10 \, \rm{Mpc}}\bigg)^{0.5} \,
\bigg(\frac{L_{\rm{coh}}}{1 \, \rm{Mpc}}\bigg)^{0.5} \,
\bigg(\frac{B}{1 \, \rm{nG}}\bigg) \,\, Z,
\end{equation}
where $L_{\rm{coh}}$ is the representative coherence length of the
extragalactic magnetic fields, $B$ is their representative magnitude
and $Z$ is the electric charge of the cosmic rays. Such deflections
result in an increase in the effective distance to a cosmic ray source
given by:
\begin{equation}
\frac{L_{\rm{eff}}}{L}(E,Z) \approx 1 + \frac{\theta^2}{2} \approx 1 +
0.065 \, \bigg(\frac{10^{20} \, \rm{eV}}{E}\bigg)^2 \,
\bigg(\frac{L}{10 \, \rm{Mpc}}\bigg) \, \bigg(\frac{L_{\rm{coh}}}{1 \,
\rm{Mpc}}\bigg) \, \bigg(\frac{B}{1 \, \rm{nG}}\bigg)^2 \,\,
\bigg(\frac{Z}{26}\bigg)^2.
\end{equation}
Thus for protons or light nuclei, nano-Gauss magnetic fields have
little impact for the high energies considered here. This is not true
for heavy nuclei, e.g. for iron nuclei propagating through nG-scale
magnetic fields, the effective distance to a source 50 Mpc away is
increased by $\sim 30\%$ at $10^{20}$ eV (alternatively, the energy
loss length is reduced by about $\sim 30\%$). Since this effect scales
with the {\em inverse square} of the cosmic ray energy, such
(plausible strength) magnetic fields would have a dramatic effect on
the propagation of lower energy heavy nuclei.

\begin{figure}[t]
\centering\leavevmode
\mbox{
\includegraphics[width=2.2in,angle=-90]{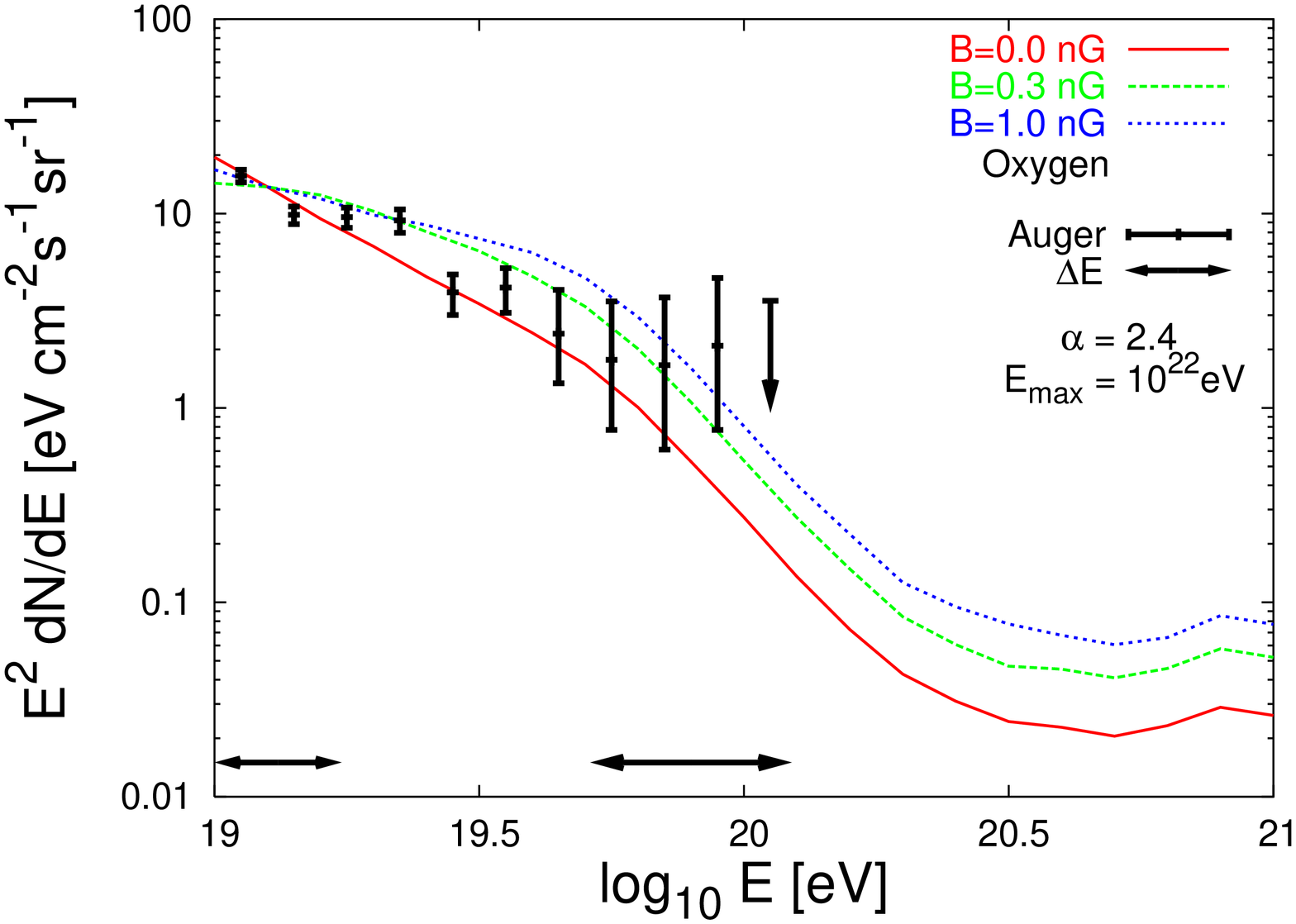}
\includegraphics[width=2.2in,angle=-90]{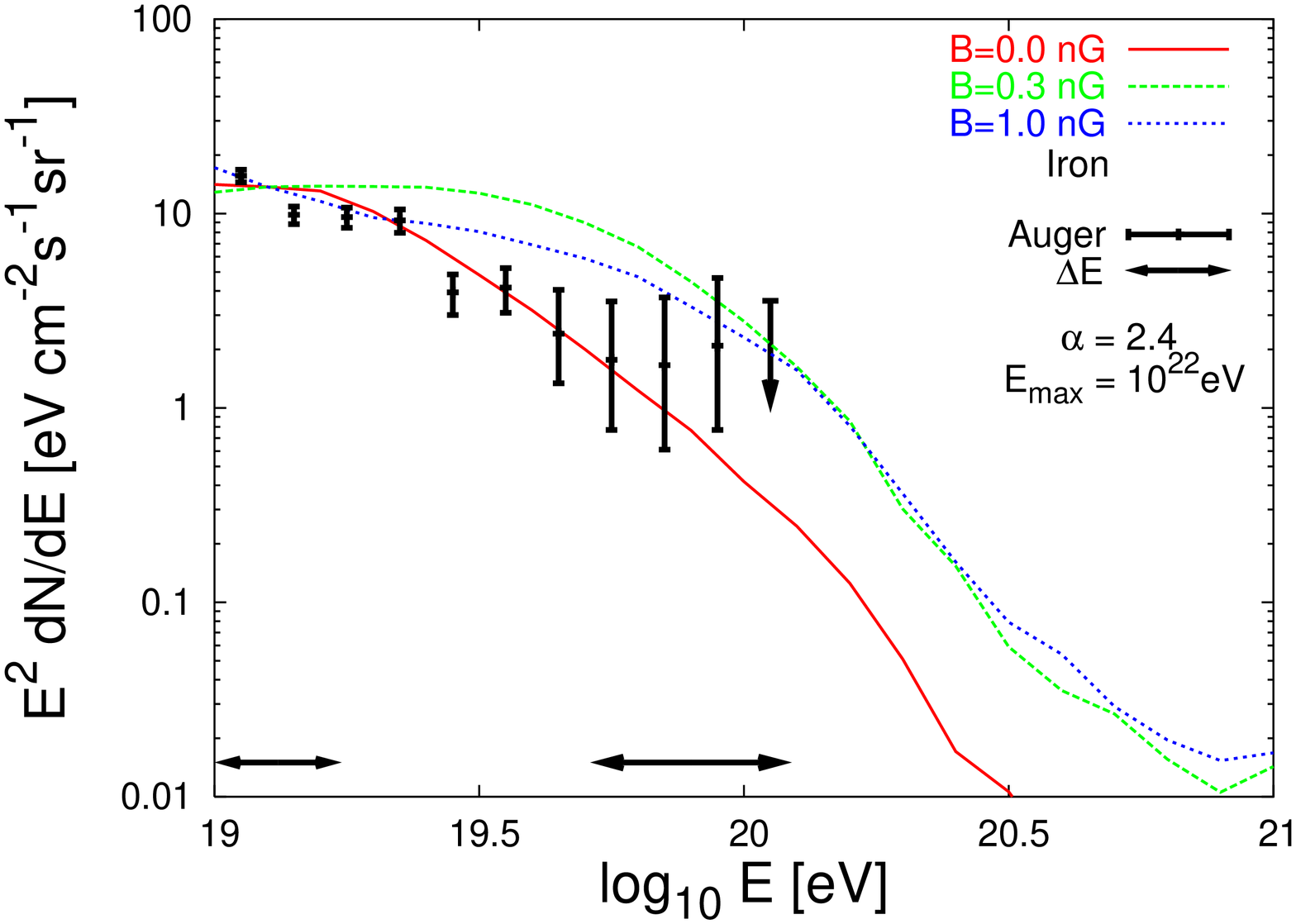}
}
\caption{The effects of nano-Gauss extragalactic magnetic fields on
the UHECR spectrum for oxygen and iron primaries with power-law
spectral index $\alpha$ = 2.4 and $E_\mathrm{max}$=10$^{22}$ eV,
assuming $L_{\rm{coh}}\sim$ 1 Mpc. The overall flux has in each case
been normalized to the Auger data \cite{Sommers:2005vs}. The Malkan \&
Stecker CIB model \cite{Malkan:2000gu} and the Lorentzian model
\cite{Khan:2004nd} for photodisintegration cross-sections have been
used.}
\label{mag}
\end{figure}

In Figure~\ref{mag} we show the effects of such extragalactic magnetic
fields on the UHECR spectrum. For oxygen primaries, the effects are
small, only becoming of any consequence at energies below a few times
$10^{19}$ eV. However the effects are more prominent for iron
primaries.

Some words of caution are called for at this point. The effects of
nG-scale magnetic fields appear to set in at an energy of roughly
$5\times 10^{19}\,\rm{eV}$ for oxygen, whose primaries can arrive from
approximately $300$ Mpc at this energy, corresponding to an effective
length of roughly $L_{\rm{eff}}/L \sim 1+0.7\times
(B/\rm{nG})^{2}$. This means that the small angle treatment is valid
for magnetic fields of strength upto $\sim0.3$~nG. On the other hand,
an iron nucleus with an energy of $5 \times 10^{19}$ eV could have
traveled hundreds or even thousands of Mpc before its arrival at
Earth, so would have been deflected by an angle of $\theta \sim
130^{\circ} \times (L/100 \, \rm{Mpc})^{0.5} \times (B/nG)$. Hence the
result in the right frame of Figure~\ref{mag} is not reliable at low
energies; to properly take into account such strong deflections, a
numerical simulation of diffusion including an appropriate description
of the magnetic field structure in the local supercluster is
required. Such a treatment is however beyond the scope of the present
study.

%%%%%%%%%%%%%%%%%%%%%%%%%%%%%%%%%%%%%
%%%%%%%%%%%%%%%%%%%%%%%%%%%%%%%%%%%%%%%%%%%%%%%%%%

\section{Mixed Ultra-High Energy Cosmic Ray Composition}
\label{mixedsec}

We now study the possibility that a mixture of protons, helium,
oxygen, and iron nuclei are injected at source in roughly the same
proportions as those observed in low-energy galactic cosmic rays
\cite{gal}. In particularl we will consider a mixture in the ratio of
H~:~He~:~O~:~Fe = 1~:~0.85~:~0.06~:~0.02 for a power-law spectral
index $\alpha=2.0$, and H~:~He~:~O~:~Fe = 1~:~1.48~:~0.19~:~0.09 for
the case of $\alpha=2.4$.

The injection spectrum is assumed to have a cutoff modelled as:
\begin{equation}
\frac{dN}{dE}\propto E^{-\alpha}
\exp\bigg[{-\bigg(\frac{E}{E_\mathrm{max}}\bigg)\bigg(\frac{26}{Z}\bigg)\bigg]}
\end{equation}
with the cutoff energy in the range: $10^{20.5}\, \rm{eV}\, \le
E_\mathrm{max} \le 10^{22}$eV. Once again, the choice of the spectrum
is motivated by the possibility that cosmic ray sources can accelerate
charged particles to a maximum energy proportional to the charge. This
leads to the expectation that such sources will preferentially
accelerate heavy nuclei to the highest energies, if such particles are
indeed able to survive the radiation fields present in the sources.

%%%%%%%%%%%%%%%%%%%%%%%%%%%%%%%%%%%%%
\begin{figure}[t]
\centering\leavevmode \mbox{
\includegraphics[width=2.2in,angle=-90]{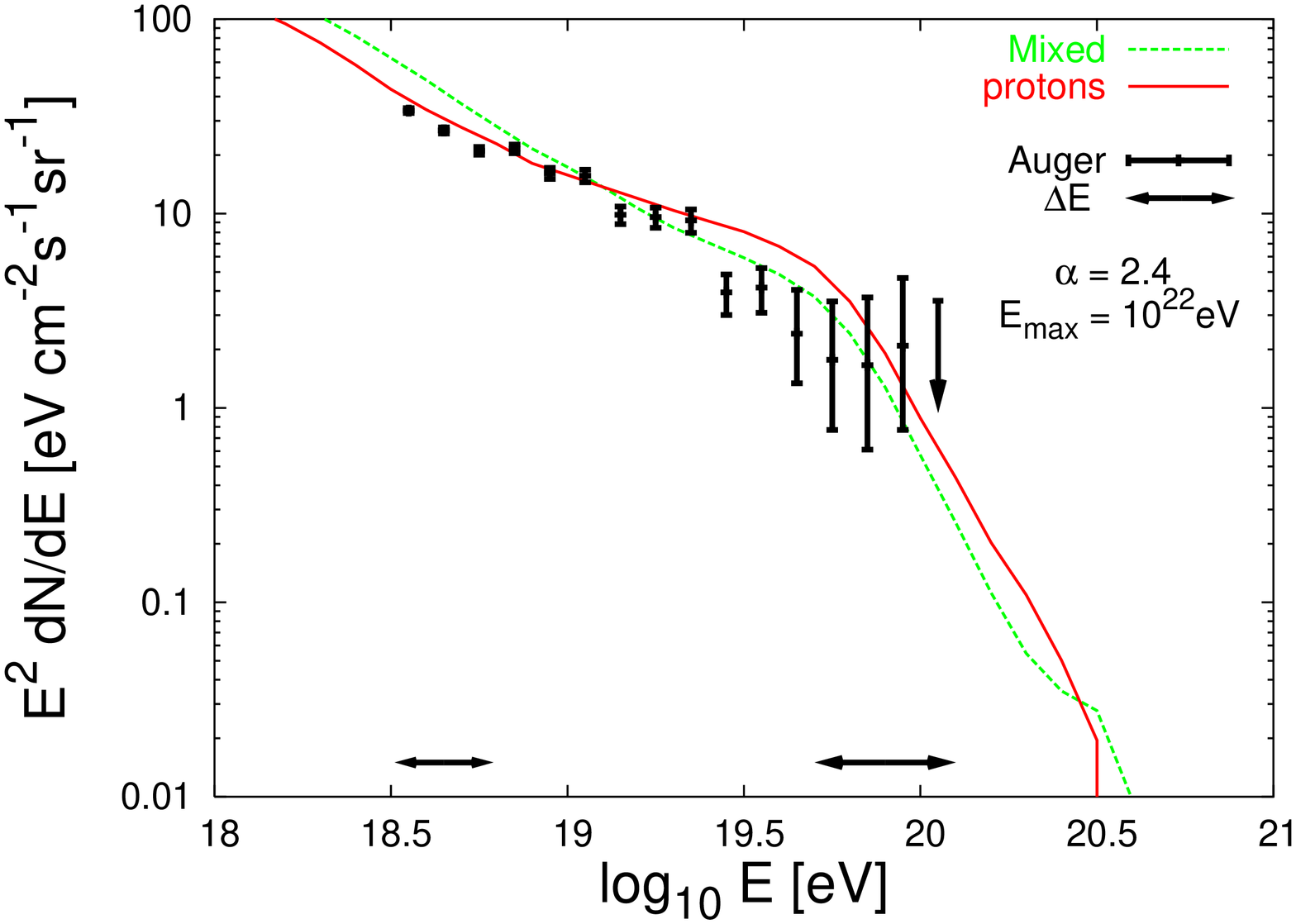}
\includegraphics[width=2.2in,angle=-90]{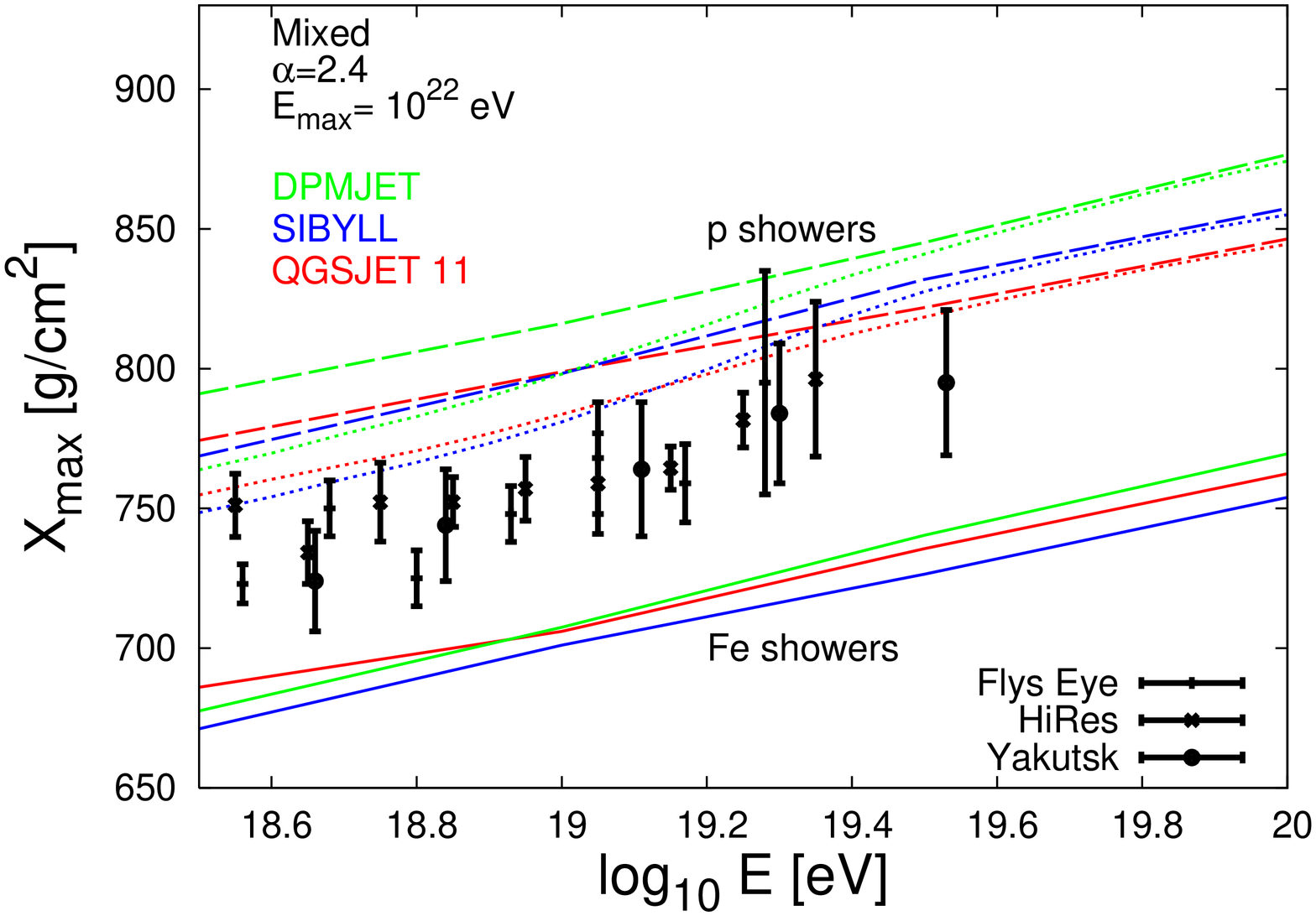}}

\mbox{
\includegraphics[width=2.2in,angle=-90]{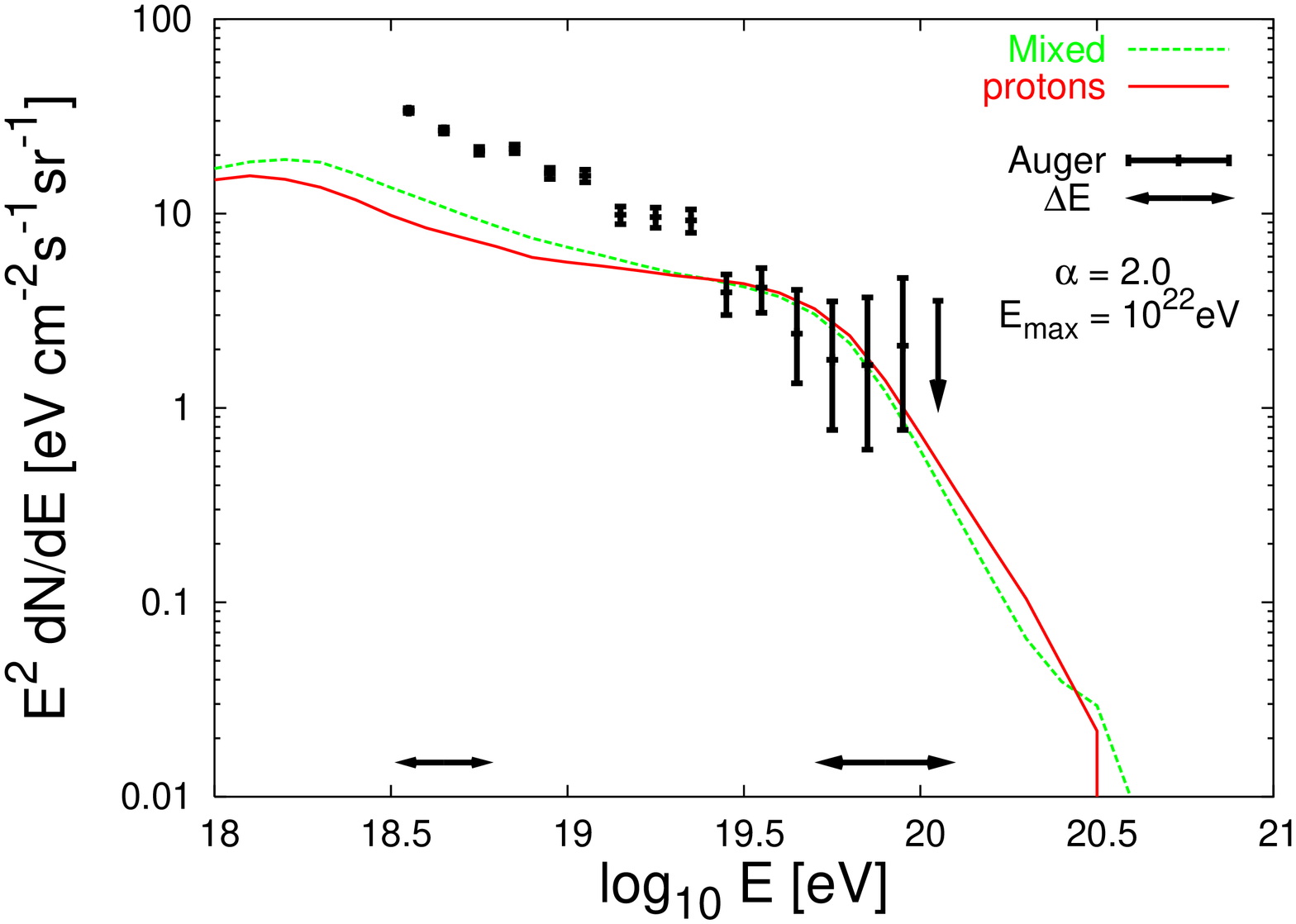}
\includegraphics[width=2.2in,angle=-90]{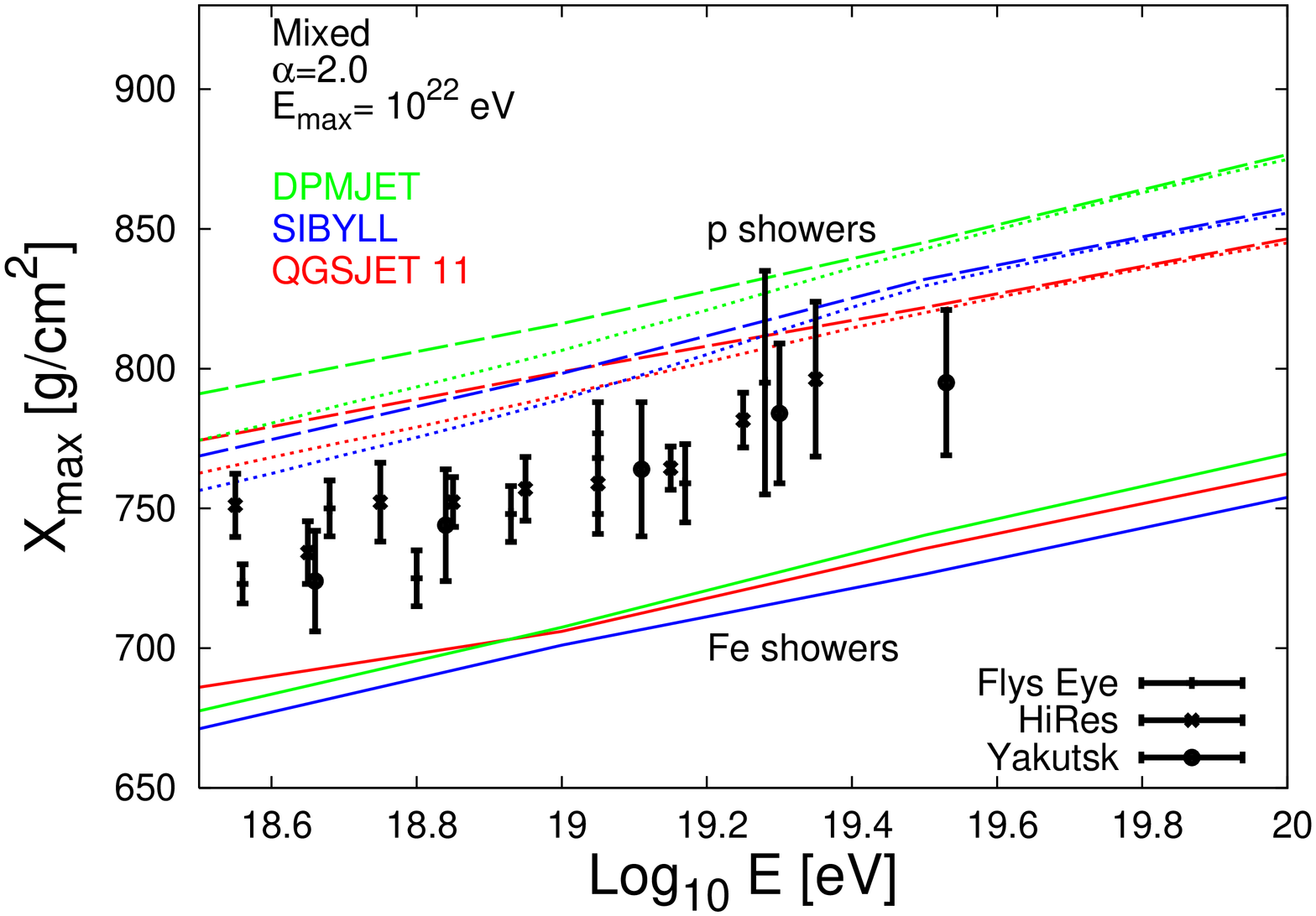}}
\caption{The spectrum (left) and depth of shower maximum,
$X_\mathrm{max}$, (right) for a mixture of protons, helium, oxygen,
and iron injected at source with proportions following that of the
Galactic cosmic rays. The solid, dashed, and dotted lines are the
predicted values of $X_\mathrm{max}$ for, respectively, a pure iron,
pure proton, and the mixed composition, obtained using three
simulation programmes --- DPMJET, SIBYLL and QGSJET. The overall flux
has in each case been normalized to the Auger data
\cite{Sommers:2005vs}. The $X_\mathrm{max}$ data are from the Fly's
Eye \cite{Bird:1993yi}, HiRes \cite{Abbasi:2004nz} and Yakutsk
\cite{yakutsk} experiments. The Malkan \& Stecker CIB model
\cite{Malkan:2000gu} and the Lorentzian model \cite{Khan:2004nd} for
photodisintegration cross-sections have been used.}
\label{mixed}
\end{figure}
%%%%%%%%%%%%%%%%%%%%%%%%%%%%%%%%%%%%%%%%%%%%%%%%%%

In the left frames of Figure~\ref{mixed}, we plot the propagated
spectrum calculated using this mixed composition at source, compared
to that for pure proton injection. It is seen that the two cases are
in fact hard to distinguish observationally.

In the right frames of Figure~\ref{mixed}, we plot the average value
of $X_\mathrm{max}$ as a function of the energy for the same mixed
composition at source. To illustrate the uncertainties in relating
$X_\mathrm{max}$ to the UHECR composition, we have shown the results
of three air shower simulation programmes: DPMJET
\cite{Kalmykov:1997te}, SIBYLL \cite{sibyll}, and QGSJET
\cite{Kalmykov:1997te}. Also shown in Figure~\ref{mixed} are the
measurements from the Fly's Eye \cite{Bird:1993yi}, HiRes
\cite{Abbasi:2004nz}, and Yakutsk \cite{yakutsk} experiments. (The
$X_{\mathrm{max}}$ values quoted in the Auger analysis which sets an
upper bound to the photon content in UHECRs \cite{Risse:2005hi} are
explicitly stated to be inappropriate for elongation rate studies.)
The data favour a composition heavier than protons, regardless of
which simulation program is adopted.

\section{Predictions for $X_\mathrm{max}$ Measurements}
\label{X_{max}}

Information on the UHECR composition at Earth can be obtained from
studies of cosmic ray shower development, in particular the lateral
distribution, muon content, and $X_\mathrm{max}$ value
\cite{Watson:2004ew}. In this Section, we will focus on the parameter
$X_\mathrm{max}$, which is the atmospheric depth at which 
the number of particles in the shower is largest. Measurements of
$X_\mathrm{max}$ are made by fluorescence detector experiments which
measure the UV light emitted by the excited atoms (mainly nitrogen) in
the air shower.

\begin{figure}[!]
\centering\leavevmode
\mbox{
\includegraphics[width=2.1in,angle=-90]{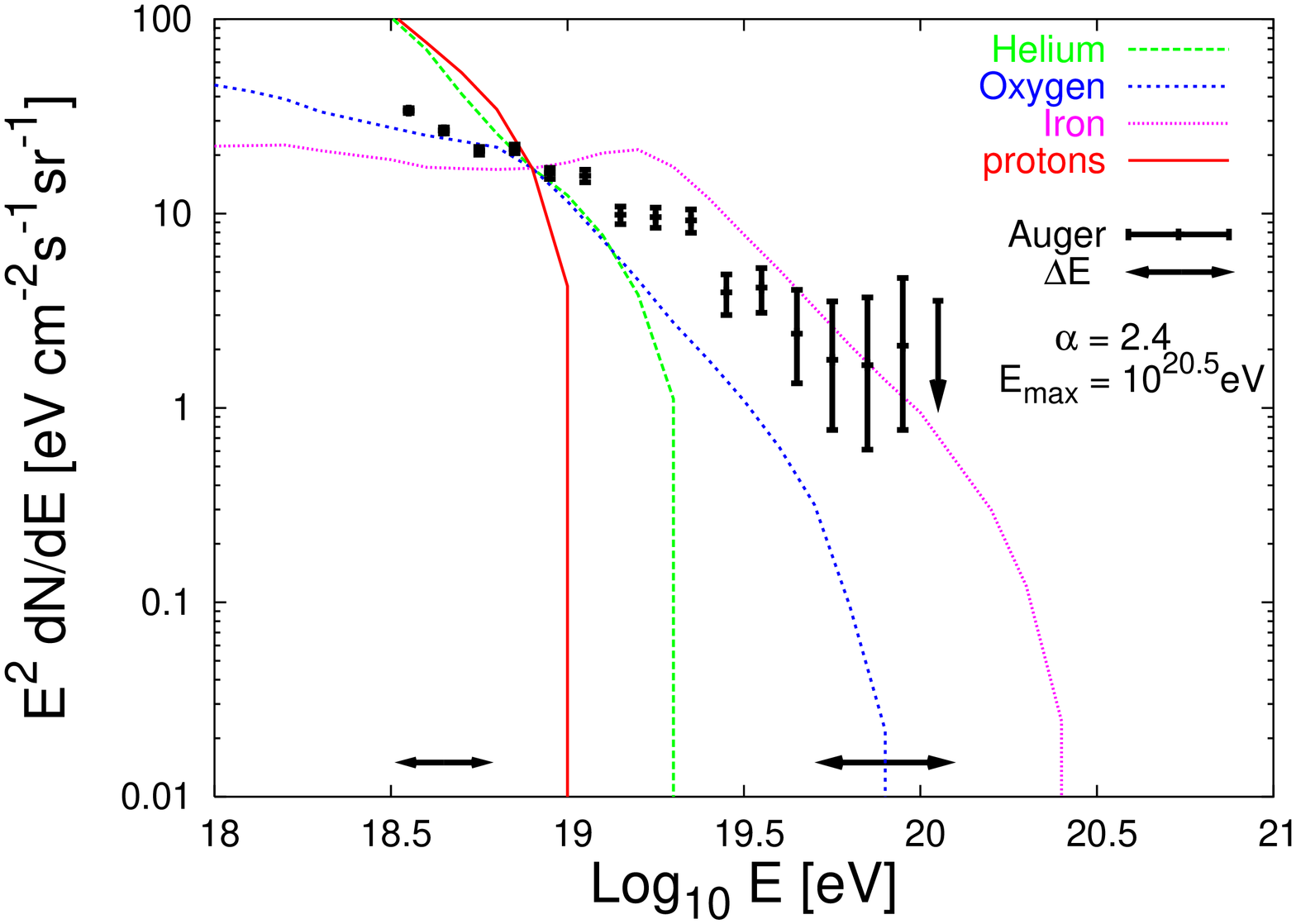}
\includegraphics[width=2.1in,angle=-90]{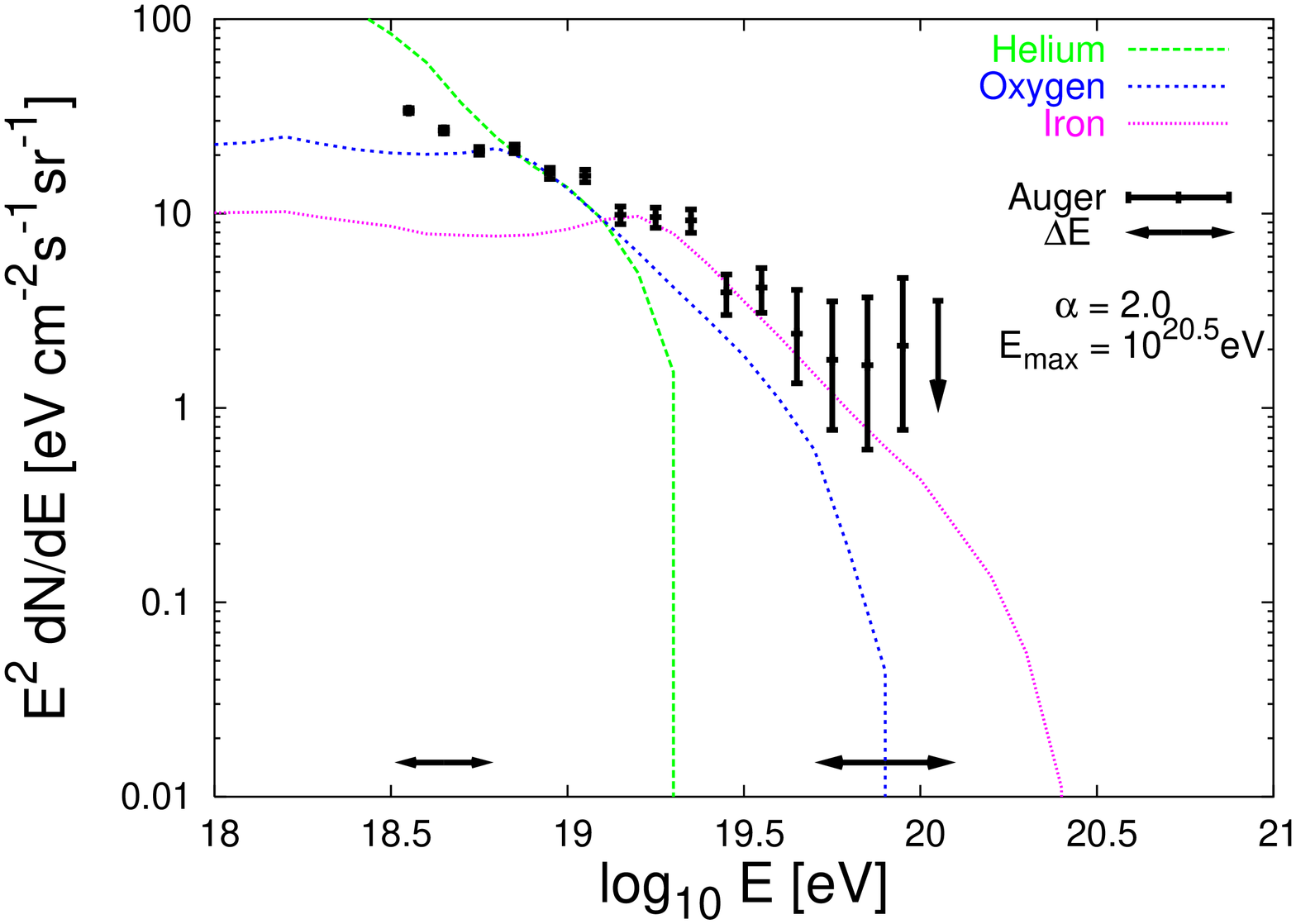}}
\mbox{
\includegraphics[width=2.1in,angle=-90]{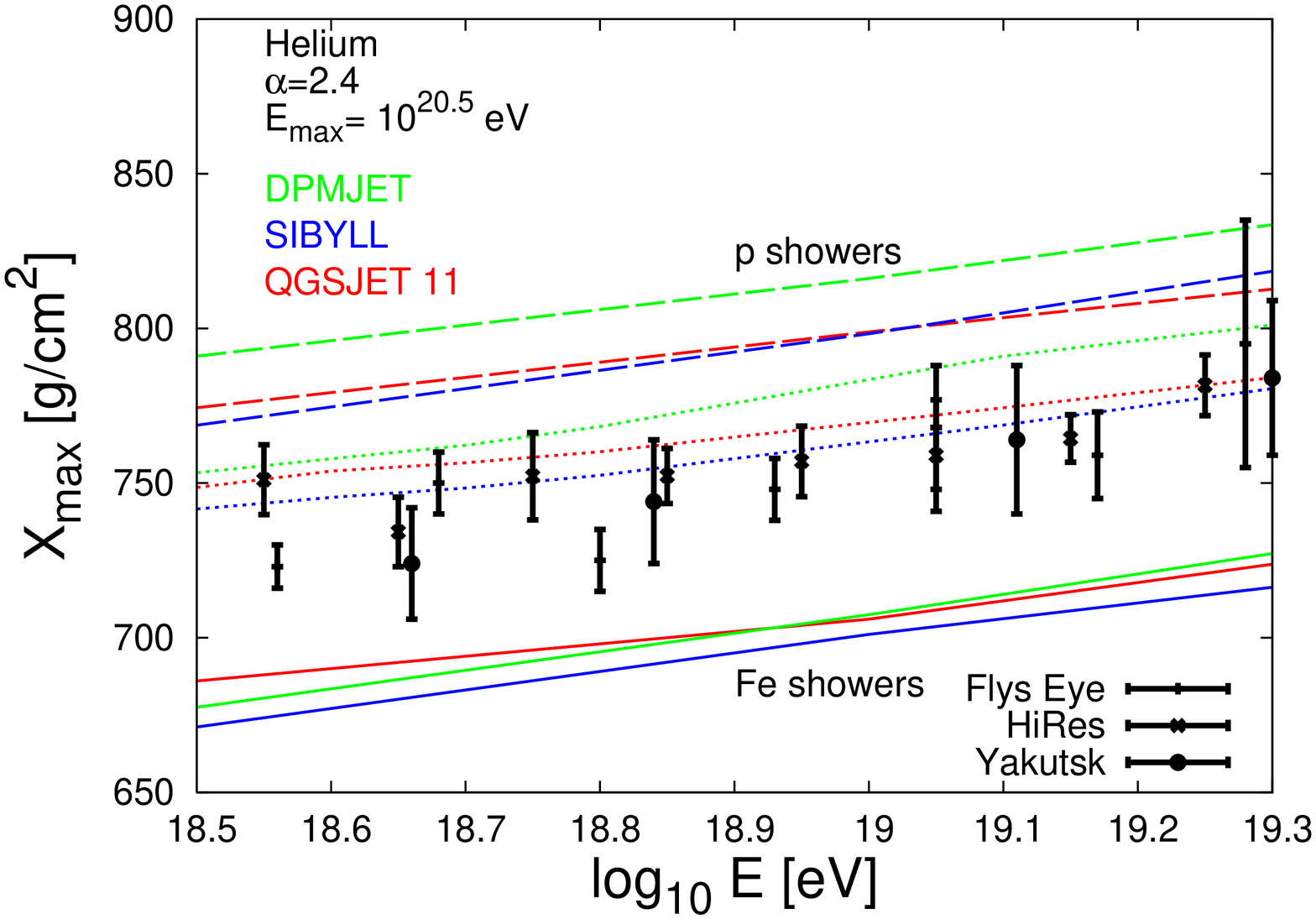}
\includegraphics[width=2.1in,angle=-90]{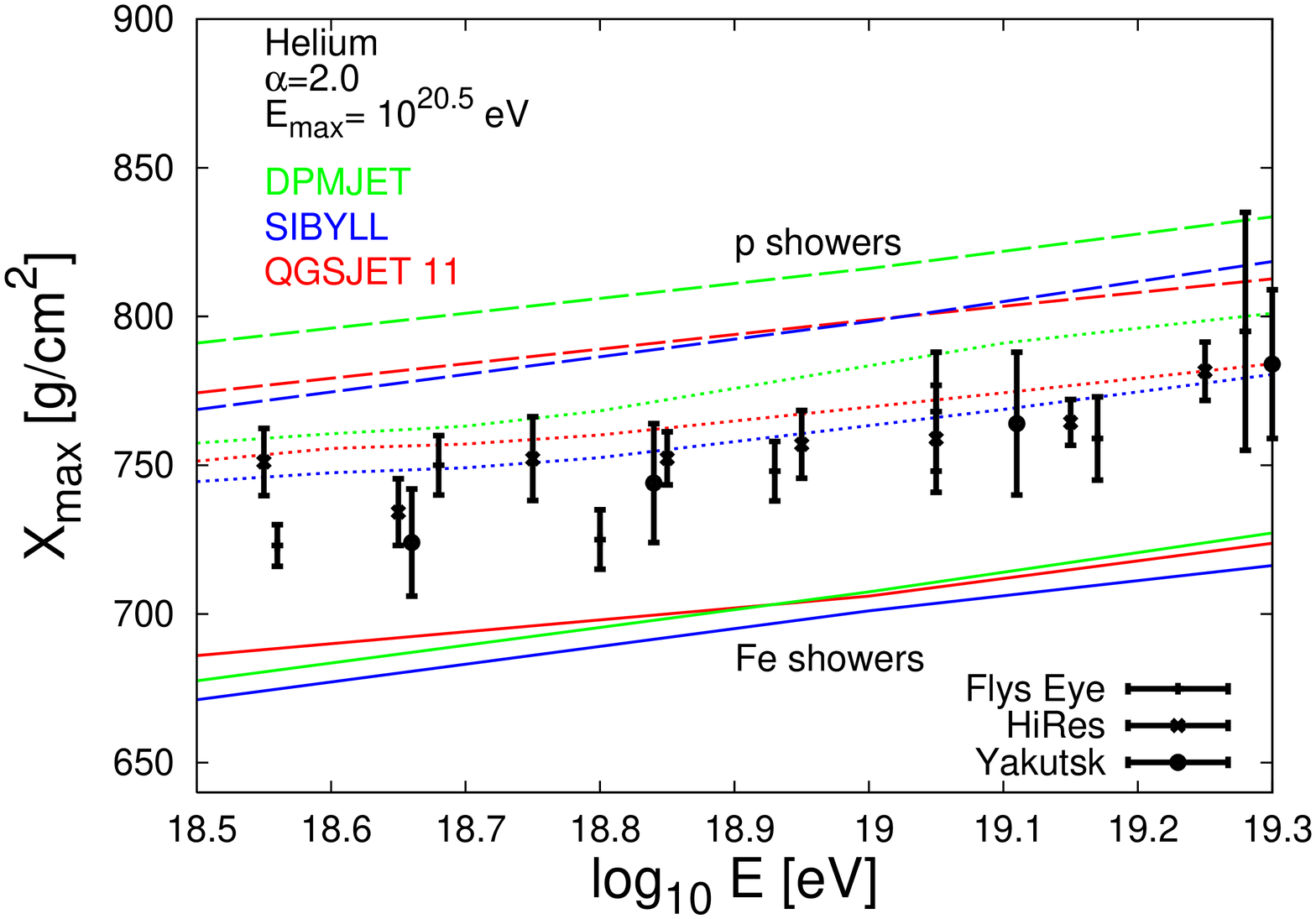}}
\mbox{
\includegraphics[width=2.1in,angle=-90]{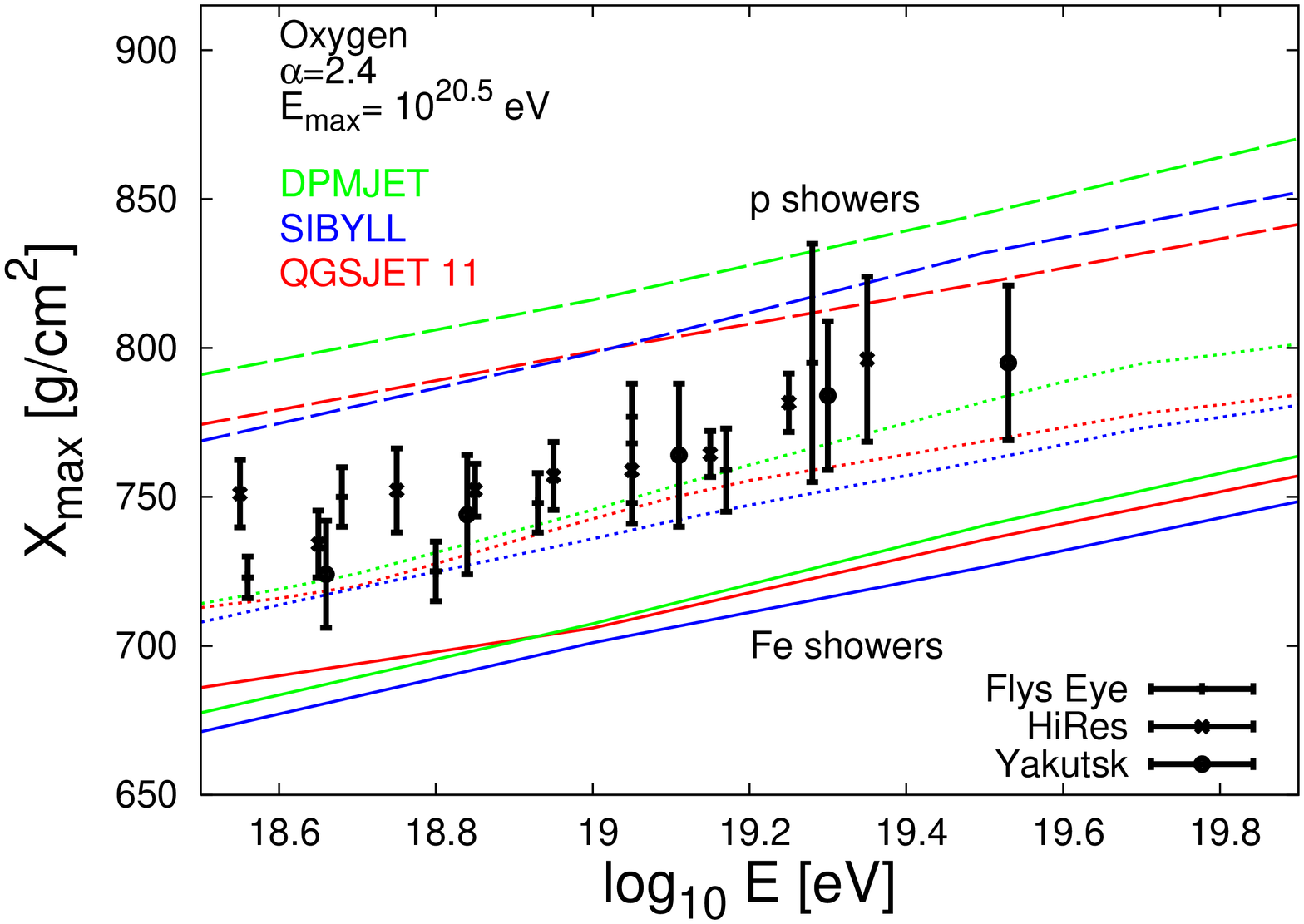}
\includegraphics[width=2.1in,angle=-90]{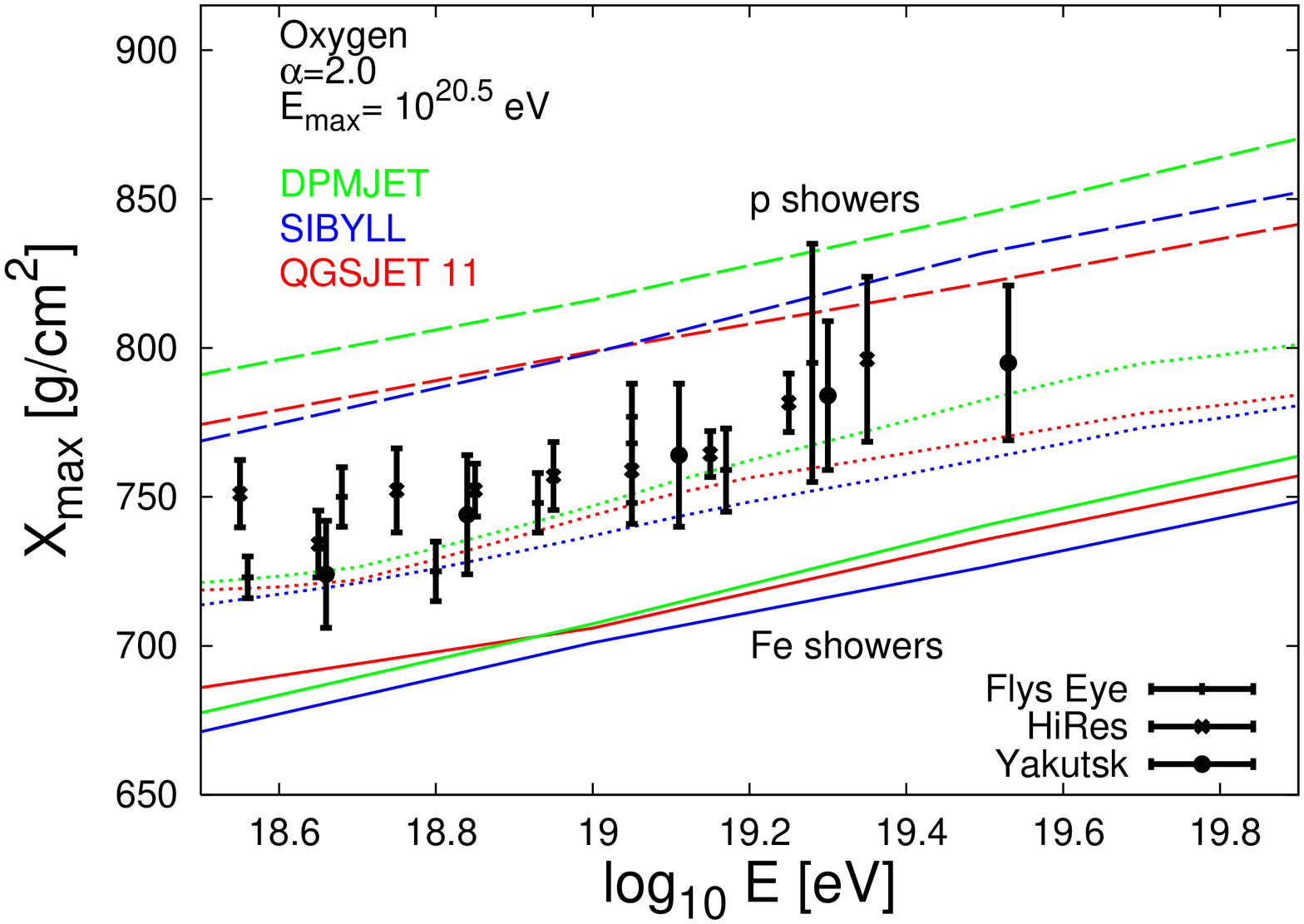}}
\mbox{
\includegraphics[width=2.1in,angle=-90]{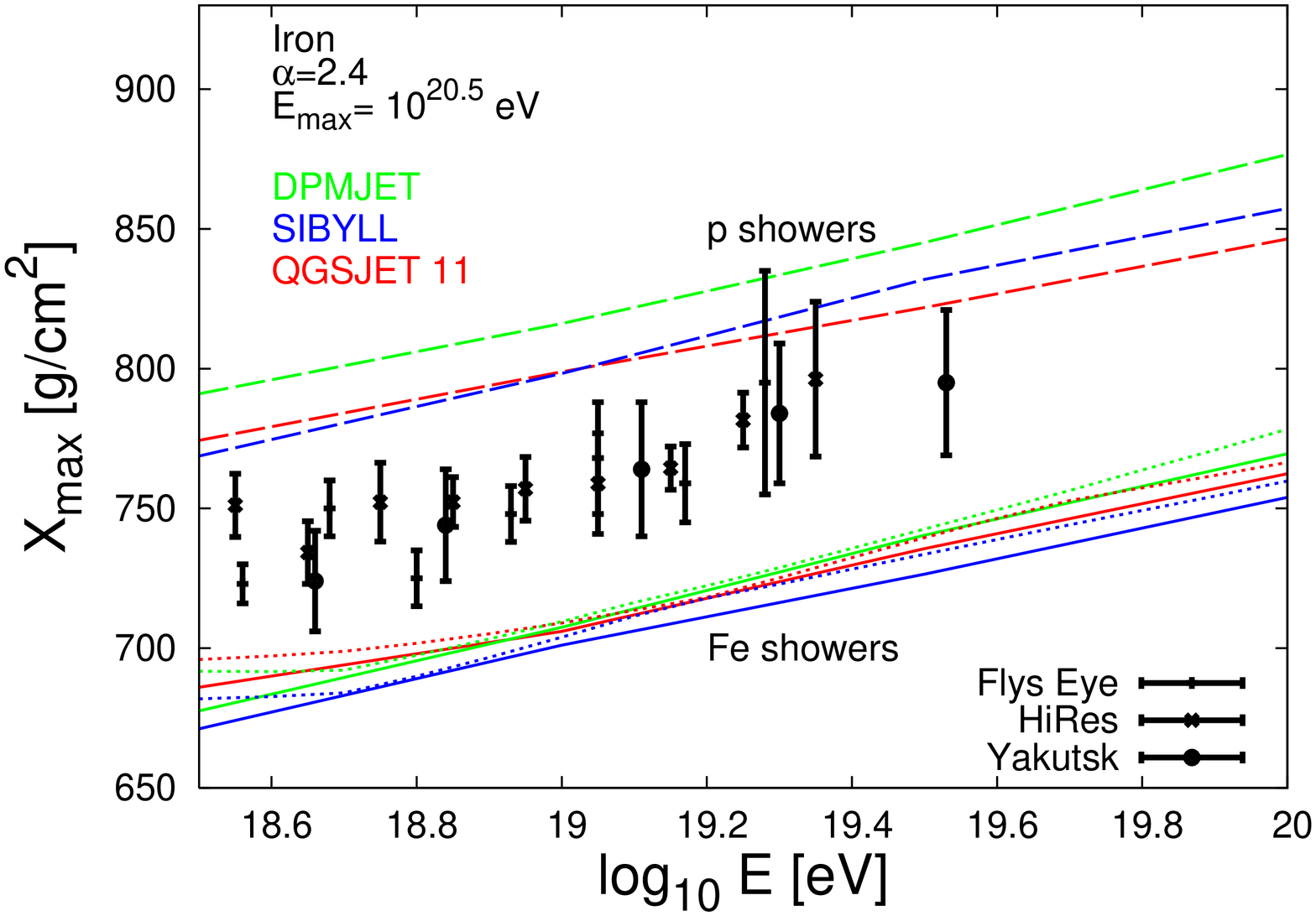}
\includegraphics[width=2.1in,angle=-90]{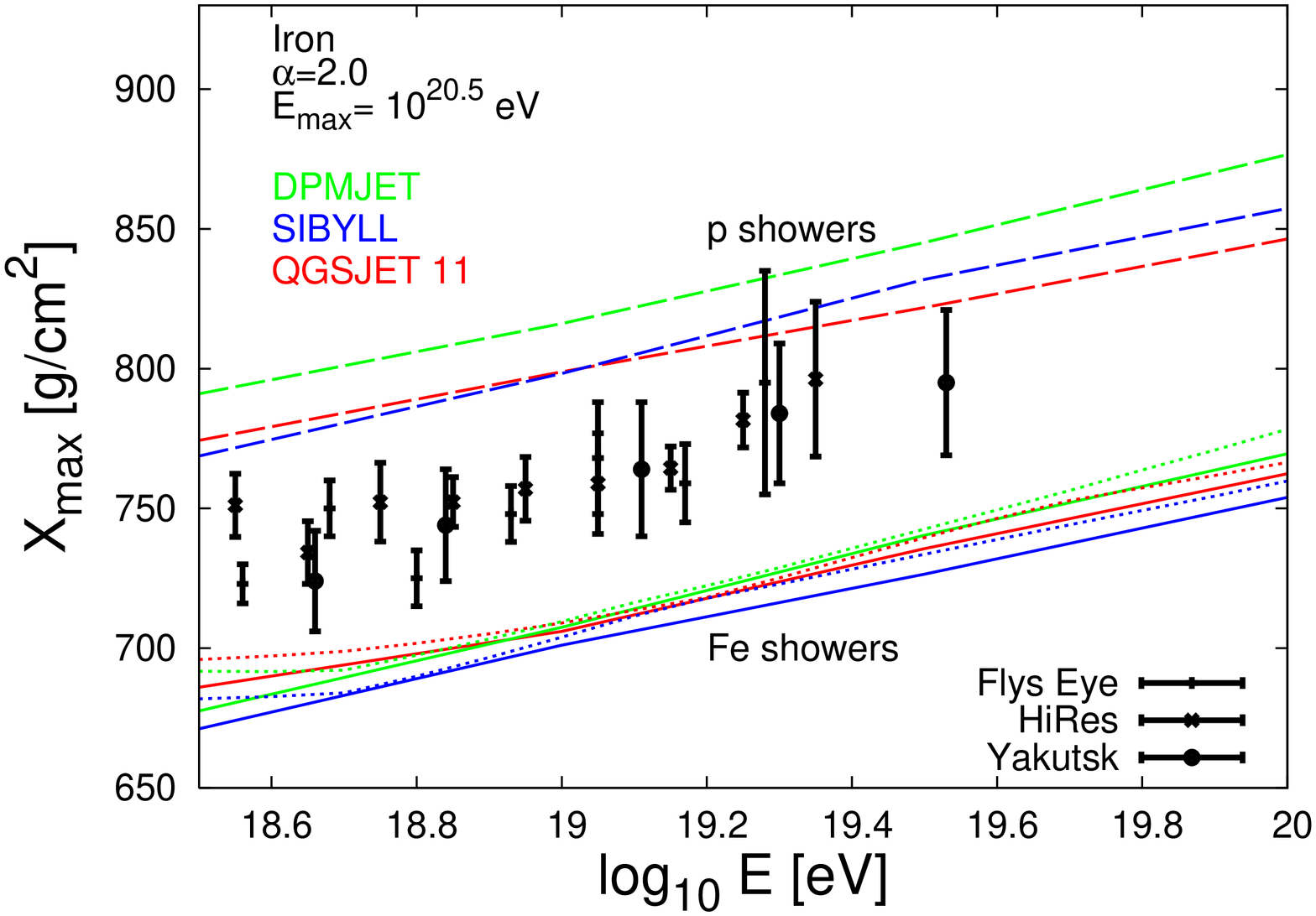}}
\caption{The expected spectrum and X$_\mathrm{max}$ for helium,
oxygen, and iron nuclei injected at source for
E$_\mathrm{max}$=10$^{20.5}$eV with $\alpha=2.4$ and 2.0. The overall
flux has in each case been normalized to the Auger data
\cite{Sommers:2005vs}. The $X_\mathrm{max}$ data are from the Fly's
Eye \cite{Bird:1993yi}, HiRes \cite{Abbasi:2004nz} and Yakutsk
\cite{yakutsk} experiments. The Malkan \& Stecker CIB model
\cite{Malkan:2000gu} and the Lorentzian model \cite{Khan:2004nd} for
photodisintegration cross-sections have been used.}
\label{Ec=20.5_alph=2.4}
\end{figure}

In the left frames of Figure~\ref{Ec=20.5_alph=2.4}, we show the UHECR
spectrum and values of $X_\mathrm{max}$ calculated for iron, oxygen
and helium nuclei injected at source, with a power-law spectral slope
of $\alpha=2.4$ and $E_\mathrm{max}=10^{20.5}$ eV. With this rather
low choice of the cutoff energy, heavy nuclei dominate the spectrum
above $\sim 10^{19}$ eV --- iron and oxygen nuclei break up rather
modestly during propagation, so the composition at Earth is similar to
that at source. In the right frames of Figure~\ref{Ec=20.5_alph=2.4},
we repeat the calculations for the case of $\alpha=2.0$ and find
similar results..

For the case of a low cutoff energy, we expect a transition from a
heavy composition at high energies to a much lighter composition at
low energies, largely independently of the ratio of species present
in the source environment. From the spectra shown in
Figure~\ref{Ec=20.5_alph=2.4}, we would expect a nearly all-iron
composition at the highest energies, and a nearly all proton/helium
composition below $\sim 10^{19}$ eV, with the value of
$X_\mathrm{max}$ changing over this range accordingly. The current
data on $X_\mathrm{max}$ is too scattered to establish whether such a
transition is present.

\begin{figure}[!]
\centering\leavevmode
\mbox{
\includegraphics[width=2.1in,angle=-90]{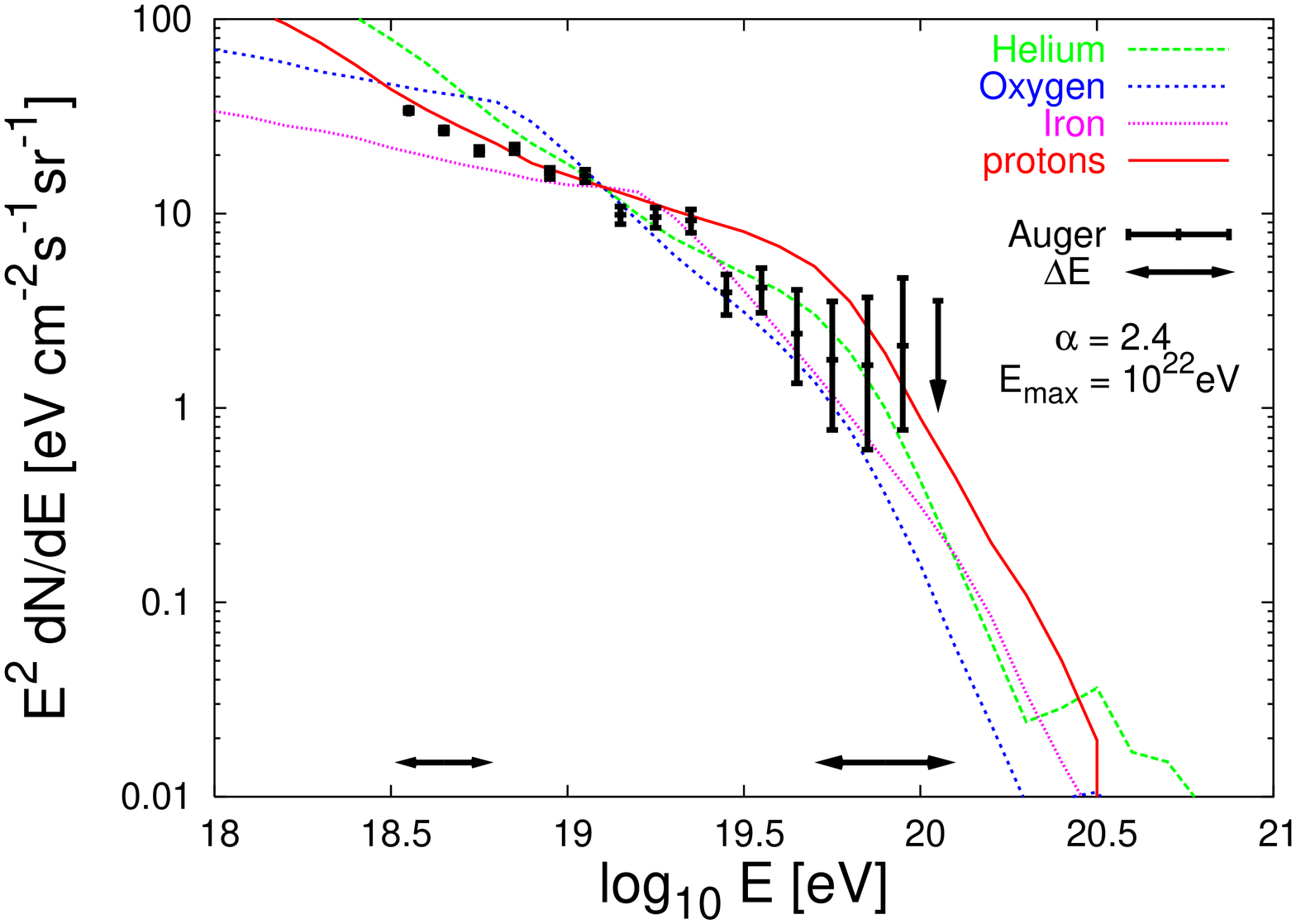}
\includegraphics[width=2.1in,angle=-90]{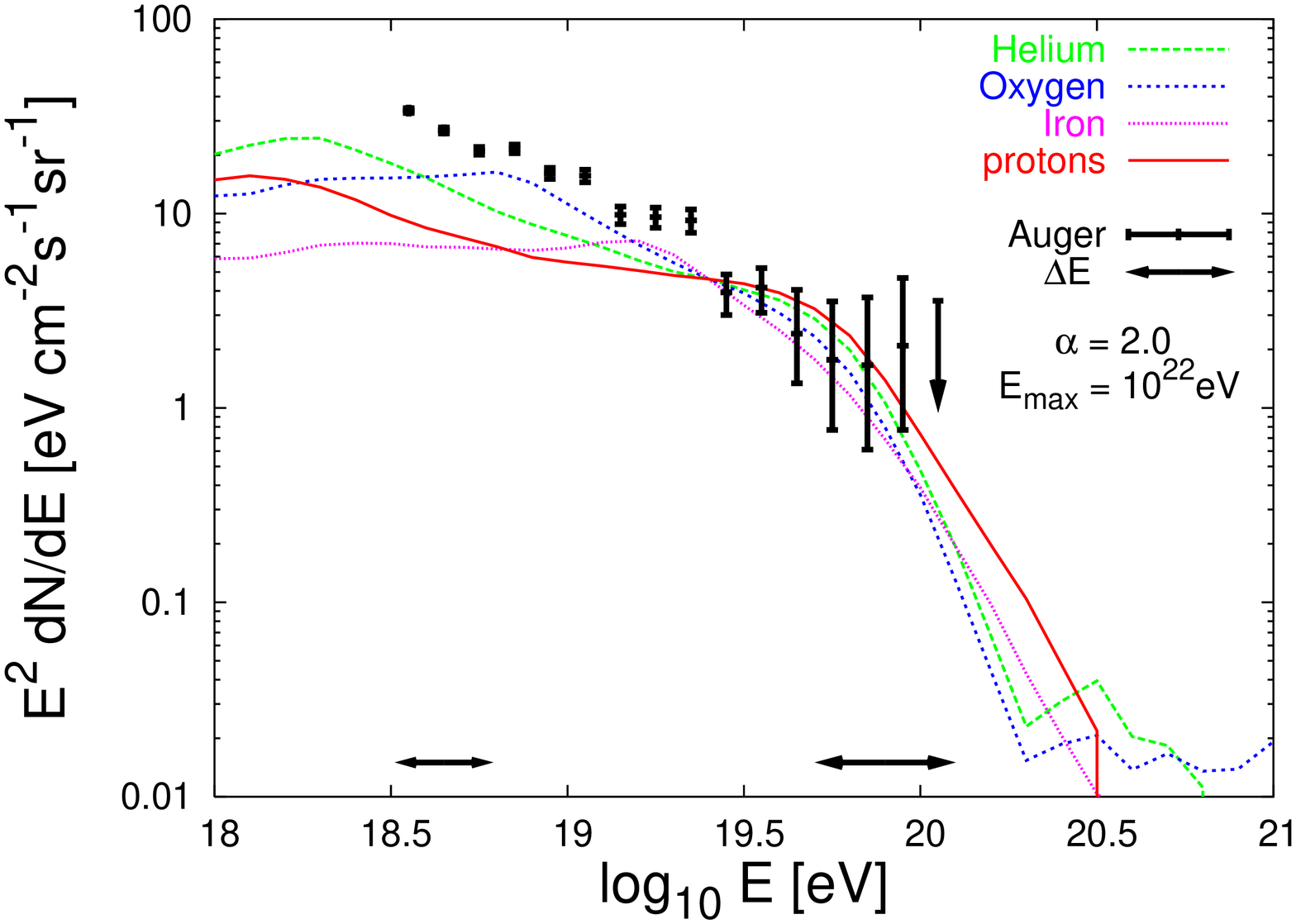}}
\mbox{
\includegraphics[width=2.1in,angle=-90]{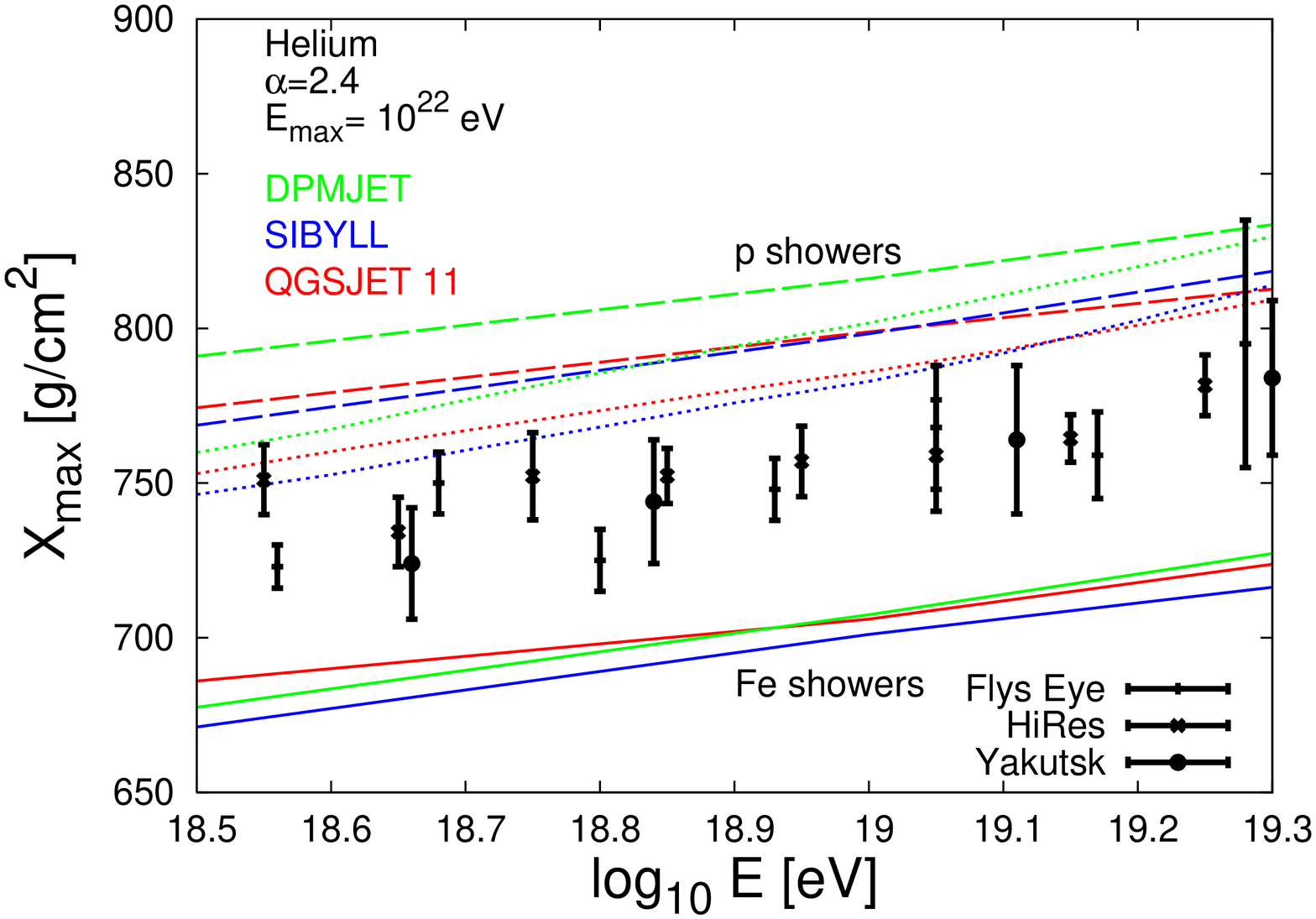}
\includegraphics[width=2.1in,angle=-90]{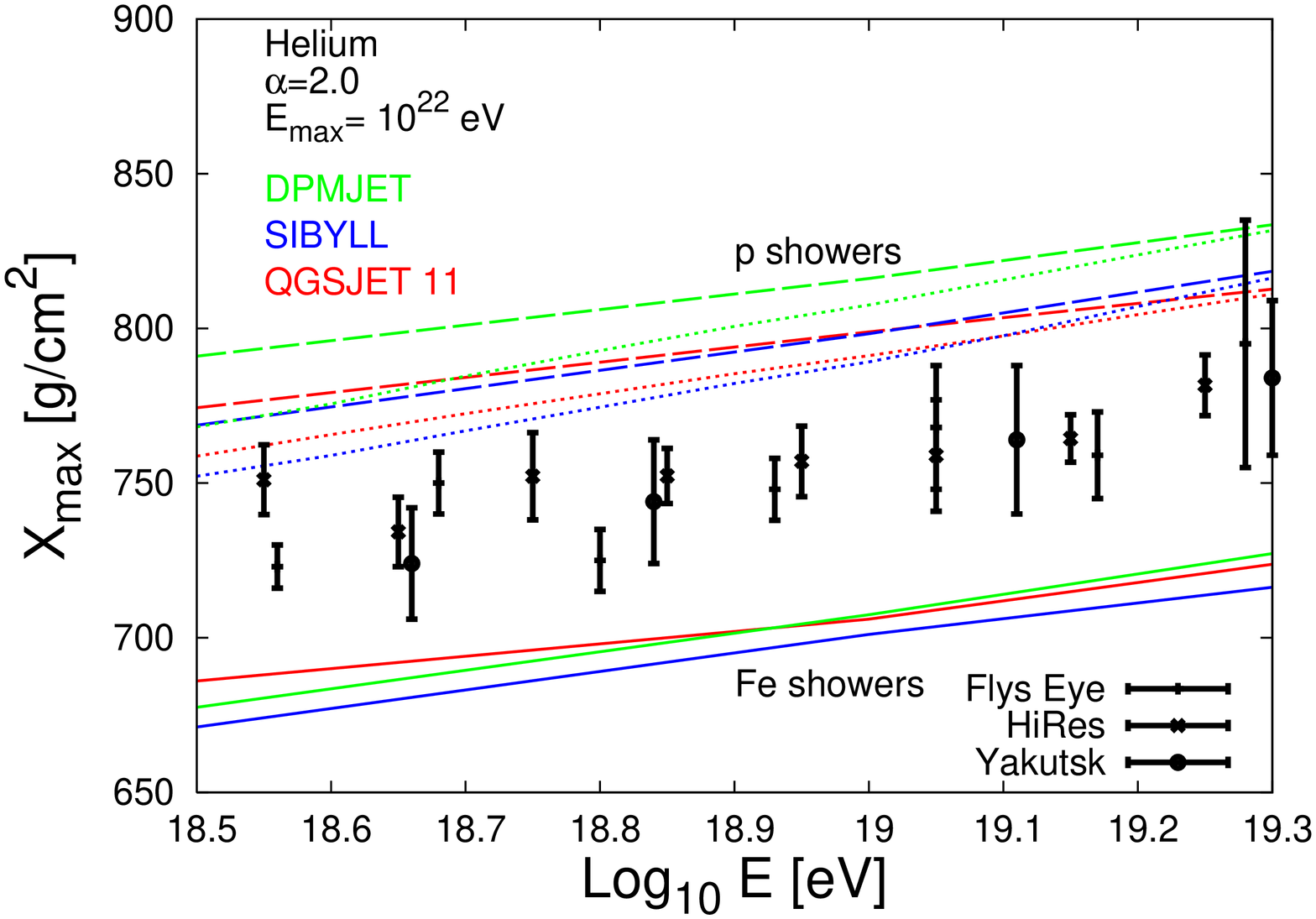}}
\mbox{
\includegraphics[width=2.1in,angle=-90]{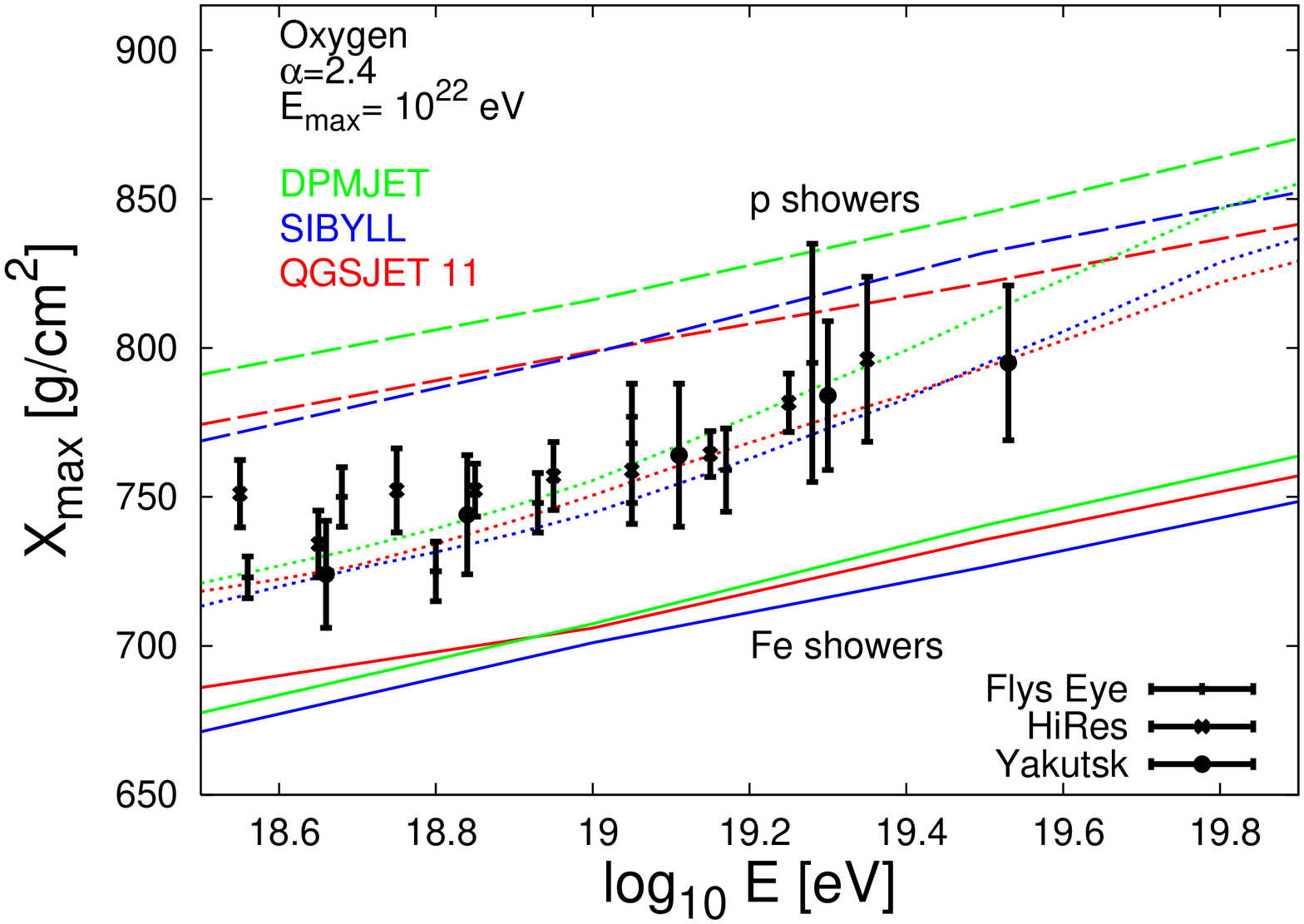}
\includegraphics[width=2.1in,angle=-90]{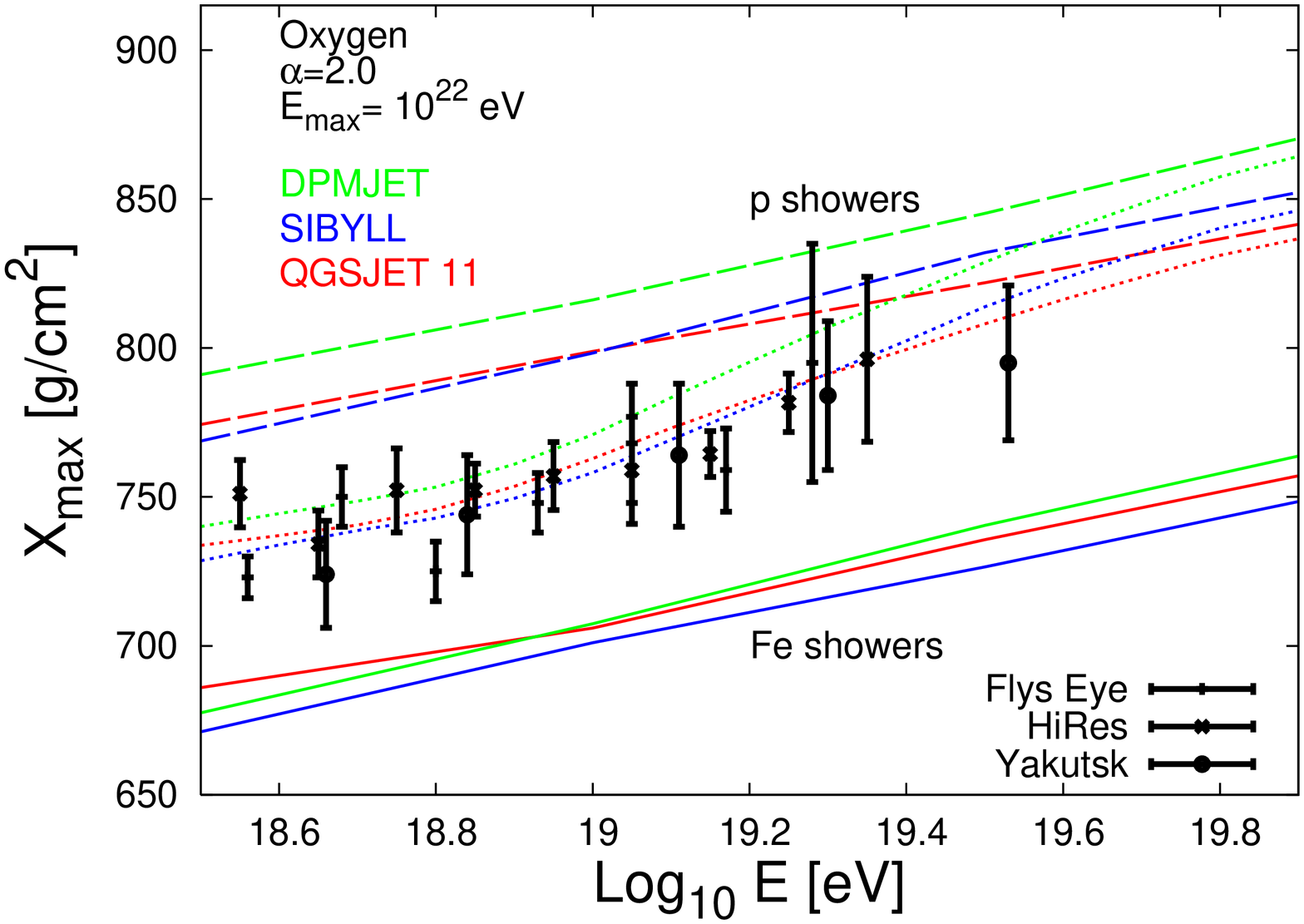}}
\mbox{
\includegraphics[width=2.1in,angle=-90]{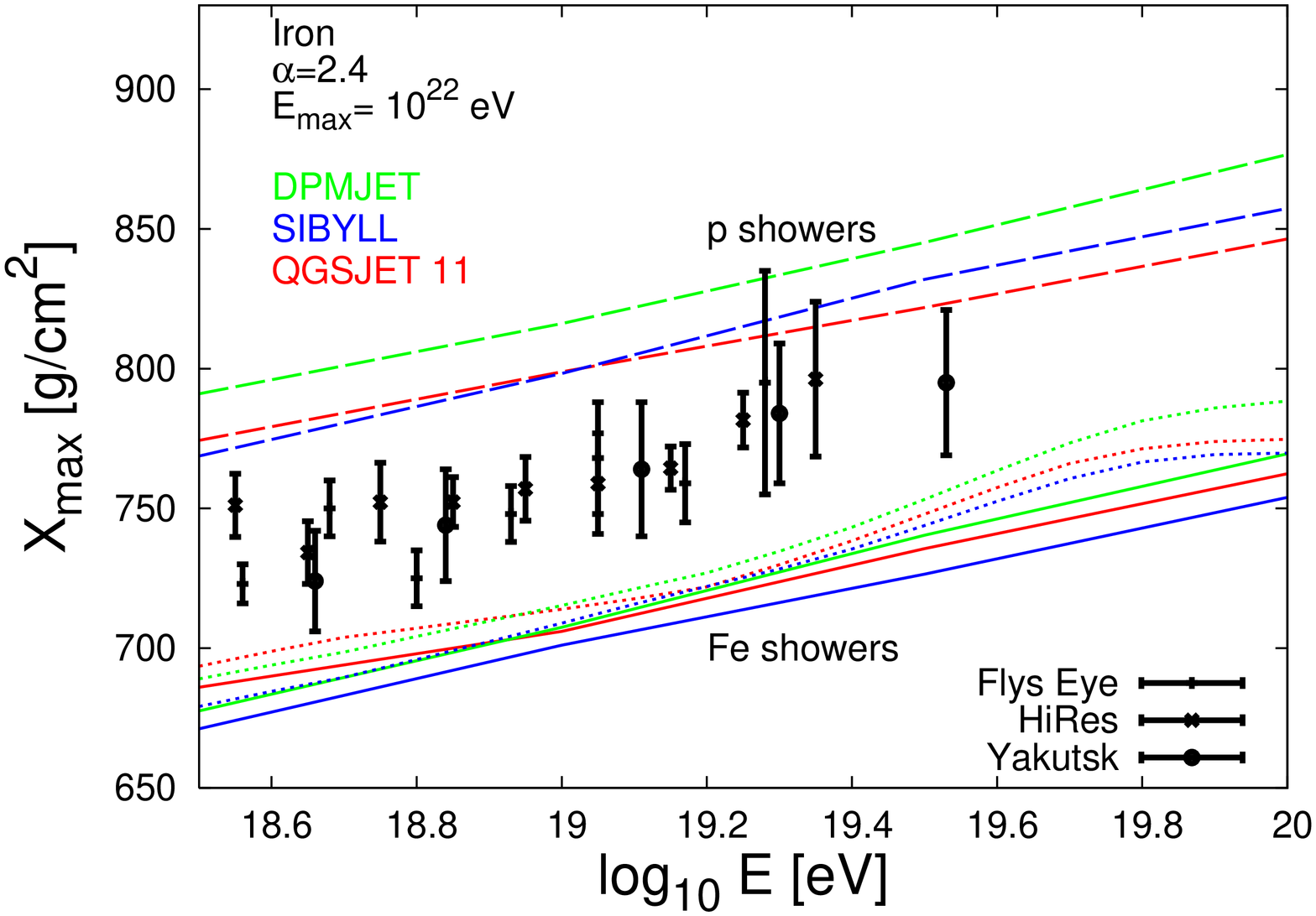}
\includegraphics[width=2.1in,angle=-90]{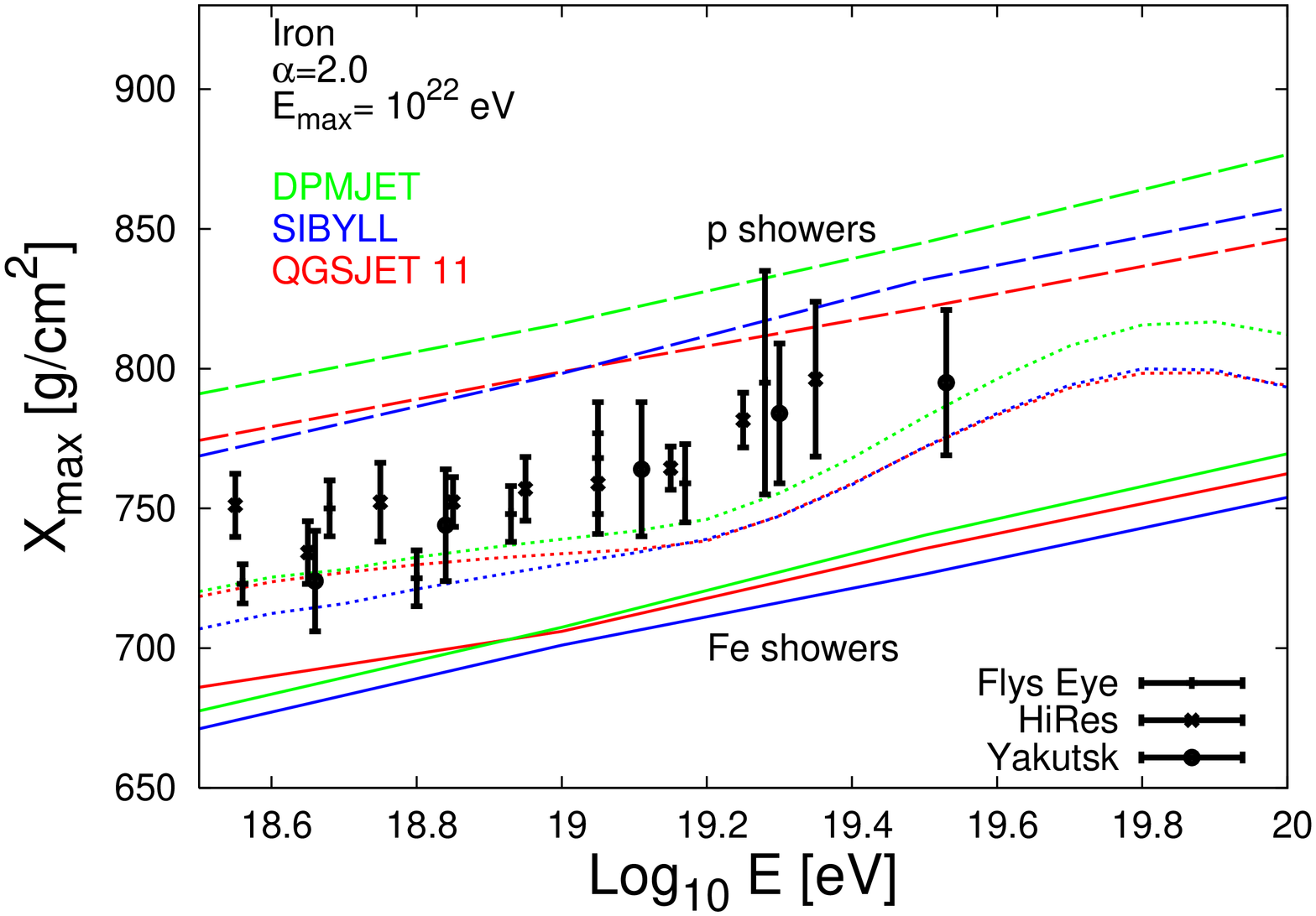}}
\caption{The expected spectrum and X$_\mathrm{max}$ for helium,
oxygen, and iron nuclei injected at source for
E$_\mathrm{max}$=10$^{22}$eV with $\alpha=2.4$ and 2.0. The overall
flux has in each case been normalized to the Auger data
\cite{Sommers:2005vs}. The $X_\mathrm{max}$ data shown are from the
Fly's Eye \cite{Bird:1993yi}, HiRes \cite{Abbasi:2004nz} and Yakutsk
\cite{yakutsk} experiments.  The Malkan \& Stecker CIB model
\cite{Malkan:2000gu} and the Lorentzian model \cite{Khan:2004nd} for
photodisintegration cross-sections have been used.}
\label{Ec=22_alph=2.4}
\end{figure}

This conclusion can be modified if the accelerators of UHECRs operate
up to a higher cutoff energy so that the heavy nuclei injected at
source are substantially photodisintegrated by the time they reach
Earth. In Figure~\ref{Ec=22_alph=2.4}, we show that for the case
$E_\mathrm{max}=10^{22}$ eV, it is not at all clear which species will
dominant the observed UHECRs --- the data is equally well fitted by an
all-proton, all-iron or mixed composition at source. What is
interesting is that the $X_\mathrm{max}$ data is reasonable well
fitted when only oxygen nuclei are accelerated by the sources. Even
pure iron injection at source matches the data for an injection
spectrum with $\alpha = 2$.

%%%%%%%%%%%%%%%%%%%%%%%%%%%%%%%%%%%%%%%%%%%%%%%%%%%

%%%%%%%%%%%%%%%%%%%%%%%%%%%%%%%%%%%%%%%%%%%%%%%%%%

%%%%%%%%%%%%%%%%%%%%%%%%%%%%%%%%%%%%%%%%%%%%%%%%%%

\section{Discussion}
\label{conc}

We have studied the intergalactic propagation of a variety of heavy
and intermediate mass cosmic ray nuclei at ultrahigh
energies. Adopting different models for the cosmic infrared background
and for the photodisintegration cross-sections has little effect on
the propagated energy spectrum and composition at Earth. Of more
significance is the choice of the source spectrum and the effect of
intergalactic magnetic fields. Our main aim was to determine the
relationship between cosmic ray composition at source and at Earth
after the effects of propagation are taken into account. Somewhat
surprisingly, extant data on the composition of UHECRs is consistent
with the injection of even pure iron nuclei by the sources.

As the Pierre Auger Observatory continues to accumulate more exposure
to UHECRs, a definitive resolution is expected soon of whether there
is indeed a GZK cutoff in the spectrum. By combining the energy
spectrum obtained using the surface detectors with $X_\mathrm{max}$
measurements by the fluorescence telescopes, information can then be
extracted on the sources of UHECRs using the results presented in the
present paper. Observations of other types of messengers associated
with the highest energy cosmic rays, such as the cosmogenic neutrino
flux, will also help to determine whether the ultra-high energy cosmic
rays are protons or heavy nuclei \cite{nu} amd bring us closer to
answering the long standing mystery of their origin.

\acknowledgements{DH is supported by the US Department of Energy and
by NASA grant NAG5-10842. SS acknowledges a PPARC Senior Research
Fellowship (PPA/C506205/1), and AT acknowledges a PPARC
Studentship. We thank Johannes Knapp for providing results on
$X_\mathrm{max}$ from air shower simulation programmes and Ralph Engel
and Alan Watson for discussions.}

\end{document}